\documentclass[a4paper,
onecolumn, unpublished,
superscriptaddress,11pt]{quantumarticle}

\pdfoutput=1
\usepackage[utf8]{inputenc}
\usepackage[english]{babel}
\usepackage[T1]{fontenc}
\usepackage{amsmath}
\usepackage[colorlinks=true, urlcolor=blue,citecolor=blue,linkcolor=blue]{hyperref}

\usepackage{amsthm}

\usepackage{tikz} 

\usepackage{xcolor}

\makeatletter 
 \hypersetup{pdftitle = {SLDsOfGraphStates},
	     pdfauthor = {Daniel Miller},
	     pdfsubject = {Quantum Information Theory},
	     pdfkeywords = {quantum, correlation, quantum correlation, classical correlation, sector length, distribution, binomial distribution, converge, Pauli, stabilizer, graph state, depolarizing, local white noise, white noise, ring graph, coloring problem, leaf, twin, leaves, twins, adjacency matrix, random graph states, quantum computer, qubits, entangled, entanglement}
	    }
\makeatother

\usepackage{cite}   
\usepackage{braket}
\usepackage{amsmath,amssymb}  
\usepackage{stmaryrd} 
\usetikzlibrary{shapes,arrows,spy,positioning,snakes}

\title{Shor-Laflamme distributions of graph states and noise robustness of entanglement}

\author{Daniel Miller} \orcid{0000-0003-2100-5612} 
\email{d.miller(at)fu-berlin.de}

\affiliation{Freie Universität Berlin, Dahlem Center for Complex Quantum Systems,  14195 Berlin, Germany. }

\affiliation{Universität Basel, Department of Physics, Klingelbergstrasse 82, CH-4056 Basel, Switzerland}

\affiliation{IBM Quantum, IBM Research Europe – Zurich, S\"aumerstrasse 4, CH-8803 R\"uschlikon, Switzerland}

\affiliation{Heinrich-Heine-Universit\"at D\"usseldorf, Institut f\"ur Theoretische  Physik III, \\ Universit\"atsstraße 1,  DE-40225  D\"usseldorf, Germany}

\author{Daniel Loss} \orcid{0000-0001-5176-3073} 
\affiliation{Universität Basel, Department of Physics, Klingelbergstrasse 82, CH-4056 Basel, Switzerland}

\author{Ivano Tavernelli} \orcid{0000-0001-5690-1981} 
\affiliation{IBM Quantum, IBM Research Europe – Zurich, S\"aumerstrasse 4, CH-8803 R\"uschlikon, Switzerland}

\author{Hermann Kampermann} \orcid{0000-0002-0659-6699}
\affiliation{Heinrich-Heine-Universit\"at D\"usseldorf, Institut f\"ur Theoretische  Physik III, \\ Universit\"atsstraße 1,  DE-40225  D\"usseldorf, Germany}

\author{\\{Dagmar} Bru\ss} 
\affiliation{Heinrich-Heine-Universit\"at D\"usseldorf, Institut f\"ur Theoretische  Physik III, \\ Universit\"atsstraße 1,  DE-40225  D\"usseldorf, Germany}

\author{Nikolai Wyderka} \orcid{0000-0003-3002-9878}
\affiliation{Heinrich-Heine-Universit\"at D\"usseldorf, Institut f\"ur Theoretische  Physik III, \\ Universit\"atsstraße 1,  DE-40225  D\"usseldorf, Germany}

\newcommand{\FF}{\mathbb{F}}
\newcommand{\RR}{\mathbb{R}}
\newcommand{\ZZ}{\mathbb{Z}}
\newcommand{\ZnZ}{\mathbb{Z}/n\mathbb{Z}}
\newcommand{\ZdZ}{\mathbb{Z}/d\mathbb{Z}}

\newcommand{\Tr}{\text{Tr}} 
\newcommand{\wt}{\text{wt}} 
\newcommand{\swt}{\text{swt}}

\newcommand{\GHZn}{{\text{GHZ}(n)}}
\newcommand{\fullysepn}{{\text{sep}(n)}} 
\newcommand{\fullysepnd}{{\text{sep}_d(n)}} 
\newcommand{\Wn}{{W(n)}}

\newcommand{\dandn}{{\text{Pust}(n)}}  
\newcommand{\Pustn}{\dandn}  
\newcommand{\RCLn}{{\text{RC}(n)}} 
\newcommand{\RCL}{{\text{RC}}}  
  
\newcommand{\LCL}{{\text{LC}}}  
\newcommand{\CL}{{\text{C}}}  
\newcommand{\AME}{{\text{AME}}}  

\newcommand{\GHZ}{{\text{GHZ}}}  

\newcommand{\TVD}{\text{TVD}}

\definecolor{Qlila}{HTML}{3f007d}   
\definecolor{Qbeige}{HTML}{daa520}   
\definecolor{PurpleBright}{HTML}{8782bc}  
\definecolor{PurpleDark}{HTML}{61409b}  
\definecolor{LavenderDark}{HTML}{b6b5d8} 
\definecolor{Lavender}{HTML}{cecee5} 
\definecolor{LavenderBright}{HTML}{e5e5ff} 
 
\definecolor{GreenDark}{HTML}{22bb22}

\definecolor{teal}{rgb}{0.0, 0.5, 0.5}

\newcounter{mycounter}

\newcounter{theorem}

\begin{document}

\begin{abstract} 
The  Shor-Laflamme distribution (SLD) of a quantum state is a collection of local unitary invariants that quantify $k$-body correlations. 
We show that the SLD of graph states can be derived by solving a graph-theoretical problem. 
In this way, the mean and variance of the SLD are obtained as simple functions of efficiently computable graph properties. 
Furthermore, this formulation enables us to derive closed expressions of SLDs for some graph state families.
For cluster states, we observe that the SLD is very similar to a binomial distribution, and we argue that this property is typical for graph states in general.
Finally, we derive an SLD-based entanglement criterion from the purity criterion and apply it to derive meaningful noise thresholds for entanglement.
Our new entanglement criterion is easy to use and also applies to the case of higher-dimensional qudits. 
In the bigger picture, our results foster the understanding both of  quantum error-correcting codes, where a closely related notion of Shor-Laflamme distributions plays an important role,
and of the geometry of quantum states, where Shor-Laflamme distributions are known as sector length distributions.
\end{abstract}


\newpage
\tableofcontents
 
\newpage
  
\section{Introduction}
In the quest toward fault-tolerant quantum computation,
\emph{quantum error-correcting codes} (QECCs) are taking the main stage.
A thorough understanding of QECCs is central to the eventual success of realizing error-corrected quantum computers.
In a landmark paper of 1997,
Peter Shor and Raymond Laflamme pointed out 
that certain numerical invariants 
are particularly useful to characterize a QECC~\cite{shor_quantum_analog_1997}.
Their idea is most easily explained for the special case of a stabilizer QECC~\cite{gottesman_phd_thesis_1997},
which is defined via a stabilizer subgroup $\mathcal{S}$ of the \emph{$n$-qubit Pauli group }$\mathcal{P}^n = \{ i^q X^\mathbf{r} Z^\mathbf{s} \ \vert \ q\in \ZZ/4\ZZ, \mathbf{r},\mathbf{s}\in \FF_2^n\}$, 
where $\FF_2= \{0,1\}$ is the binary field and a Pauli operator $X^\mathbf{r}Z^\mathbf{s}$ is defined via its action $X^\mathbf{r}Z^\mathbf{s}\ket{\mathbf{b}} = (-1)^{\mathbf{b}\cdot \mathbf{s}}\ket{\mathbf{b}+\mathbf{r}}$ on computational basis states $\ket{\mathbf{b}}\in(\mathbb C^2)^{\otimes n}$.
For a stabilizer group $\mathcal{S}$, Shor and Laflamme define for each $k\in\{0,\ldots, n\}$ the integers
\begin{align}\label{eq:shor_laflamme_a}
    A_k (\mathcal{S}) &= \left \vert \{ S \in \mathcal{S} \ \vert \ \wt(S) = k \} \right\vert \\
    \label{eq:shor_laflamme_b}
    \text{and} \hspace{5mm}
    B_k (\mathcal{S}) &= \left \vert \{ P \in \mathcal{N}(\mathcal{S})/\!\sim \ \vert \ \wt(P) = k \} \right\vert,
\end{align}
where $\mathcal{N}(\mathcal{S}) = \{P \in \mathcal{P}^n \ \vert \ \forall S\in\mathcal{S}: PSP^\dagger \in \mathcal{S}\}$ is the normalizer of $\mathcal{S}$ in the Pauli group,
and $\mathcal{N}(\mathcal{S})/\!\sim$ is the normalizer modulo global phases.
In other words, $A_k(\mathcal{S})$ counts Pauli operators $X^\mathbf{r} Z^\mathbf{s}$ acting as the logical identity on the QECC for which the \emph{Pauli weight} $\wt(X^\mathbf{r}Z^\mathbf{s}) =  \vert \{i\in\{1,\ldots,n\}\ \vert\ r_i=1 \vee s_i=1\} \vert$ is equal to $k$.
Similarly, $B_k(\mathcal{S})$ counts weight-$k$ Pauli operators that act as \emph{any} logical operations on the QECC.
In particular, we have $A_k(\mathcal{S}) \le B_k(\mathcal{S})$ for all $k$;
the smallest integer $d$ with $A_d(\mathcal{S}) < B_d(\mathcal{S})$ is the \emph{distance} of the QECC, i.e., the smallest weight of a Pauli operator that maps some codeword of the QECC onto a different one.
While the distance $d$ is one of the most important characteristics of a QECC, 
it is in general notoriously difficult to compute. 
To tackle this problem, the quantities defined in Eqs.~\eqref{eq:shor_laflamme_a} and~\eqref{eq:shor_laflamme_b} provide a powerful handle:
first note that $d = \min\{k\ge 1 \ \vert \ A_k(\mathcal{S})< B_k(\mathcal{S}) \}$ can be reconstructed if $A_k$ and $B_k$ are known for all $k$.
Strikingly, it is sufficient to know the $A_k$'s because, surprisingly, they uniquely determine the $B_k$'s
through a quantum version of the MacWilliams identity~\cite{shor_quantum_analog_1997}.
This reduces the problem of computing the distance of a stabilizer QECC to counting its weight-$k$ stabilizers, 
which is still challenging but at least it breaks down the problem.

\subsection{Relevance of our developments}
In this paper, we develop a formal approach (Theorem~\ref{thrm:puzzle}) for counting weight-$k$ stabilizers in the special case of stabilizer states~\cite{gottesman_phd_thesis_1997}.
On the one hand, our work should be understood as a first step toward tackling the challenge of computing the distance of stabilizer QECCs via the quantum MacWilliams identity.
On the other hand, 
computing $A_k(\mathcal{S})$ for a stabilizer state $\ket{\psi}$ is interesting in its own right.
For example, it is well known that $A_1=A_2=\ldots=A_m=0$ is equivalent to $\ket{\psi}$ being an $m$-uniform state, 
i.e., all $m$-body marginals of $\ket{\psi}$ being maximally mixed~\cite{scott_multipartite_entanglement_2004, PhysRevA.106.062424}.
This already shows that the SLD of a quantum state contains information about its entanglement.

As a second important contribution, we apply the purity criterion~\cite{horodecki_quantum_entanglement_2009} 
to derive a new entanglement criterion (Theorem~\ref{thrm:purity_criterion}). 
The new criterion is very general as it also applies to higher-dimensional qudits and non-stabilizer states.
Furthermore, it allows the derivation of lower bounds on the entanglement noise threshold (Corollary~\ref{cor:purity_criterion_threshold}) of quantum states for which we only need to know the SLD.
Importantly, all of this also works for the physically relevant case of local white noise,
a noise channel with many cross terms that render many other entanglement criteria inapplicable.

Among other applications, 
to showcase the effectiveness of our approach, we derive the SLD of cycle graph states. 
This enables us to improve the best previously-known lower bound on the entanglement noise threshold of cycle graph states.

\subsection{Setting the stage}

Before we begin the presentation of  our theory in Sec.~\ref{sec:2},
let us briefly review a generalization of
the definition in Eq.~\eqref{eq:shor_laflamme_a}.
For a general $n$-qubit state with density matrix $\rho$
and every $k\in\{0,\ldots, n\}$, let
\begin{align} \label{def:Ak}
    A_k[\rho] = \sum_{\substack{\mathbf{r},\mathbf{s}\in \FF_2^n\\ \swt(\mathbf{r},\mathbf{s})=k }} \left\vert \Tr[\rho X^\mathbf{r}Z^\mathbf{s}] \right \vert ^2, 
\end{align}
where $\swt(\mathbf{r},\mathbf{s})= \wt(X^\mathbf{r}Z^\mathbf{s})$ is the \emph{symplectic weight} of $(\mathbf{r},\mathbf{s}) \in \FF_2^n \oplus \FF_2^n$.
{To honor the seminal work}~\cite{shor_quantum_analog_1997} in which Eq.~\eqref{def:Ak} was first defined, we will call 
\begin{align}
     \mathbf{A}[\rho]=(A_0[\rho],\ldots,A_n[\rho]) \in \mathbb R^{n+1}
\end{align}
the \emph{Shor-Laflamme distribution} (SLD)
of the state $\rho$.
Note that $ \mathbf{A}[\rho]$ is sometimes referred to as 
\emph{sector length} (SL) \emph{distribution} in the literature~\cite{shor_quantum_analog_1997, aschauer_local_invariants_2004, de_vincente_multipartite_entanglement_2011, kloeckl_characterizing_multipartite_2015,tran_quantum_entanglement_2015, tran_correlations_between_2016, eltschka_maximum_nbody_2020, wyderka_characterizing_quantum_2020}
which conveniently has the same acronym.
Because of $\Tr[\rho^2] = \sum_{k=0}^n A_k[\rho]/2^n$, 
the \emph{normalized SLD} $\mathbf{a}=\mathbf{A}/2^n$ is a probability distribution,
provided $\rho$ is a pure state.


To develop the theory of SLDs of stabilizer states, 
we can restrict ourselves to the case of \emph{graph states}~\cite{hein_multiparty_entanglement_2004, hein_entanglement_in_2006}
\begin{align} \label{eq:graph_state}
\ket{\Gamma}
=  \frac{1}{\sqrt{2^n}} \sum_{\mathbf{r} =(r_1,\ldots, r_n)\in \mathbb F_2^n} (-1)^{\sum\limits_{i=1}^n\sum\limits _{j=i+1}^n r_i \gamma_{i,j} r_j } \ket{\mathbf r} ,
\end{align}
which are defined via the \emph{adjacency matrix} $\Gamma=(\gamma_{i,j})_{1\le i,j \le n} \in \FF_2^{n\times n}$ of a graph.
If $\Gamma'$ is a different graph (we do not distinguish between a graph and its adjacency matrix) that arises from $\Gamma$ via local complementation~\cite{bouchet_recognizing_locally_1993}, the states $\ket{\Gamma}$ and $\ket{\Gamma'}$ are local-unitary (LU) equivalent~\cite{hein_multiparty_entanglement_2004}.
Since Eq.~\eqref{def:Ak} is invariant under LU transformations~\cite{shor_quantum_analog_1997},
the SLDs of $\ket{\Gamma}$ and $\ket{\Gamma'}$ coincide.
It is well known that every stabilizer state is LU-equivalent to some graph state~\cite{vandennest_graphical_description_2004}.
For this reason, most of our results about SLDs of graph states will be directly applicable to general stabilizer states.

\subsection{Outline of our paper}
  
This paper is organized as follows.
We begin in Sec.~\ref{sec:2} by formulating and investigating a graph-theoretical color assignment problem, which the SLD of the corresponding graph state solves.
In Sec.~\ref{sec:3}, we numerically examine SLDs of cluster states and random graph states.
In Sec.~\ref{sec:4}, we generalize some of our findings to SLDs of graph states for higher-dimensional qudits,
and we present a simplified version of the purity criterion that can be tested already on the level of SLDs.
Afterward, 
in Sec.~\ref{sec:5}, we derive formulas for how SLDs change under the influence of global or local depolarizing noise,
and we investigate implications for noise thresholds of entanglement.
Finally, in Sec.~\ref{sec:conclusion}, we summarize the central results of this work and provide a short outlook about related research avenues.

\section{Graph-theoretical formulation for SLDs}
\label{sec:2}
Every $n$-qubit graph state $\ket{\Gamma}$, as defined in Eq.~\eqref{eq:graph_state}, is a stabilizer state whose stabilizer group $\mathcal{S}=\langle S_1, \ldots, S_n \rangle$ is generated by operators of the form~\cite{hein_multiparty_entanglement_2004, hein_entanglement_in_2006}
\begin{align} \label{eq:stab_gen_graph}
    S_i = X^{(i)} \prod_{j=1}^n (Z^{(j)})^{\gamma_{i,j}}.
\end{align} 
Therefore, every operator in $\mathcal{S}$ can be written as
\begin{align} \label{eq:product_of_stab_generators}
\prod_{i=1}^n S_i^{r_i} = \sigma_\Gamma(\mathbf{r}) X^\mathbf{r} Z^{\Gamma \mathbf{r}} 
\end{align}
for some bit string $\mathbf{r} = (r_1,\ldots, r_n)\in \FF_2^n$.
Note that the prefactor $\sigma_\Gamma(\mathbf{r}) = \prod_{i<j} (-1)^{r_i\gamma_{i,j}r_j}$ in Eq.~\eqref{eq:product_of_stab_generators}, which arises from the anti-commutativity relation of $X$ and $Z$, is irrelevant for our purposes as we are only interested in the Pauli weight of $X^\mathbf{r}Z^{\Gamma \mathbf{r}}$, which is equal to the symplectic weight of $(\mathbf{r}, \Gamma \mathbf{r})\in \FF_2^n\oplus \FF_2^n$.
By counting all weight-$k$ Pauli operators in $\mathcal{S}$, we obtain the $k$-body SL,
\begin{align}\label{eq:Ak_graph}
    A_k = \left\vert \left\{ \mathbf{r} \in \FF_2^n \ \vert \ \swt(\mathbf{r}, \Gamma\mathbf{r})=k  \right\} \right \vert,
\end{align}
of the graph state $\ket{\Gamma}$.
We can interpret a given bit string $\mathbf{r} \in \FF_2^n$ as a color assignment of the graph $\Gamma$ by declaring vertex $i$ to be white if $r_i=0$, and black if $r_i=1$.
The symplectic weight of $(\mathbf{r}, \Gamma \mathbf{r})$ is then given by the sum of the number of black vertices ($r_i=1$) and the number of white vertices having an odd number of black neighbors ($r_i=0$ but the $i$-th entry of $\Gamma \mathbf{r}$ is equal to $1$).
In other words, we have $\swt(\mathbf{r},\Gamma \mathbf{r}) = k$ if and only if (iff) there are exactly $n-k$ white vertices with an even number of black neighbors,
 see Tab.~\ref{tab:correspondence_example} for an illustrative example.
This shows:

\begin{table}[]
    \centering
    \begin{tabular}{c|c|c|c}
         Color assignment& $\mathbf{r}=(r_1,r_2,r_3)$ & $X^\mathbf{r}Z^{\Gamma \mathbf{r}}$ & $\swt(\mathbf{r},{\Gamma \mathbf{r}}) $  \\ \hline
\begin{tikzpicture}
 \draw[-, line width=.1em] (0.2,0) -- (0.8,0); 
 \draw[-, line width=.1em] (-0.2,0) -- (-0.8,0);  
 \draw[line width=.1em] (0,0) circle (0.2);  
 \draw[line width=.1em] (1,0) circle (0.2);  
 \draw[line width=.1em] (-1,0) circle (0.2);     
\draw (-1,0) node[text=black]{\scriptsize 1}; 
\draw ( 0,0) node[text=black]{\scriptsize 2}; 
\draw ( 1,0) node[text=black]{\scriptsize 3};  
\end{tikzpicture}  
& $(0,0,0)$ 
& $\mathbbm 1 \otimes \mathbbm 1 \otimes \mathbbm 1 $
&   $0$
\\
\begin{tikzpicture}
\fill[black] ( 1,0) circle (0.2);  
 \draw[-, line width=.1em] (0.2,0) -- (0.8,0); 
 \draw[-, line width=.1em] (-0.2,0) -- (-0.8,0);  
 \draw[line width=.1em] (0,0) circle (0.2);  
 \draw[line width=.1em] (1,0) circle (0.2);  
 \draw[line width=.1em] (-1,0) circle (0.2);     
\draw (-1,0) node[text=black]{\scriptsize 1}; 
\draw ( 0,0) node[text=black]{\scriptsize 2}; 
\draw ( 1,0) node[text=white]{\scriptsize 3};  
\end{tikzpicture}  
& $(0,0,1)$
& $\mathbbm 1 \otimes Z\otimes X $
&  $2$
\\
\begin{tikzpicture}
\fill[black] ( 0,0) circle (0.2);  
 \draw[-, line width=.1em] (0.2,0) -- (0.8,0); 
 \draw[-, line width=.1em] (-0.2,0) -- (-0.8,0);  
 \draw[line width=.1em] (0,0) circle (0.2);  
 \draw[line width=.1em] (1,0) circle (0.2);  
 \draw[line width=.1em] (-1,0) circle (0.2);     
\draw (-1,0) node[text=black]{\scriptsize 1}; 
\draw ( 0,0) node[text=white]{\scriptsize 2}; 
\draw ( 1,0) node[text=black]{\scriptsize 3};  
\end{tikzpicture}  
& $(0,1,0)$
& $Z \otimes X \otimes Z $
&  $3$
\\
\begin{tikzpicture}
\fill[black] ( 0,0) circle (0.2);  
\fill[black] ( 1,0) circle (0.2);  
 \draw[-, line width=.1em] (0.2,0) -- (0.8,0); 
 \draw[-, line width=.1em] (-0.2,0) -- (-0.8,0);  
 \draw[line width=.1em] (0,0) circle (0.2);  
 \draw[line width=.1em] (1,0) circle (0.2);  
 \draw[line width=.1em] (-1,0) circle (0.2);     
\draw (-1,0) node[text=black]{\scriptsize 1}; 
\draw ( 0,0) node[text=white]{\scriptsize 2}; 
\draw ( 1,0) node[text=white]{\scriptsize 3};  
\end{tikzpicture}  
& $(0,1,1)$
& $ Z \otimes XZ \otimes  XZ $
&   $3$
\\
\begin{tikzpicture}
\fill[black] (-1,0) circle (0.2);  
 \draw[-, line width=.1em] (0.2,0) -- (0.8,0); 
 \draw[-, line width=.1em] (-0.2,0) -- (-0.8,0);  
 \draw[line width=.1em] (0,0) circle (0.2);  
 \draw[line width=.1em] (1,0) circle (0.2);  
 \draw[line width=.1em] (-1,0) circle (0.2);     
\draw (-1,0) node[text=white]{\scriptsize 1}; 
\draw ( 0,0) node[text=black]{\scriptsize 2}; 
\draw ( 1,0) node[text=black]{\scriptsize 3};  
\end{tikzpicture}  
& $(1,0,0)$
& $ X \otimes Z \otimes \mathbbm 1 $
&    $2$
\\
\begin{tikzpicture}
\fill[black] (-1,0) circle (0.2); 
\fill[black] ( 1,0) circle (0.2); 
 \draw[-, line width=.1em] (0.2,0) -- (0.8,0); 
 \draw[-, line width=.1em] (-0.2,0) -- (-0.8,0);  
 \draw[line width=.1em] (0,0) circle (0.2);  
 \draw[line width=.1em] (1,0) circle (0.2);  
 \draw[line width=.1em] (-1,0) circle (0.2);     
\draw (-1,0) node[text=white]{\scriptsize 1}; 
\draw ( 0,0) node[text=black]{\scriptsize 2}; 
\draw ( 1,0) node[text=white]{\scriptsize 3};  
\end{tikzpicture}  
& $(1,0,1)$
& $ X \otimes \mathbbm 1 \otimes X $
& $2$
\\
\begin{tikzpicture}
\fill[black] (-1,0) circle (0.2); 
\fill[black] ( 0,0) circle (0.2); 
 \draw[-, line width=.1em] (0.2,0) -- (0.8,0); 
 \draw[-, line width=.1em] (-0.2,0) -- (-0.8,0);  
 \draw[line width=.1em] (0,0) circle (0.2);  
 \draw[line width=.1em] (1,0) circle (0.2);  
 \draw[line width=.1em] (-1,0) circle (0.2);     
\draw (-1,0) node[text=white]{\scriptsize 1}; 
\draw ( 0,0) node[text=white]{\scriptsize 2}; 
\draw ( 1,0) node[text=black]{\scriptsize 3};  
\end{tikzpicture}  
& $(1,1,0)$
& $ XZ \otimes XZ  \otimes Z $
&   $3$
\\
\begin{tikzpicture}
\fill[black] (-1,0) circle (0.2); 
\fill[black] ( 0,0) circle (0.2);
\fill[black] ( 1,0) circle (0.2);   
 \draw[-, line width=.1em] (0.2,0) -- (0.8,0); 
 \draw[-, line width=.1em] (-0.2,0) -- (-0.8,0);  
 \draw[line width=.1em] (0,0) circle (0.2);  
 \draw[line width=.1em] (1,0) circle (0.2);  
 \draw[line width=.1em] (-1,0) circle (0.2);     
\draw (-1,0) node[text=white]{\scriptsize 1}; 
\draw ( 0,0) node[text=white]{\scriptsize 2}; 
\draw ( 1,0) node[text=white]{\scriptsize 3};  
\end{tikzpicture}  
& $(1,1,1)$
& $ XZ \otimes X\otimes XZ $
&  $3$
\\

    \end{tabular}
    \caption{Correspondence between black-white color assignments, binary vectors $\mathbf{r}\in \FF_2^n$, and stabilizer operators (up to sign) $X^\mathbf{r}Z^{\Gamma \mathbf{r}}$ for the path-graph $P_{n}$ with  $n=3$ vertices. 
    White and black vertices correspond to zeros and ones in $\mathbf{r}$, respectively. 
    Thus, every black vertex is associated with an $X$-operator on the corresponding qubit.
    Likewise, a $Z$-operator is induced on each neighbor of a black vertex. 
    Since a vertex can have multiple black neighbors, the induced $Z$-operators may cancel, e.g., $\mathbf{r}=(1,0,1)$.
    The weight $\swt(\mathbf{r},  \mathbf{s})$ of a Pauli operator $\pm X^\mathbf{r} Z^{ \mathbf{s}}$ counts the number of non-identity tensor factors.
    The only possibility for $\mathbbm1$ to occur as a tensor factor of an operator $X^\mathbf{r} Z^{\Gamma \mathbf{r}}$ is if the corresponding vertex is white (no $X$) and has an even number of black neighbors (no $Z$). 
    The SLD $(A_0,A_1,A_2,A_3) = (1,0,3,4)$ coincides with the Pauli-weight distribution of the operators in the stabilizer group of $\ket{P_3}$.
    }
    \label{tab:correspondence_example}
\end{table}

\subsubsection*{\refstepcounter{mycounter}Theorem~\arabic{mycounter} (Graph-theoretical formulation of SLDs) \label{thrm:puzzle}}

\emph{Let $\ket{\Gamma}$ be an $n$-qubit graph state and $\mathbf{A} = (A_0, \ldots ,A_n)$ its SLD.  For each $k\in \{0,\ldots, n\}$, $A_k$ is equal to the number of black-white  color assignments of $\Gamma$ for which exactly $n-k$ white vertices have an even number of black neighbors.}

\subsection{General insights}
\label{sec:2.1}
 While Thrm.~\ref{thrm:puzzle} does not alleviate the exponential complexity of computing the entire SLD of an arbitrary graph state $\ket{\Gamma}$,
we can exploit it to express $A_k$ for small values of $k$ in purely graph-theoretical terms:

In the trivial case, $k=0$, the theorem only addresses the color assignment for which all $n$ vertices are white;
we obtain the well-known normalization condition $A_0=1$.

For $k=1$, the situation is more interesting: 
In order for a color assignment to contribute to $A_1$, there have to be $n-1$ white vertices that are disconnected from the black vertex. 
Thus, every color assignment with a single black, isolated vertex contributes; 
other color assignments do not contribute.
Therefore, we find
\begin{align} \label{eq:a1}
    A_1 = I,
\end{align}
where $I$ is the number of \emph{isolated vertices} of the graph. This number is efficiently computed as the number of rows of the adjacency matrix $\Gamma$ in which all entries are equal to zero.
After potentially reordering the qubits, we can write $\ket{\Gamma}=\ket{+}^{\otimes A_1} \otimes \ket{\tilde \Gamma}$ where $\tilde \Gamma$ is a graph without any isolated vertices. 

To express the $2$-body SL in graph-theoretical terms, we note that only color assignments with one or two black vertices can contribute to $A_2$. 
If there is only one black vertex, it has to be connected to exactly one other (automatically white) vertex to ensure that there are exactly $n-2$ white vertices with an even number (automatically zero) of black neighbors; thus, the black vertex has to be a \emph{leaf}.
For the color assignments with exactly two black vertices, however, all other vertices have to be connected to either both or none of the black ones. Otherwise, one of the white vertices would have an odd number of black neighbors; thus, the two black vertices must form a \emph{twin pair}, i.e., have the same neighborhood.
Therefore,
\begin{align}\label{eq:a2}
    A_2 = L + T
\end{align}
is the sum of the number of leaves $L= \vert \{ i\in \{1,\ldots,n\}\ \vert \  \exists! j\in \{1,\ldots,n\} : \gamma_{i,j}=1 \}\vert$
and the number of twin pairs 
$ T= \vert \{ \{i,j\}\subset\{1,\ldots, n\} \ \vert \ i\neq j, \forall k \in \{1,\ldots, n\} \backslash \{i,j\}: \gamma_{i,k}=\gamma_{j,k}  \}\vert 
$,
where ``$\exists!$'' denotes the unique existential quantification.
It has been pointed out before that $L+T$ is invariant under local complementation~\cite{bouchet_recognizing_locally_1993}.
Our interpretation of this number as the 2-body SL establishes the stronger~\cite{ji_the_lulc_2010,  tsimakuridze_graph_states_2017} fact that $L+T$ is an LU invariant of graph states.

In principle, one could continue in a similar manner and also express $A_k$ for $k\ge3$ in graph-theoretical terms.
By counting all color assignments contributing to $A_k$ which have exactly $b\in\{0,\ldots, k\}$ black vertices, we obtain the formal expression
\begin{align}\label{eq:Ak_coloring}
    A_k = \sum_{b=0}^k \sum_{\mathbf{r} \in \mathcal{B}_b} \delta_{\swt(\mathbf{r},\Gamma\mathbf{r}),k},
\end{align} 
where $\mathcal{B}_b\subset \FF_2^n$ is the subset of bit strings having a Hamming weight of $b$. 
For $k\ge 3$, however, the graph-theoretical interpretation of Eq.~\eqref{eq:Ak_coloring} becomes increasingly complicated.
Nevertheless, it immediately yields 
that the cumulative binomial distribution is an upper bound for the $k$-body SL, i.e.,
\begin{align} \label{eq:coarse_bound}
A_k[\rho] \le \sum_{b=0} ^k \vert \mathcal{B}_b\vert  = \sum_{b=0} ^k \binom{n}{b}
\end{align}
for every graph state $\rho = \ket{\Gamma}{\bra{\Gamma}}$.
Since every stabilizer state is LU-equivalent to a graph state and $A_k[\cdot]$ is convex and  LU-invariant, the bound in Eq.~\eqref{eq:coarse_bound} is also fulfilled for mixtures $\rho = \sum_i p_i\ket{\psi_i}\bra{\psi_i}$ of stabilizer states $\ket{\psi_i}$. 
For $k\ge 1$, we can drop the term with $b=0$ in Eq.~\eqref{eq:coarse_bound} because $\mathcal{B}_0$ only contains the trivial color assignment that contributes to $A_0$ 
but not to $A_k$ for $k>0$.

By Eq.~\eqref{eq:Ak_coloring},  $A_k$ can be computed with runtime $\mathcal{O}(n^k)$, which is efficient for small values of $k$.
In the opposite case, where $k=n$, Thrm.~\ref{thrm:puzzle} simplifies to the following problem:
\emph{``$A_n$ is equal to the number of color assignments of $\Gamma$ for which every white vertex has an odd number of black neighbors''}.
Hence, a color assignment $\mathbf{r}\in\FF_2^n$ contributes to $A_n$ iff every vertex $i\in \{1,\ldots,n\}$ is either black ($r_i = 1$) or has an odd number of black neighbors ($\sum_{j=1}^n\gamma_{i,j} r_j =1$), or both.
In other words, $\mathbf{r}$ contributes to $A_n = \vert \mathcal{V} \vert $ iff it lies in the intersection, $\mathcal{V}=\bigcap_{i=1}^n \mathcal{V}_i$,
of the $n$ quadric hypersurfaces $\mathcal{V}_i$ that are defined as
\begin{align}
    \mathcal{V}_i  =  \left\{  \mathbf{r} \in \FF_2^n \ \Bigg \vert  \  (1+r_i)\left(1+\sum_{j=1}^n \gamma_{i,j} r_j \right) =0
    \right\}.
\end{align}
Note that $\mathcal{V}$ contains the affine subspace 
\begin{align}
     \mathcal{A} = \left\{  \mathbf{r} \in \FF_2^n \ \Bigg \vert  \ \forall i\in\{1,\ldots,n\}: \sum_{j=1}^n \gamma_{i,j} r_j = 1 \right\} 
\end{align}
of the color assignments with the property \emph{``every vertex has an odd number of black neighbors''}.
For a large class of graphs, we can make the lower bound $A_n \ge \vert \mathcal{A} \vert $ explicit:

\subsubsection*{\refstepcounter{mycounter}Corollary~\arabic{mycounter} (Lower bound on the full-body SL of certain graph states) \label{cor:an_defect}}

\emph{Let $\Gamma$ be a graph that admits a color assignment with the property ``every vertex has an odd number of black neighbors''.
Then, the full-body SL of the corresponding graph state $\ket{\Gamma}$ can be lower bounded as $A_n \ge 2^{\dim(\ker(\Gamma))}$, where $\ker(\Gamma)$ is the null space of the adjacency matrix $\Gamma$.
}
\begin{proof}
Let $\mathbf{r}\in \FF_2^n$ be the color assignment with the property $\Gamma \mathbf{r} = \mathbf{1}=(1,\ldots,1)$.
Then, each of the $2^{\dim(\ker(\Gamma))}$ vectors of the form $\mathbf{r}+\mathbf{s}$ with $\mathbf{s}\in\ker(\Gamma)$ has the same property,  
 $\Gamma (\mathbf{r} + \mathbf{s}) = \mathbf{1}$, and therefore contributes to $A_n$.
\end{proof}

\subsection{Formulae for mean and variance of normalized SLDs} 
\label{sec:2.2}
 For a pure $n$-qubit state $\ket{\psi}$ the normalized SLD $\mathbf{a} = \mathbf{A}/2^n$ can be regarded as a probability distribution over the set $\{0,\ldots, n\}$.
In the special case where $\ket{\psi}$ is a stabilizer state with stabilizer group $\mathcal{S}$, the SLD coincides with the Pauli-weight distribution (PWD) for $\mathcal{S}$, i.e.,
$a_k$ is the probability that an operator $S\in \mathcal{S}$ (drawn uniformly at random) has Pauli weight $k$. 
Information about the PWD is relevant in the context of simultaneous measurements of all operators in $\mathcal{S}$~\cite{miller_hardware_tailored_2022}.
It is possible to infer mean and variance of $\mathbf{a}$ from $A_1$ and $A_2$ alone by exploiting
the MacWilliams identities
\begin{align} \label{eq:puritym}
  \sum_{k=0}^m \binom{n-k}{n-m}  A_k = 4^m \sum_{k=0}^{n}  \binom{n-k}{m}  \frac{A_k}{2^n} ,
\end{align}
which hold for all $m\in\{0,\ldots ,n\}$
and for all pure $n$-qubit states~\cite{shor_quantum_analog_1997, huber_some_ulams_2018, huber_bounds_on_2018, eltschka_maximum_nbody_2020, wyderka_characterizing_quantum_2020}.
Inserting $m=1$ into Eq.~\eqref{eq:puritym} yields the first moment of the normalized SLD,  
\begin{align} \label{eq:purity1}
 \langle k  \rangle_{\mathbf{a}}  
= \sum_{k=0}^n k a_k
= \frac{3n - A_1}{4},
\end{align}
and inserting $m=2$ yields the second moment,
\begin{align} \label{eq:purity2}
\langle k^2 \rangle_{\mathbf{a}} 
= \sum_{k=0}^n k^2 a_k
= \frac{9n^2+3n - (6n-2)A_1 +2A_2}{16}.
\end{align}
Similarly, it is possible to express $\langle k^j \rangle_{\mathbf{a}}$ in terms of $A_1,\ldots, A_j$ for all $j\le n$.
By combining Eqs.~\eqref{eq:purity1} and~\eqref{eq:purity2}, we obtain the variance of the normalized SLD,
\begin{align}\label{eq:variance}
\langle k^2 \rangle_{\mathbf{a}} - 
\langle k \rangle_{\mathbf{a}} ^2
    = \frac{3n -(A_1-2)A_1 + 2 A_2}{16}.
\end{align}
Using the bounds $0\le A_1 \le n$ and $0\le A_2 \le \binom{n}{2} $ from Ref.~\cite{wyderka_characterizing_quantum_2020}, we can infer from Eqs.~\eqref{eq:purity1}--\eqref{eq:variance} that all pure states obey 
$\langle k \rangle_{\mathbf{a}} \in [\frac{n}{2},\frac{3n}{4}]$,  
$\langle k^2 \rangle_{\mathbf{a}} \in [\frac{3n^2+5n}{16}, \frac{10n^2+2n}{16}]$, and $
\langle k^2 \rangle_{\mathbf{a}} - 
\langle k \rangle_{\mathbf{a}} ^2 \le \frac{(n+1)^2 }{16}$.
Here, the minimum mean $ \langle k  \rangle_{\mathbf{a}} = \frac n2 $ is attained iff $\ket{\psi}$ is fully separable because $A_1=n$ is equivalent to all 1-body marginals being pure~\cite{wyderka_characterizing_quantum_2020}.
Combining Eqs.~\eqref{eq:a1} and~\eqref{eq:purity1}, yields that the maximum mean $ \langle k  \rangle_{\mathbf{a}} = \frac{3n}4$ is reached for all graph states without any isolated vertices and, more generally, for all genuinely multipartite entangled (GME) stabilizer states~\cite{horodecki_quantum_entanglement_2009}.
Note that Eqs.~\eqref{eq:purity1}--\eqref{eq:variance} do not generalize to states that are not pure, e.g., the maximally mixed state $\rho=\mathbbm1 /2^n$ has $A_1=A_2=0$ but $\langle k \rangle_{\mathbf{a}}=  \langle k^2 \rangle_{\mathbf{a}}= 0$.
 
To compute the mean and variance of the normalized SLD for an arbitrary stabilizer state $\ket{\psi}$,
one can efficiently compute a graph state $\ket{\Gamma}$ that is LU-equivalent to $\ket{\psi}$ by exploiting Thrm.~1 of Ref.~\cite{vandennest_graphical_description_2004}.
Then, one can read off $I$, $L$, and $T$ from $\Gamma$ and exploit Eqs.~\eqref{eq:a1}--\eqref{eq:variance}, see Fig.~\ref{fig:examples} for an example.
This shows:
 
 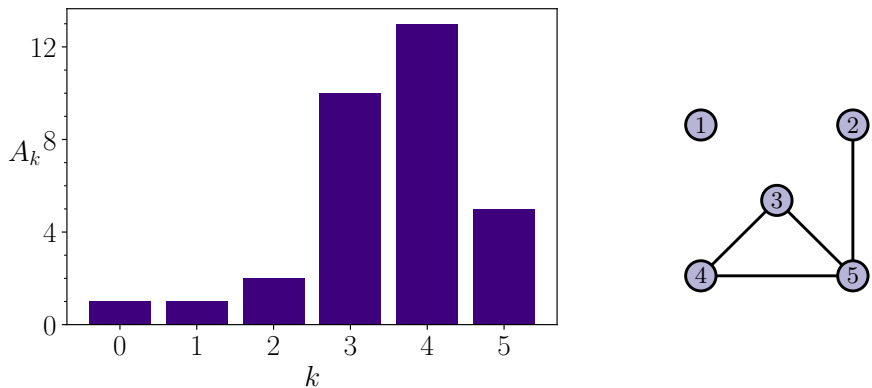
\begin{figure}[t]
\begin{center} 
\begin{minipage}{.50\textwidth}
\includegraphics[width=\textwidth]{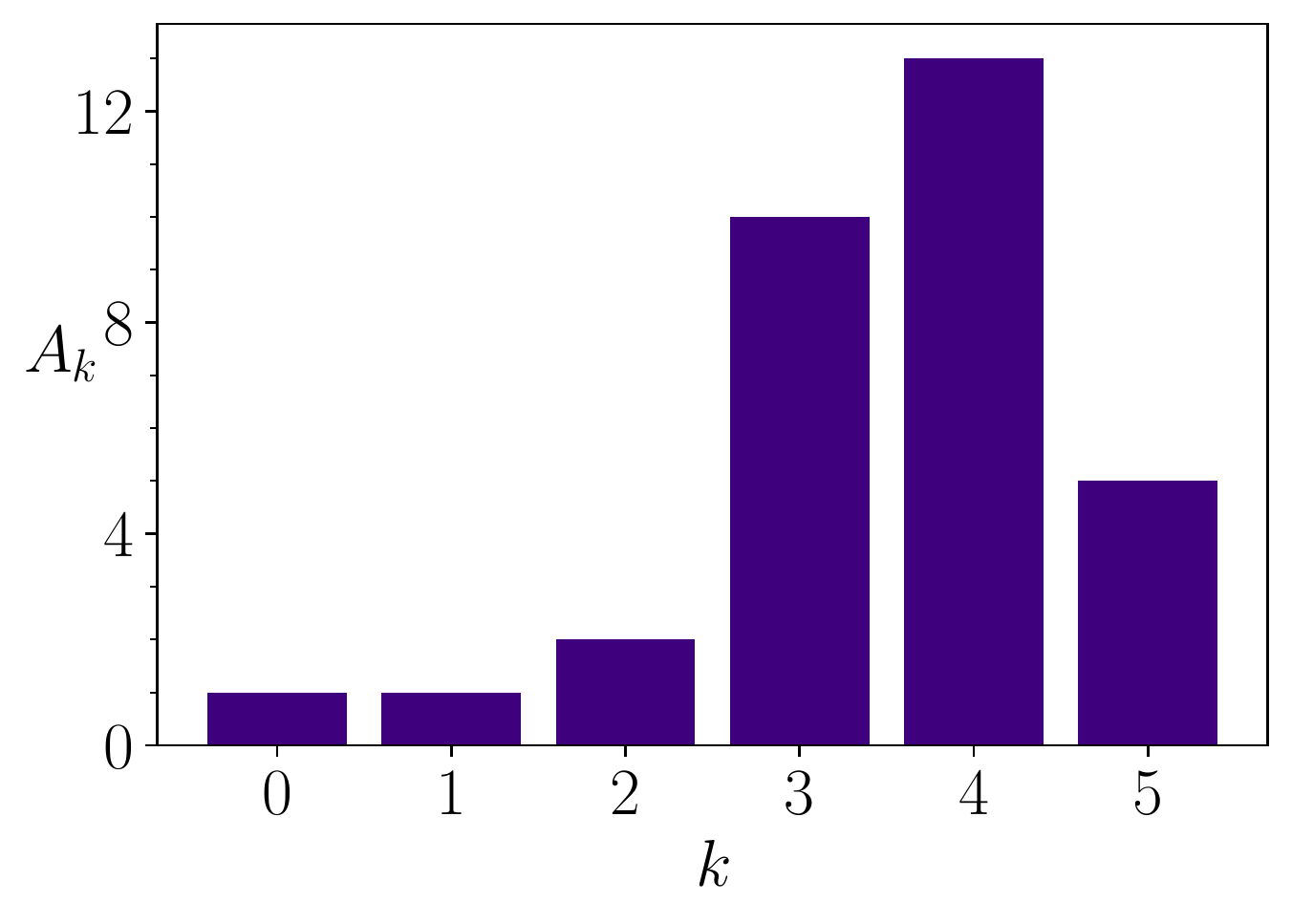}
\end{minipage}
\begin{minipage}{.35\textwidth}
\centering
 \begin{tikzpicture} 
\draw[-, line width=.1em] (1,1) -- (1,-1) -- (-1,-1) -- (0,0) -- (1,-1);
\draw[line width=.1em, fill=LavenderDark] (-1,1) circle (0.2);  
\draw[line width=.1em, fill=LavenderDark] (1,1) circle (0.2);  
\draw[line width=.1em, fill=LavenderDark] (0,0) circle (0.2);  
\draw[line width=.1em, fill=LavenderDark] (-1,-1) circle (0.2);  
\draw[line width=.1em, fill=LavenderDark] (1,-1) circle (0.2);  
\draw (-1,1) node[text=black]{\footnotesize 1};
\draw (1,1) node[text=black]{\footnotesize 2};
\draw (0,0) node[text=black]{\footnotesize 3};
\draw (-1,-1) node[text=black]{\footnotesize 4};
\draw (1,-1) node[text=black]{\footnotesize 5};
\end{tikzpicture}  
\end{minipage}
\end{center}
\caption{Example of the SLD (left)
for a graph state (right) with $n=5$ vertices and $I=L=T=1$. 
Here, vertex 1 is an isolated vertex, while vertex 2 is a leaf. 
Vertices 3 and 4 form a twin pair because they share the same neighborhood; 
this fact would not change if the edge between them was removed.
By Cor.~\ref{cor:mean_variance}, the mean of the normalized SLD $\mathbf{a}$ is given by $\langle k \rangle_{\mathbf{a}}=7/2$ and its variance is given by $\langle k^2 \rangle_{\mathbf{a}} - \langle k \rangle_{\mathbf{a}}^2 = 5/4$.
}
\label{fig:examples}
 \end{figure}
\subsubsection*{\refstepcounter{mycounter}Corollary~\arabic{mycounter} (Mean and variance of the normalized SLD of a graph state) \label{cor:mean_variance}}

\emph{Let $\Gamma$ be a graph with $n$ vertices, $I$ isolated vertices, $L$ leaves, and $T$ twin pairs. 
Then, the mean of the normalized SLD $\mathbf{a}$ of $\ket{\Gamma}$ is given by $\langle k \rangle_{\mathbf{a}} = (3n-I)/4$. 
Furthermore, its variance is given by 
 $\langle k ^2\rangle_{\mathbf{a}}  - \langle k \rangle_{\mathbf{a}} ^2 = (3n - (I-2)I + 2(L+T))/16$.}
 
\subsection{Analytical SLDs of various families of graph states}
\label{sec:2.3}
 The graph color assignment problem, as formulated in Thrm.~\ref{thrm:puzzle},
constitutes a powerful tool for understanding the geometry of quantum states.
In this section, we introduce families of graph states with certain symmetry properties which allow us to derive analytical formulas of their SLDs.

The \emph{complete graph} $K_n$ has $n$ vertices and each pair of vertices is connected by an edge, i.e., all off-diagonal entries of its adjacency matrix are equal to 1.
Its complement $\overline{K_n}$  is appropriately called the \emph{edgeless graph} and the corresponding graph state $\ket{\overline{K_n}}=\ket{+}^{\otimes n}$ is fully separable.
For every  color assignment of $\overline{K_n}$ it is vacuously true that every white vertex has zero black neighbors.
Thus, the graph-theoretical problem from Thrm.~\ref{thrm:puzzle} can be simplified as follows.
\emph{For each $k\in \{0,\ldots, n\}$}, $A_k^{\fullysepn}$ 
\emph{is equal to the number of color assignments with exactly $n-k$ white vertices.}
This immediately yields the well-known~\cite{aschauer_local_invariants_2004} SLD $A^{\fullysepn}_k =  \binom{n}{k}$ of a fully separable, pure $n$-qubit state.

The \emph{star graph} $K_{1,n-1}$ arises from $K_n$ via local complementation~\cite{bouchet_recognizing_locally_1993} at one of the vertices, say vertex 1.
Vertex 1 is then connected to all other vertices via an edge and there are no further edges.
Both $\ket{K_n}$ and $\ket{K_{1,n-1}}$ are LU-equivalent to the Greenberger-Horne-Zeilinger state $\ket{\GHZn} = \frac{1}{\sqrt{2}}(\ket0^{\otimes n} + \ket{1}^{\otimes n})$~\cite{greenberger_going_beyond_1989}.
Let us rederive its well-known~\cite{aschauer_local_invariants_2004} SLD
\begin{align} \label{eq:Ak_GHZ}
A_k^{\GHZn} = \binom{n}{k} \delta_{k,\mathrm{even}} + 2^{n-1}  \delta_{k,n},
\end{align}  
where $\delta_{k,\mathrm{even}}= \frac{1+(-1)^k}{2} $, 
by applying Thrm.~\ref{thrm:puzzle} to the star graph. 
If vertex 1 (the central vertex) is black, there are no white vertices with an even number of black neighbors. 
Thus, all of the $2^{n-1}$ color assignments $\mathbf{r}\in \FF_2^n$ with $r_1=1$ contribute to $A_n$.
Now assume that vertex 1 is white.
Then, all other vertices have zero black neighbors, 
which is even. 
There are $\binom{n-1}{k}$ color assignments for which exactly $k$ of the vertices $2,\ldots,n$ are black.
If $k$ is even, vertex 1 also has an even number of black neighbors, i.e.,
such a color assignment contributes to $A_k$ (because $n-k$ vertices are white and all of them have an even number of black neighbors).
If $k$ is odd, however, the color assignment contributes to $A_{k+1}$ as only the $n-k-1$ white vertices with index $i\in\{2,\ldots, n\}$ have an even number of black neighbors.
Therefore, the SLD of the star graph state is given by $A_k=\binom{n-1}{k-1} + \binom{n-1}{k}  = \binom{n}{k}$ if $k<n$ is even, $A_k=0$ if $k<n$ is odd, and $A_n= 2^{n-1} + \delta_{n,\mathrm{even}}$.
This proves Eq.~\eqref{eq:Ak_GHZ}.

\begin{figure}[t]
\begin{center}
\scalebox{.9}{
\begin{tabular}{cccc}
\small
\begin{tikzpicture}
\small
\draw[-, line width=.1em](0,0.2) -- (0,0.8); 
\draw[line width=.1em, fill=GreenDark] (0,0) circle (0.2);  
\draw[line width=.1em] (0,1) circle (0.2);  
\draw[-, line width=.1em] (0,-0.2) -- (0,-1.8);   
\draw[line width=.1em, fill=GreenDark] (0,-2) circle (0.2);  
\draw[-, line width=.1em] (0.1732,-1.9) -- (0.6928, -1.6);  
\draw[-, line width=.1em] (-0.1732,-1.9) -- (-0.6928, -1.6);   
\draw[line width=.1em, fill=GreenDark] ( 0.866,-1.5) circle (0.2);
\draw[line width=.1em, fill=GreenDark] (-0.866,-1.5) circle (0.2);   
\draw (0,-2) node[text=black]{\footnotesize 1}; 
 \draw (-0.866,-1.5) node[text=black]{\footnotesize 3}; 
 \draw ( 0.866,-1.5) node[text=black]{\footnotesize 4}; 
 \draw (0,0) node[text=black]{\footnotesize 2}; 
\draw (0,1) node[text=black]{\footnotesize 5}; 
 \end{tikzpicture}  
 \hspace{2em}
 &
 \begin{tikzpicture} 
\small  
\draw[-, line width=.1em] ( 0.1176,0.1618) -- ( 0.4702, 0.6472);   
\draw[-, line width=.1em] (-0.1176,0.1618) -- (-0.4702, 0.6472);  
\draw[line width=.1em, fill=GreenDark] (0,0) circle (0.2);   
\draw[line width=.1em] ( 0.5878,0.8090) circle (0.2);      
\draw[line width=.1em] (-0.5878,0.8090) circle (0.2);  
 \draw (-0.5878, 0.8090) node[text=black]{\footnotesize 5};  
 \draw ( 0.5878, 0.8090) node[text=black]{\footnotesize 6};
 
\draw[-, line width=.1em] (0,-0.2) -- (0,-1.8);   
\draw[line width=.1em, fill=GreenDark] (0,-2) circle (0.2);  
\draw[-, line width=.1em] (0.1732,-1.9) -- (0.6928, -1.6);  
\draw[-, line width=.1em] (-0.1732,-1.9) -- (-0.6928, -1.6);   
\draw[line width=.1em, fill=GreenDark] ( 0.866,-1.5) circle (0.2);
\draw[line width=.1em, fill=GreenDark] (-0.866,-1.5) circle (0.2);   
\draw (0,-2) node[text=black]{\footnotesize 1}; 
 \draw (-0.866,-1.5) node[text=black]{\footnotesize 3}; 
 \draw ( 0.866,-1.5) node[text=black]{\footnotesize 4}; 
 \draw (0,0) node[text=black]{\footnotesize 2};  
 \end{tikzpicture}  
 \hspace{1em}
 &
\begin{tikzpicture}
\small  
\draw[-, line width=.1em] (0,0.2) -- (0,0.8);  
\draw[-, line width=.1em] (0.1732,0.1) -- (0.6928, 0.4);  
\draw[-, line width=.1em] (-0.1732,0.1) -- (-0.6928, 0.4);  
\draw[line width=.1em, fill=GreenDark] (0,0) circle (0.2);   
\draw[line width=.1em] (0.866,0.5) circle (0.2);     
\draw[line width=.1em] (-0.866,0.5) circle (0.2);   
\draw[line width=.1em] (0,1) circle (0.2);  
 \draw (-0.866, 0.5) node[text=black]{\footnotesize 5};
 \draw         (0,1) node[text=black]{\footnotesize 6};     
 \draw ( 0.866, 0.5) node[text=black]{\footnotesize 7};

\draw[-, line width=.1em] (0,-0.2) -- (0,-1.8);   
\draw[line width=.1em, fill=GreenDark] (0,-2) circle (0.2);  
\draw[-, line width=.1em] (0.1732,-1.9) -- (0.6928, -1.6);  
\draw[-, line width=.1em] (-0.1732,-1.9) -- (-0.6928, -1.6);   
\draw[line width=.1em, fill=GreenDark] ( 0.866,-1.5) circle (0.2);
\draw[line width=.1em, fill=GreenDark] (-0.866,-1.5) circle (0.2);   
\draw (0,-2) node[text=black]{\footnotesize 1}; 
 \draw (-0.866,-1.5) node[text=black]{\footnotesize 3}; 
 \draw ( 0.866,-1.5) node[text=black]{\footnotesize 4}; 
 \draw (0,0) node[text=black]{\footnotesize 2};  
 \end{tikzpicture}  
 \hspace{1em}
 &
 \begin{tikzpicture}  \small 
\draw[-, line width=.1em] (0,0) -- ( 0.5, 0.866);   
\draw[-, line width=.1em] (0,0) -- (-0.5, 0.866); 
\draw[line width=.1em, fill = white] ( 0.5, 0.866) circle (0.2);     
\draw[line width=.1em, fill = white] (-0.5, 0.866) circle (0.2);   
\draw (-0.5, 0.866) node[text=black]{\footnotesize 6};
\draw ( 0.5, 0.866) node[text=black]{\footnotesize 7};

\draw[-, line width=.1em] ( 0.1902,0.0618) -- ( 0.7608, 0.2472);   
\draw[-, line width=.1em] (-0.1902,0.0618) -- (-0.7608, 0.2472);      
\draw[line width=.1em, fill = GreenDark] (0,0) circle (0.2);   
\draw[line width=.1em, fill = white] ( 0.9511,0.3090) circle (0.2);      
\draw[line width=.1em, fill = white] (-0.9511,0.3090) circle (0.2);   
 \draw (-0.9511, 0.3090) node[text=black]{\footnotesize 5}; 
 \draw ( 0.9511, 0.3090) node[text=black]{\footnotesize 8};
\draw[-, line width=.1em] (0,-0.2) -- (0,-1.8);   
\draw[line width=.1em, fill=GreenDark] (0,-2) circle (0.2);  
\draw[-, line width=.1em] (0.1732,-1.9) -- (0.6928, -1.6);  
\draw[-, line width=.1em] (-0.1732,-1.9) -- (-0.6928, -1.6);   
\draw[line width=.1em, fill=GreenDark] ( 0.866,-1.5) circle (0.2);
\draw[line width=.1em, fill=GreenDark] (-0.866,-1.5) circle (0.2);   
\draw (0,-2) node[text=black]{\footnotesize 1}; 
 \draw (-0.866,-1.5) node[text=black]{\footnotesize 3}; 
 \draw ( 0.866,-1.5) node[text=black]{\footnotesize 4}; 
 \draw (0,0) node[text=black]{\footnotesize 2}; 
  \end{tikzpicture}  
 \end{tabular}
 }
\end{center}
\caption{Pusteblume (German for dandelion) graphs with $n\in\{5,6,7,8\}$ vertices.
}
\label{fig:pusteblume}
\end{figure}
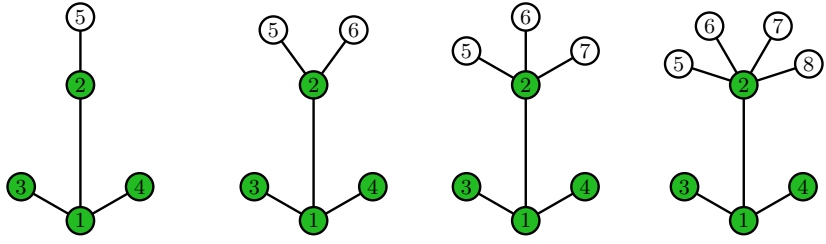

\begin{figure}
\centering
\includegraphics[width = .9\textwidth]{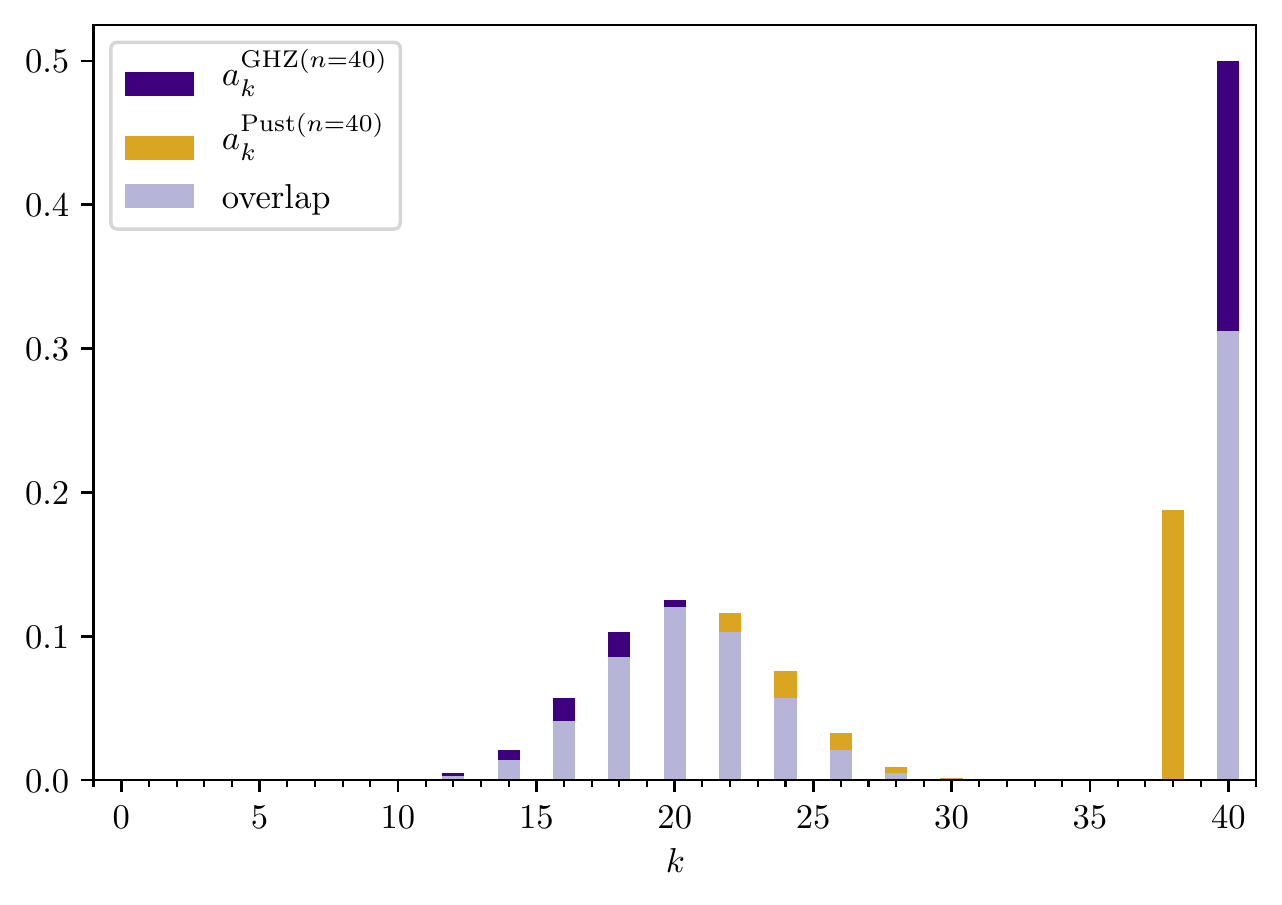}
\caption{Normalized SLDs $a_k =2^{-n} A_k$ of $\ket{\text{GHZ}(n)}$ and $\ket{\text{Pust}(n)}$ for $n=40$ qubits.
}
\label{fig:ghz_vs_pusteblume}
\end{figure}

The \emph{Pusteblume graph}~\cite{miller_graphstatevis_interactive_2021} is a close cousin of $K_{1,n-1}$, see Fig.~\ref{fig:pusteblume}. 
It has $n\ge 5$ vertices and $n-1$ edges. Vertex $1$ has three neighbors: $2,3$, and $4$. Vertex $2$ has $n-3$ neighbors: $1,5,6,\ldots$, $n$. 
An elementary but lengthy analysis, which we provide in App.~\ref{app:dandelion}, shows that the SLD of the $n$-qubit Pusteblume graph state, $\ket{\text{Pust}(n)}$, is given by 
\begin{align} \label{eq:Ak_dand}
    A_k^{\dandn} = \left(\binom{n-3}{k-3} + 3\binom{n-2}{k-2} + \binom{n-3}{k}  \right)\delta_{k,\mathrm{even}}  +  3\times 2^{n-4}\delta_{k,n-2} +   5\times 2^{n-4}\delta_{k,n},
\end{align} 
where we set $\binom{n}{k}=0$ if $n$ or $k$ is a negative integer. 
%
%
In Fig.~\ref{fig:ghz_vs_pusteblume},
we plot the normalized SLD $\mathbf{a} = \mathbf{A}/2^n$
for a GHZ state (blue) and for a Pusteblume graph state (yellow).
The two distributions have a significant amount of overlap (lavender), which we attribute to the similarity between star graphs and Pusteblume graphs.
In both cases, we observe $a_k =0$ for all odd $k$, a property that a graph state exhibits iff all of its vertices have an odd number of neighbors~\cite{huber_some_ulams_2018, miller_graphstatevis_interactive_2021}.
It is well known that $a_n$ is maximized by the GHZ state, i.e., $a_n^{\mathrm{GHZ}(n)}\ge a_n[\rho]$ for every $n$-qubit state $\rho$~\cite{tran_quantum_entanglement_2015, eltschka_maximum_nbody_2020}.
In Fig.~\ref{fig:ghz_vs_pusteblume}, we can see that the process of moving two leaves from the central vertex to one of the other leaves (a process which turns $\ket{K_{1,n-1}}$ into $\ket{\Pustn}$) has the effect that $a^{\GHZn}_n \approx 0.5$  
splits into  $a^{\Pustn}_n \approx 0.31$  and $a^{\Pustn}_{n-2} \approx 0.19$.
Since, by Cor.~\ref{cor:mean_variance}, both distributions have the same mean, $\langle k \rangle_{\mathbf{a}}=3n/4$, this splitting of $a_n$ must be compensated somehow.
Here, this compensation is ensured by $a_k^{\GHZn} < a_k^{\Pustn}$ for $n/2<k<n$, whereas $a_k^{\GHZn} > a_k^{\Pustn}$ for $k \le n/2$.
This explains why the yellow bars in Fig.~\ref{fig:ghz_vs_pusteblume} are enclosed from both sides by blue bars.

\begin{figure}[t!]
\centering
\scalebox{.9}{
 \hspace{1em}
\begin{tabular}{cccc}
 \begin{tikzpicture} 
\small
\draw[white, line width=.1em] (0,-1) circle (0.2);  \draw[white, line width=.1em] (0,1) circle (0.2);  
\draw[-, line width=.1em] (0.8268,0.1) -- (-0.3268,0.766 );     
\draw[-, line width=.1em] (0.8268,-0.1) -- (-0.3268,-0.766 );     
\draw[-, line width=.1em] (-0.5,0.666) -- (-0.5,-0.666);     
\draw[line width=.1em, fill=LavenderDark] (1,0) circle (0.2);  
\draw[line width=.1em, fill=LavenderDark] (-0.5, 0.866) circle (0.2); 
\draw[line width=.1em, fill=LavenderDark] (-0.5,-0.866) circle (0.2); 
\draw ( 1, 0) node[text=black]{\footnotesize 1};
\draw (-0.5, 0.866) node[text=black]{\footnotesize 2};
\draw (-0.5,-0.866) node[text=black]{\footnotesize 3};
\end{tikzpicture}  
 \hspace{1em}
 &
\begin{tikzpicture} 
\small
\draw[-, line width=.1em] (0.8586, 0.1414) -- (0.1414,0.8586);  
\draw[-, line width=.1em] (0.8586, -0.1414) -- (0.1414,-0.8586);  
\draw[-, line width=.1em] (-0.8586, 0.1414) -- (-0.1414,0.8586);  
\draw[-, line width=.1em] (-0.8586, -0.1414) -- (-0.1414,-0.8586);    
\draw[line width=.1em, fill=LavenderDark] (1,0) circle (0.2);  
\draw[line width=.1em, fill=LavenderDark] (-1,0) circle (0.2);     
\draw[line width=.1em, fill=LavenderDark] (0,-1) circle (0.2);  
\draw[line width=.1em, fill=LavenderDark] (0,1) circle (0.2);  

\draw ( 1, 0) node[text=black]{\footnotesize 1};
\draw ( 0, 1) node[text=black]{\footnotesize 2};
\draw (-1, 0) node[text=black]{\footnotesize 3};
\draw ( 0,-1) node[text=black]{\footnotesize 4};
 \end{tikzpicture}  
 \hspace{1em}
 &
\begin{tikzpicture} 
\small 
\draw[white, line width=.1em] (0,-1) circle (0.2);  \draw[white, line width=.1em] (0,1) circle (0.2);  
\draw[-, line width=.1em] (0.8824, 0.1618) --  (0.4266, 0.7893); 
\draw[-, line width=.1em] (0.8824,-0.1618) --  (0.4266,-0.7893);  
\draw[-, line width=.1em] (0.1188, 0.8893) -- (-0.6188, 0.6496);  
\draw[-, line width=.1em] (0.1188,-0.8893) -- (-0.6188,-0.6496);  
\draw[-, line width=.1em] (-0.8090, 0.3878)-- (-0.8090,-0.3878); 
\draw[line width=.1em, fill=LavenderDark] (1,0) circle (0.2);   
\draw[line width=.1em, fill=LavenderDark] (0.3090, 0.9511) circle (0.2);      
\draw[line width=.1em, fill=LavenderDark] (0.3090, -0.9511) circle (0.2);    
\draw[line width=.1em, fill=LavenderDark] (-0.8090, 0.5878) circle (0.2);      
\draw[line width=.1em, fill=LavenderDark] (-0.8090, -0.5878) circle (0.2);       

\draw ( 1, 0) node[text=black]{\footnotesize 1};
\draw ( 0.3090, 0.9511) node[text=black]{\footnotesize 2};
\draw (-0.8090, 0.5878) node[text=black]{\footnotesize 3};
\draw (-0.8090,-0.5878) node[text=black]{\footnotesize 4};
\draw ( 0.3090,-0.9511) node[text=black]{\footnotesize 5};
 \end{tikzpicture}  
 \hspace{1em}
 &
\begin{tikzpicture}
\small
\draw[white, line width=.1em] (0,-1) circle (0.2);  \draw[white, line width=.1em] (0,1) circle (0.2);  
\draw[-, line width=.1em] (-0.3,0.866) -- (0.3,0.866);  
\draw[-, line width=.1em] (-0.3,-0.866) -- (0.3,-0.866);  
\draw[-, line width=.1em] (0.9, 0.1732) -- (0.6,0.6928);  
\draw[-, line width=.1em] (0.9, -0.1732) -- (0.6,-0.6928);  
\draw[-, line width=.1em] (-0.9, 0.1732)  -- (-0.6,0.6928);  
\draw[-, line width=.1em] (-0.9, -0.1732) --(-0.6,-0.6928);  
 
\draw[line width=.1em, fill=LavenderDark] (1,0) circle (0.2);    
\draw[line width=.1em, fill=LavenderDark] (-1,0) circle (0.2);     
\draw[line width=.1em, fill=LavenderDark] (-0.5, 0.866) circle (0.2);     
\draw[line width=.1em, fill=LavenderDark] (-0.5,-0.866) circle (0.2);    
\draw[line width=.1em, fill=LavenderDark] (0.5, 0.866) circle (0.2);     
\draw[line width=.1em, fill=LavenderDark] (0.5,-0.866) circle (0.2);   

\draw ( 1  , 0    ) node[text=black]{\footnotesize 1};
\draw ( 0.5, 0.866) node[text=black]{\footnotesize 2};
\draw (-0.5, 0.866) node[text=black]{\footnotesize 3};
\draw (-1  , 0    ) node[text=black]{\footnotesize 4};
\draw (-0.5,-0.866) node[text=black]{\footnotesize 5};
\draw ( 0.5,-0.866) node[text=black]{\footnotesize 6};
 \end{tikzpicture}  
 \end{tabular}
 }
\caption{Cycle graphs $C_n$ with $n\in\{3,4,5,6\}$ vertices.}
\label{fig:cycle}
\end{figure}
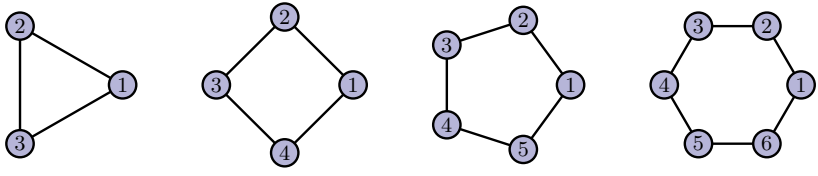 
\begin{figure}
\centering
\includegraphics[width = .9\textwidth]{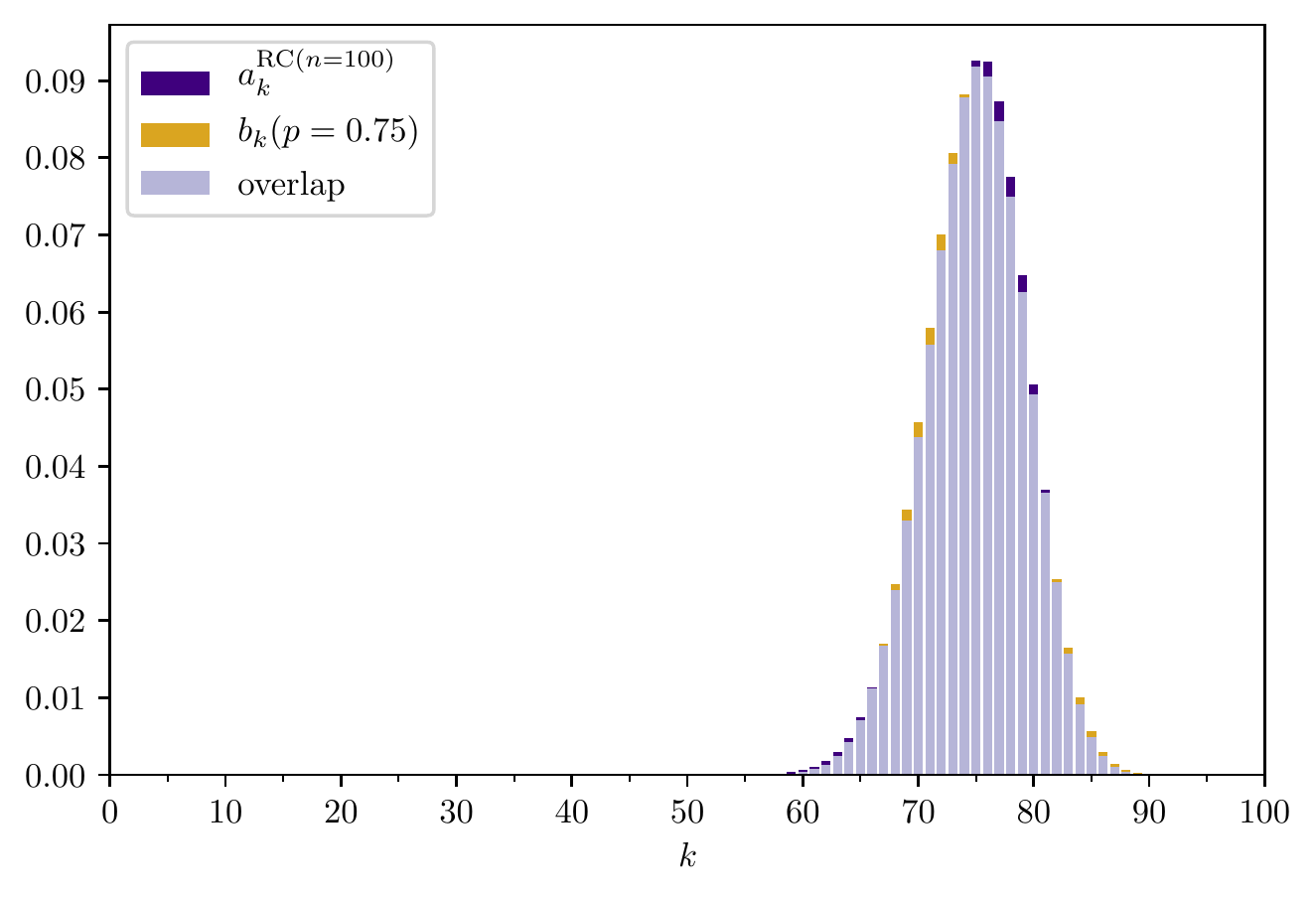} 
\caption{Comparison of the normalized SLD $a_k^\RCLn =2^{-n} A_k^\RCLn $ of the ring cluster state from Eq.~\eqref{eq:Ak_RCL} to an asymmetric binomial distribution $b_k(p=3/4)=\binom{n}{k} 3^k4^{-n}$  for $n=100$.}
  \label{fig:RCL_vs_binom}
\end{figure}

The \emph{cycle graph} $C_n$ has $n$ vertices, where vertex $i\in \{1,\ldots,n\} $ is connected to vertices $i-1$ (mod $n$) and $i+1$ (mod $n$), see Fig.~\ref{fig:cycle}. 
The corresponding cycle graph state, $\ket{C_n}=\ket{\RCLn}$, is also known as the ring cluster ($\RCL$) state~\cite{jungnitsch_entanglement_witnesses_2011}, 
which is a prototypical resource state for measurement-based quantum computation~\cite{raussendorf_measurement_based_2003, briegel_measurement_based_2009}.
By exploiting periodic boundary conditions of $C_n$, we show in App.~\ref{app:ring} its SLD is given by
$A^\RCLn_0=1$ and 
\begin{align} \label{eq:Ak_RCL}
A_k^\RCLn = \frac{n}{k}\binom{k}{n-k}  + \sum_{m=1}^{\left\lfloor \frac{k-1}{2} \right \rfloor } \frac{n}{m} \binom{k-2m-1}{m-1} \sum_{l=0}^{n-k}  \binom{k-3m}{n-k-l}
\binom{l + m-1 }{l}
\end{align}  
for all $n\ge 3$  and all $k \in \{1,\ldots, n\}$. 
Note that $A_n^\RCLn$ is minimal among all $n$-qubit graph states with a connected graph of $n\le8$ vertices~\cite{miller_small_quantum_2019}; 
for $n=9$, however, there is already a graph state with an even lower $n$-body SL~\cite{royle_graph_that_2020}.
As we portray in Fig.~\ref{fig:RCL_vs_binom} for the example of $n=100$ qubits, 
the normalized SLD $\mathbf{a}^\RCLn$ (blue) has a very large overlap (lavender) with the asymmetrical binomial distribution $\mathbf{b}(p)$
for success probability $p = 3/4$ (yellow).
By applying Cor.~\ref{cor:mean_variance}, we find that the mean $\langle k \rangle =  3n/4$
and the variance $\langle k^2 \rangle  - \langle k \rangle^2  = 3n/16$ coincide for both distributions.
Still, there are minor differences between them: 
At the left tail, $k\in \{0,\ldots, 66\}$, the normalized SLD of the ring cluster state dominates, with the exception of $a_1^{\RCL(100)} = a_2^{\RCL(100)}=0$. At the right tail, $k\in\{82, \ldots, 100\}$, the binomial distribution takes larger values.
Around the peak, the behavior is reversed: 
The binomial distribution dominates left of the peak, $k\in\{67,\ldots,74\}$, whereas the SLD of the ring cluster state is larger for $k\in\{75, \ldots 81\}$. 
These minor differences have profound implications on the robustness of the entanglement in $\ket{\RCLn}$, 
see Sec.~\ref{sec:5.2} and App.~\ref{app:noise_thresholds}.
 
\section{Numerical investigation of SLDs of graph states}
\label{sec:3}
In the previous section (see Fig.~\ref{fig:RCL_vs_binom}), we noticed how 
the normalized SLD $\mathbf{a}=(a_0,\ldots, a_n)$ of an  $n$-qubit $\RCL$ state visually matches a binomial distribution $\mathbf{b}(p=0.75)$, where $b_k(p)= \binom{n}{k} p^k (1-p)^{n-k}$.
In this section, we turn  such qualitative statements into quantitative ones by investigating the difference of the distributions in terms of 
the total variation distance 
\begin{align}\label{eq:TVD}
    \TVD({\mathbf{a}}, \mathbf{b}(p)) =
    \frac{1}{2}
    \sum_{k=0}^n
\vert a_k - b_k(p) \vert  \hspace{4mm}  \in\, [0,1].
\end{align} 
The TVD is equal to 0 iff the two probability distributions coincide, and equal to 1 iff the supports of $\mathbf{a}$ and $\textbf{b}(p)$ are disjoint.

\subsection{SLDs of cluster states}
\label{sec:3.1}
An important family of well-studied graph states are \emph{cluster states},
which are crucial resource states for measurement-based quantum computation~\cite{raussendorf_measurement_based_2003, briegel_measurement_based_2009}.
For example, the 2D cluster state $\ket{\CL(l,m)}$ has an $l\times m$ grid as its graph, see Fig.~\ref{fig:CLlm}.
To contribute to the theoretical understanding of cluster states, we now investigate their SLDs as this provides new insights about their stabilizer groups.
Furthermore, when applied to Cor.~\ref{cor:purity_criterion_threshold} in Sec.~\ref{sec:5.2}, this will yield insights into the noise robustness of the entanglement that is exhibited by these states.

\begin{figure}[t]
\begin{center} 
 
\begin{tikzpicture}     
\draw[line width=.1em, decoration={brace, mirror,
raise=0.5cm}, decorate] (-.2,-2.75) -- (5.2,-2.75);
\draw[line width=.1em, decoration={brace, mirror,
raise=0.5cm}, decorate] (-.05,.2 ) -- (-.05,-2.9);
\draw (-1.1,-1.35) node{$l$}; 
\draw (2.5,-3.6) node{$m$}; 
\draw[-, line width=.1em] (0,0) -- (0,-1.55);  
\draw[-, line width=.1em] (1,0) -- (1,-1.55); 
\draw[-, line width=.1em] (2,0) -- (2,-1.55); 
\draw[-, line width=.1em] (3,0) -- (3,-1.55);  
\draw[-, line width=.1em] (5,0) -- (5,-1.55);      
\draw (0, -1.75) node{$\vdots$}; 
\draw (1, -1.75) node{$\vdots$}; 
\draw (2, -1.75) node{$\vdots$}; 
\draw (3, -1.75) node{$\vdots$}; 
\draw (5, -1.75) node{$\vdots$}; 
\draw[-, line width=.1em] (0,-2.15) -- (0,-2.5);
\draw[-, line width=.1em] (1,-2.15) -- (1,-2.5);
\draw[-, line width=.1em] (2,-2.15) -- (2,-2.5);
\draw[-, line width=.1em] (3,-2.15) -- (3,-2.5);
\draw[-, line width=.1em] (5,-2.15) -- (5,-2.5);
\draw[-, line width=.1em] (0,0)    -- (3.55,0);  
\draw[-, line width=.1em] (0,-1)   -- (3.55,-1);  
\draw[-, line width=.1em] (0,-2.7) -- (3.55,-2.7);  
\draw (4,-0.01) node{$\cdots$}; 
\draw (4,-1.01) node{$\cdots$}; 
\draw (4,-2.71) node{$\cdots$}; 
\draw[-, line width=.1em] (4.45,0)    -- (5,0);  
\draw[-, line width=.1em] (4.45,-1)   -- (5,-1);  
\draw[-, line width=.1em] (4.45,-2.7) -- (5,-2.7);    
\draw[line width=.1em, fill=LavenderDark] (0,0) circle (0.2); 
\draw[line width=.1em, fill=LavenderDark] (0,-1) circle (0.2); 
\draw[line width=.1em, fill=LavenderDark] (0,-2.7) circle (0.2); 
\draw[line width=.1em, fill=LavenderDark] (1,0) circle (0.2); 
\draw[line width=.1em, fill=LavenderDark] (1,-1) circle (0.2); 
\draw[line width=.1em, fill=LavenderDark] (1,-2.7) circle (0.2); 
\draw[line width=.1em, fill=LavenderDark] (2,0) circle (0.2); 
\draw[line width=.1em, fill=LavenderDark] (2,-1) circle (0.2); 
\draw[line width=.1em, fill=LavenderDark] (2,-2.7) circle (0.2); 
\draw[line width=.1em, fill=LavenderDark] (3,0) circle (0.2); 
\draw[line width=.1em, fill=LavenderDark] (3,-1) circle (0.2); 
\draw[line width=.1em, fill=LavenderDark] (3,-2.7) circle (0.2); 
\draw[line width=.1em, fill=LavenderDark] (5,0) circle (0.2); 
\draw[line width=.1em, fill=LavenderDark] (5,-1) circle (0.2); 
\draw[line width=.1em, fill=LavenderDark] (5,-2.7) circle (0.2); 
 \end{tikzpicture}  
 \vspace{-5mm}
 
\end{center} 
\caption{Graph of a 2D cluster state $\ket{\CL(l,m)}$ with $n=l\times m $ qubits. 
}
\label{fig:CLlm}
\end{figure}
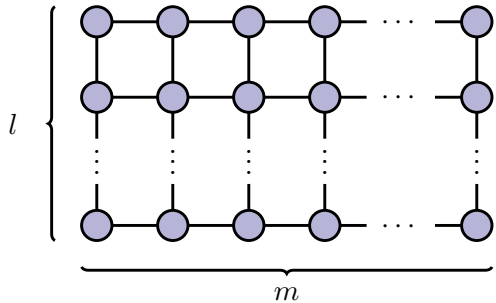
\begin{figure}
\centering
\begin{minipage}{5em}
\vspace{0.0em}

 $\TVD(\mathbf{a}, \mathbf{b}(p))$
\end{minipage}
\begin{minipage}{33em} 
\includegraphics[width = \textwidth]{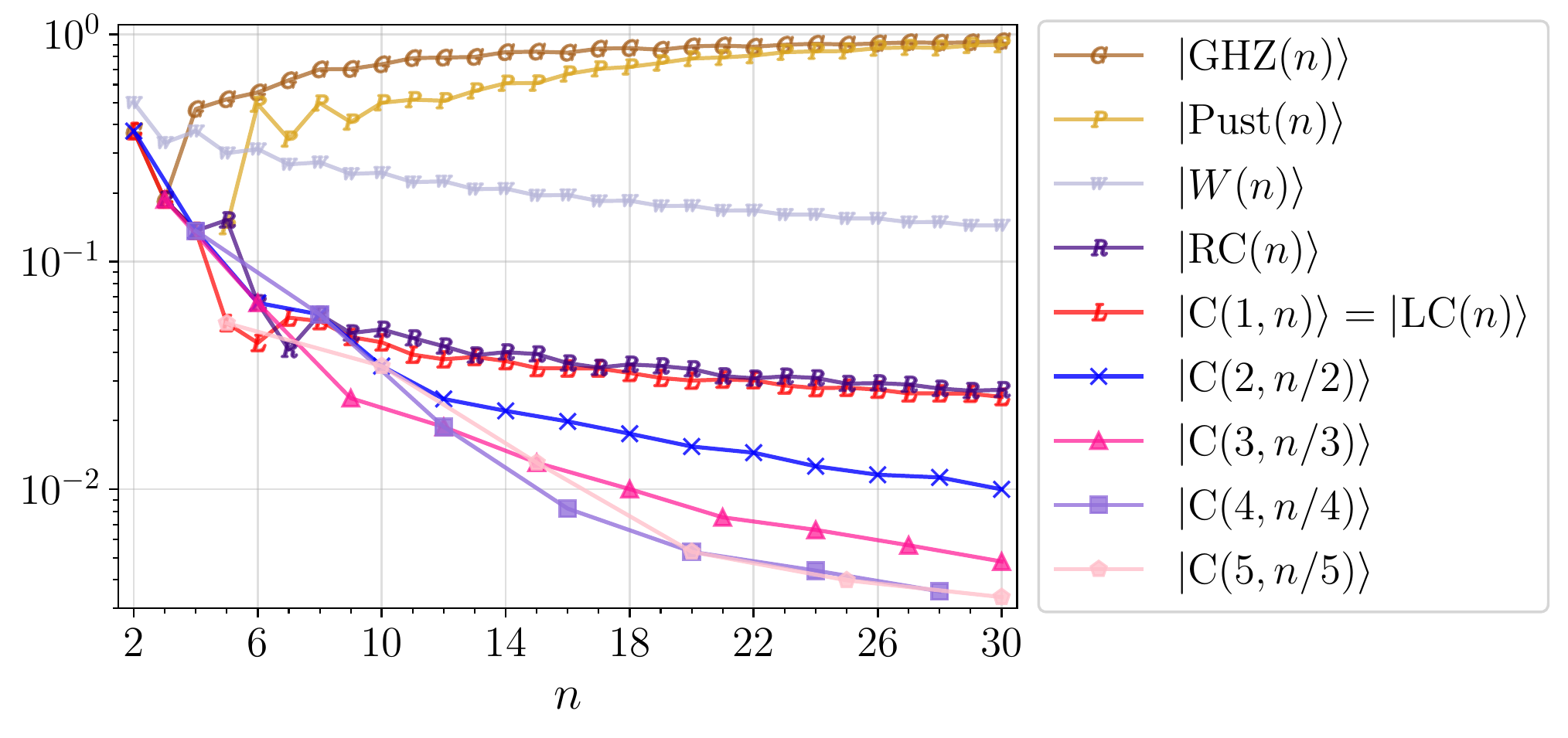} 
\end{minipage}
  \caption{Total variation distance, see Eq.~\eqref{eq:TVD}, between the normalized SLD $\mathbf{a}$ of various families of $n$-qubit states and a corresponding binomial distribution $\mathbf{b}(p)$ with success probability $p=3/4$ for all states with the exception of $\ket{W(n)}$, where we instead use $p=1/2$ for reasons explained in App.~\ref{app:W}.}
  \label{fig:TVD_VS_n}
\end{figure}
In Fig.~\ref{fig:TVD_VS_n},
we plot the TVD between the normalized SLD $\mathbf{a}$ of certain $n$-qubit states and the binomial distribution $\mathbf{b}(p)$ for an appropriately chosen probability $p$.
For the $W$ state (lavender $W$'s), we choose $p=0.5$ 
as this causes $\TVD(\mathbf{a}^{W(n)}, \mathbf{b}(p=0.5))\rightarrow 0$  for $n \rightarrow \infty$;
for readers that are interested in the important case of SLDs of non-stabilizer states,
we provide more details in App.~\ref{app:W}.
For all other states, we use $p=0.75$ as this ensures that $\mathbf{a}$ and $\mathbf{b}(p)$ have the same mean, recall Cor.~\ref{cor:mean_variance}.
We observe in Fig.~\ref{fig:TVD_VS_n} that the TVD converges to 1 for GHZ states (brown $G$'s) and Pusteblume graph states (yellow $P$'s).
This is because their normalized SLDs are far from being binomial distributions, recall Fig.~\ref{fig:ghz_vs_pusteblume}. 
As expected, the TVD for $\RCL$ states (blue $R$'s) is smaller than for GHZ states and Pusteblume graph states.  
With Eq.~\eqref{eq:Ak_RCL} at hand (the solution of the graph-theoretical problem for cycle graphs), 
we compute $\TVD(\mathbf{a}^{\RCLn}, \mathbf{b}(p=0.75))$ for all $n\le 1000$ and find that
$0.15/\sqrt{n}$ fits the data very well for large $n$.

For the broader class of general cluster states (red to blue), 
we lack the solution of the graph-theoretical problem, 
thus,
we are limited to $n\le 30$.
Aside from finite size effects,
we can see that the TVDs decrease with $n$.
Hereby, the TVD is smaller for 2D cluster states $\ket{\CL(l,m)}$ than for $\ket{\RCLn}$ and  1D \emph{linear cluster} ($\LCL$) states $\ket{\LCL(n)}=\ket{\CL(1,n)}$ (red $L$'s).
For 2D cluster states, the TVD tends to be smaller for broader cluster patches, e.g., for $n=30$ qubits, the TVD of the width-2 cluster state (blue crosses) is three times as large as that of the width-5 cluster state (pink pentagons).
We also compute the SLD of an analogously-defined
3D cluster state $\ket{\CL(3,3,3)}$ for  $n= 27$ qubits and find an even smaller TVD of $0.00138$ (not plotted).\footnote{Digital feature: For the graph of $\ket{\CL(3,3,3)}$ and its SLD, please click on this \href{https://graphvis.uber.space/?graph=27_444000141000080800088800141000404002020020200004011100A08080811105042022024040188A1046A5}{link}.}  

In conclusion, the normalized SLD is very similar to a binomial distribution for some graph states (cluster states), while for others (GHZ, Pusteblume) it is not.
To identify which of the two is the exception and which is the norm, we will next investigate random graph states.

\subsection{SLDs of random graph states}
To further solidify our understanding of the geometry of quantum states, we now illustrate the behavior of random graphs states.
To this end, we employ the Erdős-Rènyi graph model~\cite{erdos_on_the_1960},
however, we expect that similar results hold true for other common random graph models as well.
Given a probability $q\in [0,1]$, a random Erdős-Rényi graph with $n$ vertices is created as follows:
For each $i\in\{1,\ldots,n\}$ and $j\in\{i+1,\ldots,n\}$, an edge between vertex $i$ and $j$ is created with probability $q$.
We denote the resulting random variable as 
$\Gamma_n^{(q)}$.
The probability of drawing a specific graph $\Gamma \sim \Gamma_n^{(q)}$ only depends on its number of edges $e(\Gamma)$
and is given by
\begin{align}
     \Pr[\Gamma_n^{(q)}= \Gamma] = q^{e(\Gamma)}(1-q)^{\binom{n}{2} - e(\Gamma)}.
\end{align}
In particular, $\langle A_1 \rangle_q=n(1-q)^{n-1}$ is the expected number of isolated vertices.
Thus, by Cor.~\ref{cor:mean_variance}, the expected mean of the SLD of a random graph state is given by
\begin{align}
 \langle  \langle k  \rangle_{\mathbf{a}} \rangle_q  =
 \frac{3n}{4} - \frac{n(1-q)^{n-1}}{4},
\end{align}
which is approximately equal to $3n/4$ if $q$ and $n$ are large enough. 

\begin{figure}[t]
\centering
\begin{minipage}{7em}
\vspace{5mm}

 $\langle \TVD(\mathbf{a}, \mathbf{b}(p)) \rangle_q$
\end{minipage}
\begin{minipage}{29em} 
\includegraphics[width = \textwidth]{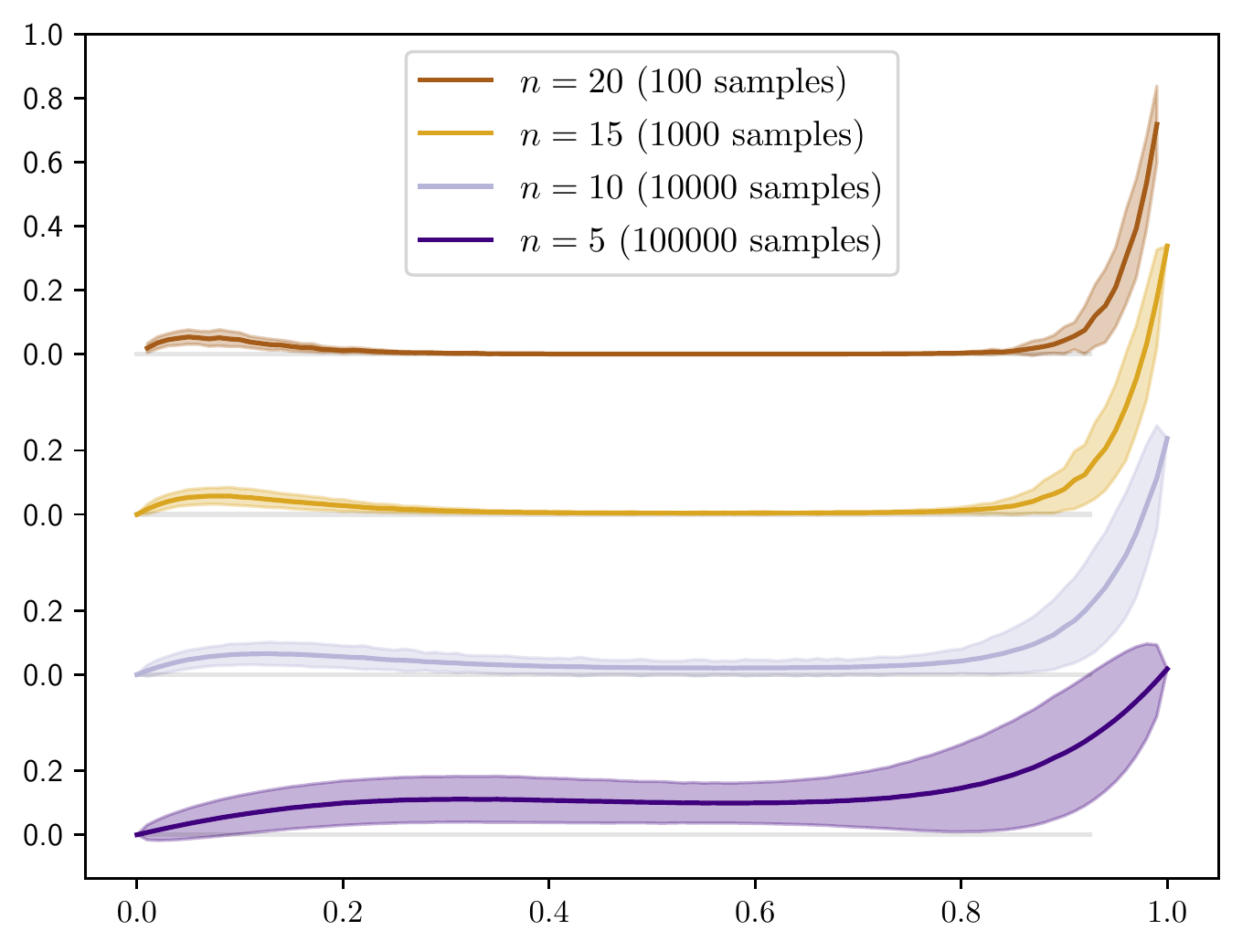} 

\hspace{14.5em}
$q$
\end{minipage} 
  \caption{Total variation distance, see Eq.~\eqref{eq:TVD}, between the normalized SLD $\mathbf{a}$ of random graph states and a corresponding binomial distribution $\mathbf{b}(p)$, where $p$ is selected for every individual graph such that the two distributions have the same mean.
  In the generation of the random graphs, an edge between each pair of qubits is created with probability $q$. 
  The shaded region marks the range in which the TVD lies with probability 68\% (1 sigma).
  For better readability, the curves have an offset spacing. 
  }
  \label{fig:TVD_random}
\end{figure}

In a numerical experiment, we sample Erdős-Rényi graphs with $n\in\{5,10,15,20\}$ vertices and compute the normalized SLD $\mathbf{a}$ of the corresponding random graph states.
Then, we calculate the TVD between  $\mathbf{a}$ and a binomial distribution $\mathbf{b}(p)$ with the same mean, i.e., for each sample holds $\langle k \rangle_{\mathbf{a}} = np$.  
In Fig.~\ref{fig:TVD_random},
we plot the result over the whole interval $q\in[0,1]$ with a step size of $0.01$. For complexity reasons, we vary the number of samples from $10^5$ for $n=5$ (blue) to $10^2$ for $n=20$ (brown).
Overall, the curves show a similar behavior albeit less pronounced for $n=5$ due to finite size effects. 
For $q=0$, there are no edges and every sampled graph state is equal to $\ket{\Gamma}=\ket{+}^{\otimes n}$.
Since the normalized SLD of such a fully separable state is equal to a symmetric binomial distribution, the TVD between the two distributions vanishes.
As $q\gtrsim0$ grows, the TVD first begins to increase before it drops again and stagnates at a very small value over a wide range of $q$.
The latter observation implies that random graph states with $\langle \TVD(\mathbf{a}, \mathbf{b}(p)) \rangle_q \approx0$ are abundant.
We attribute the small initial peak at small $q>0$ to the existence of tensor factors that are LU-equivalent to $\ket{\text{GHZ}(m)}$ for small values of $m<n$.\footnote{If the edge probability $q$ is small, there will be many graph components with a small number $m$ of vertices. For $m=2$ and $m=3$, every graph state is LU-equivalent to a GHZ state, and for $m=4$, GHZ and cycle graph are the only LU equivalence classes. The SLDs of GHZ states are far from the binomial distribution, see Fig.~\ref{fig:TVD_VS_n}. 
This (total variational) distance is inherited by SLDs of product states that contain a considerable amount of GHZ states.}
The position of the peak is consistent with $q\approx\ln(n)/n$, which is the threshold below (above) which $\Gamma \sim \Gamma^{(q)}_n$ is almost surely disconnected (connected)~\cite{erdos_on_the_1960}.
Around $q\approx 0.8$ the TVD suddenly starts to grow again and eventually, at $q=1$, the complete graph is reached and $\ket{\Gamma}=\ket{K_n}$ is LU-equivalent to $\ket{\GHZn}$, which has a very large TVD to the corresponding binomial distributions; recall Fig.~\ref{fig:TVD_VS_n}.
For larger $n$, we observe a decline of TVD at intermediate values of $q$, e.g.,
at $q=0.5$ we find the values
$\langle \langle k\rangle_{\mathbf{a}} \rangle_{q}=0.10(6)$ for $n=5$, 
$\langle \langle k\rangle_{\mathbf{a}} \rangle_{q}=0.02(2)$ for $n=10$,
$\langle \langle k\rangle_{\mathbf{a}} \rangle_{q}=0.004(3)$ for  $n=15$, and 
$\langle \langle k\rangle_{\mathbf{a}} \rangle_{q}=0.0007(4)$ for $n=20$.
Also, the plateau of small TVD values is broader for larger $n$ as both the small initial peak and the final steep are sharpened.

To explain the emergence of the plateaus in Fig.~\ref{fig:TVD_random}, we show in App.~\ref{app:sld_random} that the expected $k$-body SL of an $n$-vertex Erdős-Rényi graph state with edge-probability $q$ is given by
\begin{align}\label{eq:Ak_random_unsolved_3}
    \langle A_k \rangle_q  
    &= \binom{n}{k}2^{-n} \sum_{b=0}^k \binom{k}{b} 2^{b} (1+(1-2q)^b)^{n-k}(1-(1-2q)^b)^{k-b}.
\end{align}
For $n\gg 1$, we can use the approximation $\binom{k}{b} 2^{b} (1+(1-2q)^b)^{n-k}(1-(1-2q)^b)^{k-b} \approx \binom{k}{b} 2^{b}$
at a wide range around $q\approx 1/2$.
This allows us to simplify Eq.~\eqref{eq:Ak_random_unsolved_3} using the binomial theorem, which yields
\begin{align}
    \label{eq:Ak_random_approximation}
    \langle a_k \rangle_q  
\ \approx \  \binom{n}{k} 4^{-n} \sum_{b=0}^k \binom{k}{b} 2^b 
\ = \   \binom{n}{k}3^k 4^{-n} 
\ = \    B_k(p=3/4).
\end{align}
The plateaus in Fig.~\ref{fig:TVD_random} show for which values of $n$ and $q$ the approximation in Eq.~\eqref{eq:Ak_random_approximation} is valid.

A direct physical consequence of the results in Fig.~\ref{fig:TVD_random}
is the following:
if we prepare the state $\ket{+}^{\otimes n}$, where $n \gg 5$,
and apply to each pair of qubits a controlled-$Z$ gate with probability $0 \ll q \ll 1$ (and do nothing with probability $1-q$),
then we should expect that the SLD of the resulting state is approximately given by 
$A_k \approx \binom{n}{k}3^k2^{-n}$.
However, this approximation should be applied with care, see footnote~3 in App.~\ref{app:noise_thresholds} for an example of what can go wrong otherwise.

Now, we are finally in the position to answer the question raised at the end of Sec.~\ref{sec:3.1}:
There is an abundance of random graph states for which the normalized SLD is remarkably close to an asymmetrical binomial distribution; we call SLDs with this property \emph{generic}.
In that sense, cluster graphs have generic SLDs (recall Fig.~\ref{fig:RCL_vs_binom}), whereas star, complete, and  Pusteblume graphs do not (recall Fig.~\ref{fig:ghz_vs_pusteblume}).
Hence, we will say that SLDs of the latter graph states are \emph{special}.
Also note that among the $4^n$ Pauli operators of the form $X^\mathbf{r}Z^\mathbf{s}$, there are exactly $\binom{n}{k}3^k$ operators with $\wt(X^\mathbf{r}Z^\mathbf{s})=k$.
This shows that the PWD of the stabilizer group of a graph state with a generic SLD closely resembles the PWD of the full Pauli group.

\section{Generalization to higher-dimensional qudits}
\label{sec:4}
In this section, we extend the scope of our investigation to the case of 
$n$-qudit states, where every qudit has a Hilbert space dimension of $d\ge 2$.
For studying such states, we find it convenient to label the computational basis by elements of the free module $ (\ZdZ)^n$ over the ring $\ZdZ = \{0,1,\ldots, d-1\}$ of integers modulo $d$.
In this way, any pure state can be written as a superposition of states of the form $\ket{\mathbf{j}}$, where $\mathbf{j}\in (\ZdZ)^n$.
Moreover, a general mixed state for $n$ qudits can be written as 
\begin{align}  \label{eq:PauliDecomposition_qudit}
 \rho = \frac{1}{d^n} \sum_{\mathbf{r,s}\in (\ZdZ)^n} \rho_{\mathbf{r},\mathbf{s}} X_d^\mathbf{r} Z_d^\mathbf{s}
 \end{align}
for unique coefficients $\rho_{\mathbf{r},\mathbf{s}}\in \mathbb{C}$, where the \emph{generalized Pauli operators} can be defined as
\begin{align} \label{eq:PauliXZ_qudit}
X_d^\mathbf{r}Z_d^\mathbf{s}= \sum_{\textbf{j}\in (\ZdZ) ^n} \omega_d^{\mathbf{j}\cdot \mathbf{s}}\ket{\mathbf{j}+\mathbf{r}}\bra{\mathbf{j}}
\end{align} 
and $\omega_d=\exp(2\pi \text{i}/d)$~\cite{knill_non_binary_1996}.
Then, the $n$-qudit $k$-body SL of $\rho$ is defined as
\begin{align} \label{def:Ak_qudit}
    A_k[\rho] = \sum_{\substack{\mathbf{r},\mathbf{s}\in (\ZdZ)^n\\ \swt_d(\mathbf{r},\mathbf{s})=k }} \left\vert \Tr[\rho X_d^\mathbf{r}Z_d^\mathbf{s}] \right \vert ^2
 = \sum_{\substack{\mathbf{r,s}\in (\ZdZ)^n\\ \swt_d({\mathbf{r},\mathbf{s}})=k}}
 \vert  \rho_{\mathbf{r},\mathbf{s}} \vert^2,
\end{align} 
where $\swt_d(\mathbf{r},\mathbf{s}) = \vert\{ i \in\{1,\ldots,n\}\ \vert \ r_i\neq0 \vee s_i \neq 0\} \vert $
is the symplectic weight for qudits~\cite{eltschka_maximum_nbody_2020}.
Note the similarity between Eq.~\eqref{def:Ak_qudit} and Eq.~\eqref{def:Ak}.
If $\rho$ is a pure state, the normalized SLD $\mathbf{a} = \mathbf{A}/d^n$ is a probability distribution, 
and Eq.~\eqref{eq:puritym} generalizes to
\begin{align} \label{eq:puritym_qudit}
    \sum_{k=0}^m \binom{n-k}{m-k} A_k 
    = 
    d^{2m} \sum_{k=0}^n  \binom{n - k}{m} a_k
\end{align}
for all $m\in\{0,\ldots,n\}$~\cite{huber_bounds_on_2018, eltschka_maximum_nbody_2020, wyderka_characterizing_quantum_2020}. 
After inserting $m=1$ and $m=2$ into Eq.~\eqref{eq:puritym_qudit}, and after a little algebra, 
we find 
\begin{align}\label{eq:purity1_qudit}
    \langle k \rangle_{\mathbf{a}} &= \frac{(d^2-1)n-A_1}{d^2} \\
\label{eq:purity2_qudit}
\text{and } \hspace{2mm}   \langle k^2 \rangle_{\mathbf{a}} &= \frac{d^4n^2 - d^2 (2n-1)n + n(n-1) \ + (2(n-1)- d^2(2n-1))A_1
+ 2 A_2}{d^4}.
\end{align}
Note that Eqs.~\eqref{eq:purity1_qudit}--\eqref{eq:purity2_qudit} generalize Eqs.~\eqref{eq:purity1}--\eqref{eq:purity2}
to the case of pure $n$-qudit states; to the best of our knowledge, both results are new.

\subsection{Known results about SLDs of qudit states}

For every Abelian subgroup $\mathcal{S} \subset \mathcal{P}^n_d$ of the $n$-qudit Pauli group
\begin{align}\label{eq:Pauli_group_qudit}
\mathcal{P}^n_d=
\left \{ \omega_{2d}^q X^\mathbf{r}Z^\mathbf{s}\ \big \vert \ q\in \ZZ/2d\ZZ, \ \mathbf{r},\mathbf{s} \in (\ZdZ)^n \right \}
\end{align}
with $\vert \mathcal{S}\vert = d^n$ and $z\mathbbm 1 \not \in \mathcal{S}$ for $z\in \mathbb{C}\backslash\{1\}$,
there exists a unique \emph{stabilizer state} $\ket{\psi}\in (\mathbb{C}^d)^{\otimes n}$ which, by definition, fulfills $S\ket{\psi} = \ket{\psi}$ for all $S \in \mathcal{S}$~\cite{gottesman_phd_thesis_1997, gheorghiu_standard_form_2014}.
In this case, the $k$-body SL is equal to the number of \emph{stabilizer operators} $S\in \mathcal{S}$ which have a Pauli weight of  $k$~\cite{kloeckl_characterizing_multipartite_2015}.
For example, the $n$-qudit GHZ state
\begin{align} \label{eq:GHZ_qudit}
\ket{\text{GHZ}_d(n)} = \frac{1}{\sqrt{d}}(\ket{0}^{\otimes n} + \ldots + \ket{d-1}^{\otimes n})
\end{align}
is a stabilizer state for which $\mathcal{S}$ is generated by $X_d\otimes\ldots\otimes X_d$ and $Z_d^{(i)}Z_d^{(i+1)\dagger}$ for all $i\in \{1,\ldots, n-1\}$.
In Prop.~10 of Ref.~\cite{miller_small_quantum_2019}, we have derived its SLD
\begin{align} \label{eq:Ak_GHZ_qudit}
A_k^{\text{GHZ}_d(n)} =   \binom{n}{k} \frac{(d-1)^k + (-1)^k(d-1)}{d} + 
     \delta_{k,n}(d-1)d^{n-1}
\end{align}
by counting all weight-$k$ operators in $\mathcal{S}$ (see Ref.~\cite{eltschka_maximum_nbody_2020} for an alternative proof).
For every symmetric matrix $\Gamma=(\gamma_{i,j})\in (\ZdZ)^{n\times n}$ with zeros on the diagonal, a \emph{qudit graph state}
\begin{align} \label{eq:graph_state_qudit}
    \ket{\Gamma} =  \frac{1}{\sqrt{d^n}} \sum_{\mathbf{r} \in (\ZdZ)^n} \omega_d^{\sum\limits_{i=1}^n\sum\limits _{j=i+1}^n r_i \gamma_{i,j} r_j } \ket{\mathbf r}
\end{align}
is defined~\cite{grassl_graphs_quadratic_2002, bahramgiri_graph_states_2006, looi_tripartite_entanglement_2011}.
An important example is  $\ket{+_d}^{\otimes n} = \tfrac{1}{\sqrt{d^n}}\sum_{\mathbf{k}\in (\ZdZ)^n}\ket{\mathbf{k}}$, which is the graph state with the trivial adjacency matrix $\Gamma=0$.
As the stabilizer group of $\ket{+_d}^{\otimes n}$ is given by $\mathcal{S}=\{X_d^{\mathbf{r}} \ \vert \ \mathbf{r}\in (\ZdZ)^n \}$, 
its SLD follows as  $A_k^\fullysepnd = \binom{n}{k}(d-1)^k$.

Since SLs are convex and LU-invariant, the $k$-body SL of a fully separable state cannot exceed  $A_k^\fullysepnd $.
In other words,
every $n$-qudit state with 
\begin{align} \label{eq:kSL_criterion}
    A_k[\rho] > \binom{n}{k}(d-1)^k
\end{align}
is entangled~\cite{tran_quantum_entanglement_2015, tran_correlations_between_2016}.
We refer to Eq.~\eqref{eq:kSL_criterion} as the \emph{$k$-body SL criterion}.
In all examples we know of, the $n$-body SL criterion is stronger than other $k$-body SL criteria.
To experimentally verify that a state $\rho$ is entangled, it is therefore sufficient to measure the expectation values $\Tr[\rho P_i]$ for an increasing number $N$ of weight-$n$ Pauli operators $P_1,\ldots, P_N \in \mathcal{P}_d^n$ until $\sum_{i=1}^N \vert \Tr[\rho P_i] \vert ^2$ 
exceeds the full-separability bound $(d-1)^n$ with high confidence.
This approach is particularly promising for qubits, where estimating only $N=2$ Pauli expectation values can be sufficient.
For $D>2$, on the other hand, this entanglement test is not scalable as an exponential (in $n$) number of Pauli expectation values would need to be estimated experimentally.
Note that for every ideal, i.e., noise-free, graph state $\ket{\Gamma}$, the $n$-body SL is lower bounded as
\begin{align}\label{eq:an_bound_qudit}
    A_n\left[\ket{\Gamma } \bra{\Gamma} \right] \ge    A_n^\fullysepnd  = (d-1)^n
\end{align}
because (up to a global phase)  $X^\mathbf{r}Z^{\Gamma \mathbf{r}}$ is a weight-$n$ stabilizer operator of $\ket{\Gamma}$ for every $\mathbf{r}\in \{1,\ldots, d-1\}^n$.
For a technical discussion how the bound in Eq.~\eqref{eq:an_bound_qudit} can be improved, 
see App.~B in Ref.~\cite{miller_small_quantum_2019}.
It is well known that in the case of qubits, $A_n$ is maximized by the GHZ state,
but for higher-dimensional qudits, $A_n$ is maximized by a biseparable state~\cite{tran_quantum_entanglement_2015, eltschka_maximum_nbody_2020}. 
Furthermore, some GME states with $A_n=0$ have been identified~\cite{kaszlikowski_quantum_correlation_2008}.
These two facts demonstrate that  $A_n$
only contains limited information about the entanglement of a state.
If the SLD is considered as a whole, 
however, 
it is possible to establish that a state is GME in a few cases~\cite{de_vincente_multipartite_entanglement_2011}. 
For these reasons, here we will also adopt the mindset that the SLD should be considered as a whole.

\subsection{A novel entanglement criterion for multi-qudit states based on SLDs}

By exploiting the purity criterion~\cite{horodecki_quantum_entanglement_2009}, 
we can derive the following entanglement criterion.

\subsubsection*{\refstepcounter{mycounter}Theorem~\arabic{mycounter} (Purity criterion applied to SLDs)\label{thrm:purity_criterion}} 
\emph{Let $\mathbf{A}=(A_0,\ldots, A_n)$ be the SLD of an $n$-qudit state, $\rho$, with qudit dimension  $d\ge 2$.  If}
\begin{align}\label{eq:new_criterion}
\sum_{k=0}^n \left( (d-1){n} - dk\right)A_k[\rho] < 0,
\end{align}
\emph{then $\rho$ is entangled.}
\vspace{1em}

The proof of a generalized version of this theorem is stated in App.~\ref{app:proof_purity}.
To apply Thrm.~\ref{thrm:purity_criterion}, the only information needed about a state is its SLD.
In the special case where $\ket{\Gamma}$ is an $n$-qudit graph state,
we can relate the SLD to a graph-theoretical problem that generalizes Thrm.~\ref{thrm:puzzle}.
For this, we associate every element  $r\in\ZdZ$ with a color. 
Then, each color assignment (with $d$ colors) of $\Gamma$ corresponds to a stabilizer operator $X^\mathbf{r}Z^{\Gamma \mathbf{r}}$ (up to phase) and contributes to $A_k$  iff exactly $n-k$ white ($r_i=0$) vertices $i\in\{1,\ldots,n\}$ have the property 
\begin{align} \label{eq:puzzle_qudit}
\sum_{j=1}^n \gamma_{i,j} r_j = 0.
\end{align}
Only for qubits, Eq.~\eqref{eq:puzzle_qudit} simplifies to the property \emph{``the number of vertices $j$ with $\gamma_{i,j}=1$ and $r_j=1$ is equal to 0 modulo 2, i.e., even''.}
In the qudit case, the situation is more involved.
This is because computing $A_k$ amounts to counting solutions to equations in modular arithmetic,
see App.~\ref{app:qudits} for a
more detailed treatment.
Here, we restrict ourselves to presenting only some of our less-technical results:
If $d$ is a prime number, we find 
\begin{align}\label{eq:a1_qudit_prime}
    A_1 = (d-1) I
\end{align}
as a generalization of Eq.~\eqref{eq:a1}, 
where $I$ again denotes the number of isolated vertices of $\Gamma$.
Furthermore, we find for $d$ prime that the $2$-body SL obeys
\begin{align} \label{eq:a2_qudit_prime_both_bounds}
T_0(d-1)^2 + (L+T_1)  (d-1)
\hspace{1mm}
\le 
\hspace{2mm}
A_2 
\hspace{2mm}
\le 
\hspace{1mm}
T_0(d-1)^2 + \left(L + \sum_{m=1}^{n-2}T_m\right) (d-1),
\end{align}
where $L$ is the number of leaves and $T_m$ denotes the number of (twin) vertex pairs with exactly $m$ common neighbors and zero non-shared neighbors, e.g., $T_0=\binom{I}{2}$.
For a given graph state $\ket{\Gamma}$, one can efficiently compute the exact value of $A_2$ by exploiting the formula 
\begin{align} \label{eq:Ak_coloring_qudit}
    A_k = \sum_{b=1}^k \sum_{\mathbf{r} \in \mathcal{D}_{b}} \delta_{\swt_d(\mathbf{r},\Gamma\mathbf{r}),k},
\end{align} 
which generalizes Eq.~\eqref{eq:Ak_coloring} and holds for arbitrary $d\ge 2$ and $k\ge 1$.
Here, $\mathcal{D}_{b}\subset (\ZdZ)^n$ denotes the subset of ``dit'' strings with exactly $b$ nonzero entries. 
The evaluation runtime of Eq.~\eqref{eq:Ak_coloring_qudit} is given by $\mathcal{O}((dn)^k)$, which is efficient for small values of $k$.
This also enables the efficient computation of the mean and the variance of the normalized SLD of a qudit graph state for arbitrary $d$
via Eqs.~\eqref{eq:purity1_qudit} and~\eqref{eq:purity2_qudit}. 
Finally note that, if $d$ is prime, every $n$-qudit stabilizer state is LU-equivalent to a qudit graph state~\cite{bahramgiri_graph_states_2006},
which further extends the applicability of our results.

\section{SLDs of noisy states}
\label{sec:5}
Until this point, we have exclusively focused on SLDs of \emph{pure} quantum states.
In reality, however, experimental imprecision and decoherence always lead to some uncertainty about the state of a quantum system.
This necessitates that we extend our discussion to the more general case of \emph{mixed} states.
In Sec.~\ref{sec:5.1}, we investigate the impact of noise on the SLD of a general $n$-qudit state.
Then, in Sec.~\ref{sec:5.2}, we apply our insights for the derivation of noise levels below which entanglement is preserved.
 
\subsection{The impact of noise on qudit SLDs}
\label{sec:5.1}
The $n$-qudit depolarizing channel of strength $p\in [0,1]$, which is defined via
\begin{align}\label{eq:glob_white_noise} 
\mathcal{E}^{(p)}_\text{glob}[\rho]  = (1-p)\rho + p \frac{\mathbbm{1}}{d^n},
\end{align}    
is a very simplistic model that describes \emph{global white noise} acting on all qudits simultaneously.
Since there are only two terms in Eq.~\eqref{eq:glob_white_noise},
global white noise is easy to treat theoretically and, therefore, often used as a first approximation.
A more realistic error channel, which takes spatial separation of qudits into account, is the \emph{local white noise channel} 
\begin{align} \label{eq:loc_white_noise}
	 \mathcal{E}^{(p)}_\text{loc} [\rho] &=  \left(\mathcal{E}^{(p)}\right)^{\otimes n}[\rho], 
\end{align} 
where $\mathcal{E}^{(p)}$ denotes the single-qudit depolarizing channel of strength $p$.
Both the global and the local white noise channel are generalized Pauli channels,
\begin{align}
\mathcal{E}^{(p)}_\text{glob/loc} [\rho] 
 =  \sum_{\mathbf{r},\mathbf{s}\in (\ZdZ)^n} p^\text{glob/loc}_{\mathbf{r},\mathbf{s}}  (X_d^\mathbf{r} Z_d^\mathbf{s})\, \rho\, (X_d^\mathbf{r} Z_d^\mathbf{s})^\dagger,
\end{align}
where a discrete Pauli error $X_d^\mathbf{r}Z_d^\mathbf{s}$
occurs with probability
\begin{align}\label{eq:p_glob}
 p^\text{glob}_{\mathbf{r},\mathbf{s}} = \begin{cases}
1- p +\frac{p}{d^{2n}}, & \text{ if } \mathbf{r}=\mathbf{s} = (0,\ldots,0) \\ 
\frac{p}{d^{2n}},  & \text{ otherwise}
 \end{cases}
\end{align}
and 
\begin{align}\label{eq:p_loc}
p^\text{loc}_{\mathbf{r},\mathbf{s}} = \left( \frac{p}{d^2} \right)^{\swt_d(\mathbf{r},\mathbf{s})}\left(1-p+ \frac{p}{d^2} \right)^{n-\swt_d(\mathbf{r},\mathbf{s})},
\end{align}
respectively~\cite{miller_propagation_of_2018}.
To establish the influence of any given quantum channel $\mathcal{E}$ 
on the Bloch decomposition of an $n$-qudit state $\rho$ as in Eq.~\eqref{eq:PauliDecomposition_qudit}, 
it suffices to compute how $\mathcal{E}$ acts on individual Pauli operators.
This is because $\mathcal{E}$ is a linear map, 
\begin{align}
    \mathcal{E}\left[ \frac{1}{d^n} \sum_{\mathbf{r},\mathbf{s}\in(\ZdZ)^n} \rho_{\mathbf{r},\mathbf{s}} X_d^\mathbf{r}Z_d^\mathbf{s} \right] = \frac{1}{d^n} \sum_{\mathbf{r},\mathbf{s}\in(\ZdZ)^n}
  \rho_{\mathbf{r},\mathbf{s}} \,
  \mathcal{E}\left[  X_d^\mathbf{r}Z_d^\mathbf{s} \right].
\end{align}
By exploiting $\Tr[X_d^\mathbf{r}Z_d^\mathbf{s}]= \delta_{\mathbf{r},0} \delta_{\mathbf{s},0} d^n$, we find 
\begin{align}
     \mathcal{E}^{(p)}_\text{glob} [X_d^\mathbf{r}Z_d^\mathbf{s}] = (1-p)X_d^\mathbf{r} Z_d^\mathbf{s} + p \Tr[X_d^\mathbf{r}Z_d^\mathbf{s}] \frac{\mathbbm 1}{d^n} = \begin{cases}
\mathbbm 1, & \text{ if } \mathbf{r}=\mathbf{s}=0 \\
(1-p)X_d^\mathbf{r} Z_d^\mathbf{s}, & \text{ otherwise}
     \end{cases}
\end{align}
for the global white noise channel and
\begin{align} 
 \mathcal{E}^{(p)}_\text{loc} [X_d^\mathbf{r}Z_d^\mathbf{s}] 
& = \bigotimes_{i=1}^n \left(  (1-p) X_d^{r_i}Z_d^{s_i} + p \Tr[X^{r_i}Z^{s_i}]\frac{\mathbbm 1}{d}  \right) = (1-p)^{\swt_d(\mathbf{r},\mathbf{s})} X_d^\mathbf{r}Z_d^\mathbf{s}.
\end{align}
for local white noise.
Inserting this into Eq.~\eqref{def:Ak_qudit} yields the $k$-body SLs,
\begin{align} \label{eq:Ak_p_glob}
A_k \left[ \mathcal{E}^{(p)}_\text{glob} [\rho]   \right] 
 &= (1-p)^{2 } A_k \left[ \rho \right] \\
	\text{and } \hspace{5mm}
 A_k\left[ \mathcal{E}^{(p)}_\text{loc} [\rho]   \right] 
 &= (1-p)^{2k} A_k \left[ \rho \right] 
 \label{eq:Ak_p_loc}
 \end{align}
of the noisy states $ \mathcal{E}^{(p)}_\text{glob} [\rho] $ and  $ \mathcal{E}^{(p)}_\text{loc} [\rho]$.
Since the prefactor $(1-p)^{2k}$ is exponentially suppressed,
the correlations between large numbers of subsystems are strongly diminished in the presence of local white noise.
This is unsurprising because, by Eq.~\eqref{eq:p_loc},
Pauli errors $X_d^\mathbf{r}Z_d^\mathbf{s}$ that jointly affect a large number $k=\swt_d(\mathbf{r},\mathbf{s})$ of subsystems are very unlikely to occur.
In a recent work~\cite{quek_exponentially_tighter_2022},
where Eq.~\eqref{eq:Ak_p_loc} was independently derived, 
this insight played a role in establishing stringent limitations on the experimental feasibility of quantum error mitigation protocols on near-term quantum computers.

\subsection{Lower bounds on entanglement noise thresholds}
\label{sec:5.2}
 Quantum entanglement is a crucial resource for many quantum information protocols, especially in quantum communication~\cite{horodecki_quantum_entanglement_2009, graselli_quantum_cryprography_2020}.
Here, we address the question
\emph{``how much noise can an entangled state tolerate before it becomes fully separable?''}.

For example, if $\ket{\psi}$ is an $n$-qudit stabilizer state that is not fully separable, 
then the noisy state $\mathcal{E}^{(p)}_\text{glob} \big[\ket{\psi}\bra{\psi} \big]$ is also entangled for all values of $p$ that are smaller than
\begin{align}\label{eq:threshold_stab_global_PPT}
p_\text{PPT,glob}^\text{stab} = 1-\frac{1}{d^{n-1}+1},
\end{align}
as we show in Ref.~\cite{miller_small_quantum_2019} for arbitrary $d$ and $n$ by exploiting the positive partial transpose (PPT) criterion~\cite{peres_separability_criterion_1996, horodecki_separability_of_1996}. 
While Eq.~\eqref{eq:threshold_stab_global_PPT} is both simple and general, 
its physical relevance is questionable since $p_\text{PPT,glob}^\text{stab} \rightarrow 1$ for $n \rightarrow \infty$.
In other words, entanglement can be preserved arbitrarily well by adding more and more qudits in state $\ket{0}$ to a system that is affected by global white noise.
This behavior is clearly unphysical in a quantum communication setting, in which the qudits are spatially separated.
For this setting, the local white noise model is more appropriate.

Luckily, it is also possible to derive noise thresholds for the case of local white noise, e.g., 
we can insert Eq.~\eqref{eq:Ak_p_loc} into Eq.~\eqref{eq:kSL_criterion} and solve for $p$.
For every $n$-qudit state $\rho$, this yields that the noisy state $\mathcal{E}^{(p)}_\text{loc} [\rho]$ is entangled for all values of $p$ below
\begin{align} \label{eq:nSL_bound}
    p _{n\text{SL},\text{loc}} = 1- \sqrt[2n]{ \frac{(d-1)^n}{A_n[\rho]}}.
\end{align} 
Recall from Eq.~\eqref{eq:an_bound_qudit} that every $n$-qudit graph state $\ket{\Gamma}$ has $A_n \ge (d-1)^n$.
Hence, for a nontrivial threshold $p _{n\text{SL},\text{loc}}>0$ it is sufficient that the graph $\Gamma$ admits a color assignment contributing to $A_n$ with at least one white vertex.
Furthermore, we can exploit Eq.~\eqref{eq:Ak_p_loc} in combination with our new entanglement criterion from Thrm.~\ref{thrm:purity_criterion} to derive the following result:

\subsubsection*{\refstepcounter{mycounter}Corollary~\arabic{mycounter} (Local-white-noise threshold for entanglement)\label{cor:purity_criterion_threshold}} 

\emph{Let $\rho$ be an $n$-qudit state that is entangled by Thrm.~\ref{thrm:purity_criterion}. Then,  the polynomial function}
\begin{align}\label{eq:purity_criterion_threshold}
f: [0,1] \longrightarrow \mathbb R, \hspace{5mm} 
p \longmapsto \sum_{k=0}^n \left( (d-1){n} - dk\right)(1-p)^{2k}A_k[\rho] 
\end{align}
\emph{has a root $p_\mathrm{pur,loc} \in (0,1)$ at which the sign of $f$ changes from minus to plus.
Moreover, every such solution is a lower bound on the local-white-noise threshold for entanglement, i.e.,  $\mathcal{E}^{(p)}_\mathrm{loc} [\rho] $ is entangled for every value of $p< p_\mathrm{pur,loc}$}.

\begin{proof}
By assumption, we have $f(0)<0$. 
Because of $f(1) = (d-1)n > 0$, a solution of $f(p_\text{pur,loc})=0$ with $0<p_\text{pur,loc}<1$ and  the desired sign change is guaranteed by the intermediate value theorem.
Without loss of generality, $p_\text{pur,loc}$ is the largest (polynomials have finitely many roots) such solution.
Now, let $p < p_\text{pur,loc}$ and consider the state $\rho' = \mathcal{E}^{(p)}_\text{loc} [\rho]$.
By construction, it is possible to select $p' \in [p, p_\text{pur,loc})$ with $f(p')<0$.
Because of $0 \le p \le p' < 1$, we have $ 0 \le \tfrac{p'-p}{1-p}< 1$.
Thus, we can apply a depolarizing channel of strength $q = \tfrac{p'-p}{1-p}$ to every qudit of $\rho'$.
Because of $\mathcal{E}^{(q)}_\text{loc} \left [ \mathcal{E}^{(p)}_\text{loc} \left [ \rho \right] \right] = 
\mathcal{E}^{(p+q-pq)}_\text{loc} \left [\rho \right]$, this results in the state 
$\mathcal{E}^{(q)}_\text{loc} \left [ \rho' \right] = \mathcal{E}^{(p')}_\text{loc} \left [ \rho \right]$, which is entangled by $f(p')<0$.
Since local operations cannot create entanglement, the initial state $\rho'$ must have been entangled as well.
\end{proof}

Note that Cor.~\ref{cor:purity_criterion_threshold} is constructive as $p_\text{pur,loc}>0$ can always be found algorithmically, for instance via the bisection method. 
For example, we can apply Cor.~\ref{cor:purity_criterion_threshold} to the logical states $\ket{0}_\text{L}$, $\ket{1}_\text{L}$, $\ket{+}_\text{L}$, and $\ket{-}_\text{L}$ of the $\llbracket n,1,\sqrt{n} \rrbracket $ rotated surface code~\cite{bombin_optimal_resources_2007, bravyi_correcting_coherent_2018}, all of which have the same SLD $\mathbf{A}^{\text{surf}(n)}$ because the logical operators $X_\text{L}$ and $Z_\text{L}$ are transversal.
For the smallest nontrivial instance of the rotated surface code, we find
\begin{align} 
\mathbf{A}^\text{surf(9)} = (1,\ 0,\ 4,\ 12,\ 22,\ 52,\ 100,\ 148,\ 129,\ 44),
\end{align}
which yields $p_\text{pur,loc}^\text{surf(9)} \approx 0.28$, whereas the bound  $p _{n\text{SL},\text{loc}}^\text{surf(9)} \approx 0.19$ from Eq.~\eqref{eq:nSL_bound}, which is based on the previously-known $n$-body SL criterion, is weaker.
Both bounds show that
the entanglement-noise threshold for $\ket{0}_\text{L}$ etc., lies well above the error-correcting threshold $\sim 1 \%$ of the surface code~\cite{fowler_surface_codes_2012}.
Similarly, we compute 
\begin{align}
\nonumber
\mathbf{A}^\text{surf(25)}  
= (&
1,\	
0,\	
8,\	
0,\	
72,\	
80,\	
534,\	
984,\	
3715,\	
8776,\	
25816,\	
62160,\	
158448,	
\\ &
386416,\	
782532,\	
1561984,\	
2726047,\	
3951328,\	
5115376,\	
5666352,	
\\ &
5136632,\	
3919936,\	
2437206,\	
1141160,\	
390829,\	
78040	
),
\nonumber
\end{align}
which yields 
$p_\text{pur,loc}^\text{surf(25)} \approx 0.31$ and 
$p _{n\text{SL},\text{loc}}^\text{surf(25)} \approx 0.20$. 
This indicates that entanglement is better preserved in states of quantum error-correcting codes with larger code distances.
We find similar results for other families of stabilizer states and refer the interested reader to App.~\ref{app:noise_thresholds}.

\section{Conclusion and outlook}
\label{sec:conclusion}
 In this paper, we developed the theory of Shor-Laflamme distributions (SLDs) of graph states, 
which has its historical origins in Refs.~\cite{vandennest_finite_set_2005, cabello_compact_set_2009}.
The starting point of our exploration was Thrm.~\ref{thrm:puzzle}, which relates SLDs to a graph color assignment problem. 
By solving this problem in the special case of Pusteblume graph states and ring cluster (RC) states, 
we derived explicit formulas for their SLDs.
In this way, we extended the list of analytically known SLDs, which to our knowledge was hitherto limited to fully separable states, Greenberger-Horne-Zeilinger (GHZ) states, Dicke states, and tensor products thereof~\cite{aschauer_local_invariants_2004}. 
For graph states based on random Erdős-Rényi graphs, we discovered that the normalized SLD is remarkably close to an asymmetrical binomial distribution; hence, we proposed to call such SLDs \emph{generic}.
This discovery was spurred by a visualization tool that we report separately in Ref.~\cite{miller_graphstatevis_interactive_2021}.
While SLDs of GHZ states and alike are not generic, those of cluster states are.
Hence, RC states now constitute the only family of states with analytically-known, generic SLDs.
Further consequences of Thrm.~\ref{thrm:puzzle} are captured in Cor.~\ref{cor:an_defect} and
Cor.~\ref{cor:mean_variance}, 
which provide simple formulas for a bound on the full-body sector length for certain graph states 
and for the mean and the variance of the normalized SLD for arbitrary graph states, respectively.
Additionally, we formulated similar results for the more general case of higher-dimensional qudits.

While our theoretical developments have their own intrinsic academic relevance, 
we can also apply them to tackle difficult relevant problems in other branches of quantum information theory.
To accomplish this, in Thrm.~\ref{thrm:purity_criterion} we reformulate the purity criterion~\cite{nielsen_separable_states_2001} such that (a potentially weaker form of) it can be tested based on knowledge of the SLD alone.
After having derived formulas for the decline of SLDs in the presence of noise,
we deduce Cor.~\ref{cor:purity_criterion_threshold} 
and apply it to compute lower bounds on noise thresholds for entanglement.
In some cases, this approach allows us to outperform the best previous results based on other criteria~\cite{hein_entanglement_properties_2005}.

By definition, the SLDs are invariants of degree two in the quantum state. Analogous to other hierarchies of entanglement criteria~\cite{navascues2008convergent}, we expect more information on the state to be embodied in higher-degree invariants.
Thus, further research could focus on 
investigating higher-degree 
generalizations of SLDs. 
Similarly, it could be fruitful to extend the discussion of qudit graph states to the case of continuous variable systems~\cite{adesso2007entanglement, sun2012entanglement},
and search for easily-applicable entanglement criteria that are similar to our Thrm.~\ref{thrm:purity_criterion}.

We envision that our graph-theoretical formulation of the SLD problem will facilitate the discovery of SLDs for a wider range of quantum states, 
e.g., for certain logical states of quantum error-correcting (QEC) codes 
or for cluster states that appear in measurement-based quantum computation (MBQC).
As QEC and MBQC are fields that heavily rely on the stabilizer formalism, 
we anticipate that our results will find applications there.
Last but not least, we hope that our work will stimulate the investigation of SLDs in a more general setting, e.g., for Dicke states for which the SLDs are available~\cite{aschauer_local_invariants_2004}, 
or for qubit (or qudit) hypergraph states for which a theory of SLDs is not  developed yet~\cite{rossi_quantum_hypergraph_2013}.

\section*{Acknowledgments}
We thank Lennart Bittel, Felix Huber, Matthias Miller, and Gordon Royle for fruitful discussions.
This research is part of a project that has received funding from the European Union’s Horizon 2020 research and innovation programme under the Marie Skłodowska-Curie grant agreement No 847471.
This work was supported as a part of NCCR SPIN, a National Centre of Competence (or Excellence) in Research, funded by the Swiss National Science Foundation (grant number 51NF40-180604).
This work received financial support from the Munich Quantum Valley (K-8), the BMBF (HYBRID, REALISTIQ, MUNIQC-Atoms), and the EU Quantum Technology Flagship (MILLENION).
N.~W.~acknowledges support by the QuantERA project QuICHE via the German Ministry of Education and Research (BMBF Grant No. 16KIS1119K).
IBM, the IBM logo, and ibm.com are trademarks of International Business Machines Corp., registered in many jurisdictions worldwide. Other product and service names might be trademarks of IBM or
other companies. The current list of IBM trademarks is available at \url{https://www.ibm.com/legal/copytrade}.

\appendix

\section{Solution of the graph-theoretical problem for the Pusteblume graph}  
\label{app:dandelion}
Here we solve the color assignment problem for Pusteblume graphs, see Fig.~\ref{fig:pusteblume}, which will prove Eq.~\eqref{eq:Ak_dand} from the main text.
To accomplish this, we distinguish the four cases where vertices $1$ and $2$ are black and white, respectively.
Note that, in each case, there are $2^{n-2}$ color assignments that contribute to certain SLs. 

\begin{figure}[t]

\begin{center}
\begin{tabular}{cccc}
\small
 \hspace{1em}
\begin{tikzpicture}
\small  
\draw[-, line width=.1em] (0,0.2) -- (0,0.8);  
\draw[-, line width=.1em] (0.1732,0.1) -- (0.6928, 0.4);  
\draw[-, line width=.1em] (-0.1732,0.1) -- (-0.6928, 0.4);  
\draw[-, line width=.1em] (0,-0.2) -- (0,-1.8);   
\draw[-, line width=.1em] (0.1732,-1.9) -- (0.6928, -1.6);  
\draw[-, line width=.1em] (-0.1732,-1.9) -- (-0.6928, -1.6);

\draw[line width=.1em, fill=black] (0,-2) circle (0.2);  
\draw[line width=.1em, fill=black] (0,0) circle (0.2);   
\draw[line width=.1em, fill=white] (-0.866,-1.5) circle (0.2);  
\draw[line width=.1em, fill=white] ( 0.866,-1.5) circle (0.2);  
\draw[line width=.1em, fill=white] (-0.866,0.5) circle (0.2);   
\draw[line width=.1em, fill=white] (0,1) circle (0.2);  
\draw[line width=.1em, fill=white] (0.866,0.5) circle (0.2);   

\draw (0,-2) node[text=white]{\tiny 1}; 
 \draw (0,0) node[text=white]{\tiny 2};  
 \draw (-0.866,-1.5) node[text=black]{\tiny 3}; 
 \draw ( 0.866,-1.5) node[text=black]{\tiny 4}; 
 \draw (-0.866, 0.5) node[text=black]{\tiny 5};
 \draw         (0,1) node[text=black]{\tiny 6};     
 \draw ( 0.866, 0.5) node[text=black]{\tiny 7};
 \end{tikzpicture}  
 \hspace{1em}
 & 
 \hspace{1em}
\begin{tikzpicture}
\small  
\draw[-, line width=.1em] (0,0.2) -- (0,0.8);  
\draw[-, line width=.1em] (0.1732,0.1) -- (0.6928, 0.4);  
\draw[-, line width=.1em] (-0.1732,0.1) -- (-0.6928, 0.4);  
\draw[-, line width=.1em] (0,-0.2) -- (0,-1.8);   
\draw[-, line width=.1em] (0.1732,-1.9) -- (0.6928, -1.6);  
\draw[-, line width=.1em] (-0.1732,-1.9) -- (-0.6928, -1.6);

\draw[line width=.1em, fill=white] (0,-2) circle (0.2);  
\draw[line width=.1em, fill=black] (0,0) circle (0.2);   
\draw[line width=.1em, fill=black] (-0.866,-1.5) circle (0.2);  
\draw[line width=.1em, fill=white] ( 0.866,-1.5) circle (0.2);  
\draw[line width=.1em, fill=white] (-0.866,0.5) circle (0.2);   
\draw[line width=.1em, fill=white] (0,1) circle (0.2);  
\draw[line width=.1em, fill=white] (0.866,0.5) circle (0.2);   

\draw (0,-2)         node[text=black]{\tiny 1}; 
 \draw (0,0)         node[text=white]{\tiny 2};  
 \draw (-0.866,-1.5) node[text=white]{\tiny 3}; 
 \draw ( 0.866,-1.5) node[text=black]{\tiny 4}; 
 \draw (-0.866, 0.5) node[text=black]{\tiny 5};
 \draw         (0,1) node[text=black]{\tiny 6};     
 \draw ( 0.866, 0.5) node[text=black]{\tiny 7};
 \end{tikzpicture}  
 \hspace{1em}
 & 
 \hspace{1em}
\begin{tikzpicture}
\small  
\draw[-, line width=.1em] (0,0.2) -- (0,0.8);  
\draw[-, line width=.1em] (0.1732,0.1) -- (0.6928, 0.4);  
\draw[-, line width=.1em] (-0.1732,0.1) -- (-0.6928, 0.4);  
\draw[-, line width=.1em] (0,-0.2) -- (0,-1.8);   
\draw[-, line width=.1em] (0.1732,-1.9) -- (0.6928, -1.6);  
\draw[-, line width=.1em] (-0.1732,-1.9) -- (-0.6928, -1.6);

\draw[line width=.1em, fill=black] (0,-2) circle (0.2);  
\draw[line width=.1em, fill=white] (0,0) circle (0.2);   
\draw[line width=.1em, fill=white] (-0.866,-1.5) circle (0.2);  
\draw[line width=.1em, fill=white] ( 0.866,-1.5) circle (0.2);  
\draw[line width=.1em, fill=black] (-0.866,0.5) circle (0.2);   
\draw[line width=.1em, fill=white] (0,1) circle (0.2);  
\draw[line width=.1em, fill=white] (0.866,0.5) circle (0.2);   

\draw (0,-2)         node[text=white]{\tiny 1}; 
 \draw (0,0)         node[text=black]{\tiny 2};  
 \draw (-0.866,-1.5) node[text=black]{\tiny 3}; 
 \draw ( 0.866,-1.5) node[text=black]{\tiny 4}; 
 \draw (-0.866, 0.5) node[text=white]{\tiny 5};
 \draw         (0,1) node[text=black]{\tiny 6};     
 \draw ( 0.866, 0.5) node[text=black]{\tiny 7};
 \end{tikzpicture}  
 \hspace{1em}
 & 
 \hspace{1em}
\begin{tikzpicture}
\small  
\draw[-, line width=.1em] (0,0.2) -- (0,0.8);  
\draw[-, line width=.1em] (0.1732,0.1) -- (0.6928, 0.4);  
\draw[-, line width=.1em] (-0.1732,0.1) -- (-0.6928, 0.4);  
\draw[-, line width=.1em] (0,-0.2) -- (0,-1.8);   
\draw[-, line width=.1em] (0.1732,-1.9) -- (0.6928, -1.6);  
\draw[-, line width=.1em] (-0.1732,-1.9) -- (-0.6928, -1.6);

\draw[line width=.1em, fill=white] (0,-2) circle (0.2);  
\draw[line width=.1em, fill=white] (0,0) circle (0.2);   
\draw[line width=.1em, fill=white] (-0.866,-1.5) circle (0.2);  
\draw[line width=.1em, fill=white] ( 0.866,-1.5) circle (0.2);  
\draw[line width=.1em, fill=black] (-0.866,0.5) circle (0.2);   
\draw[line width=.1em, fill=black] (0,1) circle (0.2);  
\draw[line width=.1em, fill=white] (0.866,0.5) circle (0.2);   

\draw (0,-2) node[text=black]{\tiny 1}; 
 \draw (0,0) node[text=black]{\tiny 2};  
 \draw (-0.866,-1.5) node[text=black]{\tiny 3}; 
 \draw ( 0.866,-1.5) node[text=black]{\tiny 4}; 
 \draw (-0.866, 0.5) node[text=white]{\tiny 5};
 \draw         (0,1) node[text=white]{\tiny 6};     
 \draw ( 0.866, 0.5) node[text=black]{\tiny 7};
 \end{tikzpicture}
 \hspace{1em}  \\
 (a) & (b) & (c) & (d)
 \end{tabular}
\end{center}

\caption{Some black-white color assignments of a Pusteblume graph with $n=7$ vertices. 
(a) All color assignments where vertex 1 and 2 are black contribute to $A_n$.
(b) If vertex 1 is white, vertex 2 is black, and vertex 3 is black, there are exactly two white vertices (1 and 4) with an even number of black neighbors, i.e., such a color assignment contributes to $A_{n-2}$.
(c) If vertex~1 is black and vertex 2 is white, every white neighbor of vertex~2 has an even number (zero) of white neighbors.
(d) If vertex~1 and 2 are white, the white vertices with an even number of black neighbors are vertex 1, 3, 4, all white neighbors of vertex 2, and possibly vertex~2.
}
\label{fig:pusteblume_colorings}
\end{figure}
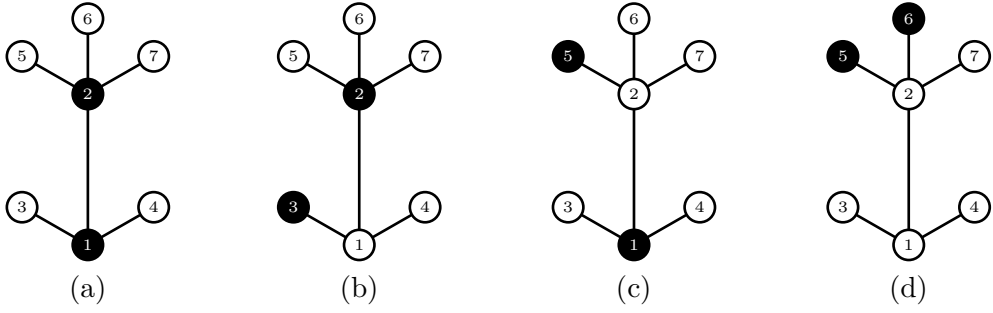 
\begin{itemize}
    \item 
If both vertex~$1$ and~$2$ are black,
see Fig.~\ref{fig:pusteblume_colorings} (a),
there cannot be a white vertex with an even number of black neighbors because all remaining vertices are leaves with a black neighbor. Thus, all $2^{n-2}$ color assignments of the leaves contribute to $A_n$.

\item
If vertex~$1$ is white and vertex~$2$ is black,
see Fig.~\ref{fig:pusteblume_colorings} (b),
the colors of the $n-4$ neighbors of vertex~$2$ do not influence the number of white vertices having an even number of black neighbors; only the colors of vertex~$3$ and~$4$ do.
If both of these vertices are black, vertex~1 has three black neighbors, and all $2^{n-4}$ color assignments of vertex $5,\ldots, n$ contribute to $A_n$. 
Otherwise, there are two white vertices with an even number of black neighbors, i.e., the remaining $3\times 2^{n-4}$ color assignments contribute to $A_{n-2}$.

\end{itemize}

The first of these four cases has a contribution of $2^{n-2}\delta_{k,n}$ to $A_k$, while the second case contributes $ 3\times2^{n-4}\delta_{k,n-2} +   2^{n-4}\delta_{k,n} $.
Thus, together they yield the term 
\begin{align}
3\times2^{n-4}\delta_{k,n-2} +   5\times2^{n-4}\delta_{k,n},
\end{align} 
in Eq.~\eqref{eq:Ak_dand} from the main text.
We continue with the remaining two cases.
\begin{itemize}
\item
If vertex~$1$ is black and vertex~$2$ is white, 
see Fig.~\ref{fig:pusteblume_colorings} (c),
all four color assignments of vertex~$3$ and~$4$ (neighbors of vertex $1$) contribute to the same SL. 
There are $\binom{n-4}{b} $ color assignments of the neighbors of vertex~$2$ with $b$ black neighbors.
Since vertex~$2$ is white, all its $n-4-b$ white neighbors have an even number (zero) of black neighbors. If $b$ is odd, vertex~$2$ also has an even number of black neighbors. In that case, all $4 \binom{n-4}{b}$ color assignments contribute to 
$A_{n-(n-4-b+1)}=A_{b+3}$.
If $b$ is even, however, we have  $4 \binom{n-4}{b}$ color assignments contributing to $A_{b+4}$ because the white vertex~$2$ has an odd  number of black neighbors.

\item 
If both vertex~$1$ and~$2$ are white,
see Fig.~\ref{fig:pusteblume_colorings} (d),
we look at vertex~$3$ and~$4$ first.
If they are also white, we already have three white vertices ($1,3$, and $4)$ having an even (zero) number of black neighbors.
Otherwise, there is exactly one such white vertex among $1,3$, and $4$.
Again, we distinguish between the $\binom{n-4}{b}$ color assignments of vertex $5,\ldots n$ with exactly $b$ black vertices.
If $b$ is odd, the total number of white vertices with an even number of black neighbors is either $(n-4-b)+3$ (if all $1,3$, and $4$ are among them) or $(n-4-b)+1$ (if only one vertex among $1,3$, and $4$ is white and has an even number of black neighbors).
Thus, if $b$ is odd, we have $\binom{n-4}{b}$ color assignments which contribute to $A_{b+1}$ and $3 {\binom{n-4}{b}}$ contributing to $A_{b+3}$.
If $b$ is even, however, the total number of white vertices with an even number of black neighbors is either $(n-4-b)+3+1$   or $(n-4-b)+1+1$ because now vertex~2 is also one of them.
Thus, we have $\binom{n-4}{b}$ color assignments contributing $A_{b}$ and $3 {\binom{n-4}{b}}$ contributing to $A_{b+2}$.
\end{itemize}  
We have distinguished between the number $b\in\{0,\ldots,n-4\}$ of the dandelion seed head vertices being black.
We find that we only get a contribution to $A_k$ if $k$ is even.
From case three, $A_k$ gets a contribution of $4\binom{n-4}{k-4}+ 4\binom{n-4}{k-3}= 4\binom{n-3}{k-3} $, where the first and second terms come from the color assignments where $b=k-4$ and $b=k-3$, respectively.
Similarly, from case four, $A_k$ gets a contribution of $3\binom{n-4}{k-3}+3\binom{n-4}{k-2} + \binom{n-4}{k-1}+\binom{n-4}{k} = 3 \binom{n-3}{k-2} + \binom{n-3}{k}$, if $k$ is even.
In total, case three and four have a contribution of
\begin{align} \left( \binom{n-3}{k-3} + 3\binom{n-2}{k-2} + \binom{n-3}{k} \right)\delta_{k,\mathrm{even}} 
\end{align} 
to $A_k$ for each $k$. 
This finishes the derivation of 
Eq.~\eqref{eq:Ak_dand} stated in the main text.

\section{Solution of the graph-theoretical problem for the cycle graph}  
\label{app:ring}
In this appendix, we derive the SLD of the $n$-qubit ring cluster state, which is stated in Eq.~\eqref{eq:Ak_RCL}.
According to the color assignment problem, $A_k$ is the number of color assignments of the $n$-vertex cycle graph, see Fig.~\ref{fig:cycle}, such that $n-k$ white vertices have an even number of black neighbors.
All possible color assignments are parameterized by the set $\FF_2^n$ where $\mathbf{r} \in \FF_2 ^n$ corresponds to the color assignment where vertex $i\in V$ is white if $r_i =0$ and black if $r_i=1$.
To capture the periodicity of the cycle, we use  integers modulo $n$ as the vertex set $V=\ZnZ$.
Since each vertex of the cycle graph has exactly two neighbors, the condition of having an even number of black neighbors is equivalent to the condition of both neighbors having the same color.
Thus, the number of white vertices fulfilling this condition for a given color assignment $\mathbf{r}\in \FF_2^n$ 
can be expressed as
\begin{align}
x_1(\mathbf{r}) = \left \vert \left \{ i \in V \ \big \vert \ r_{i}=0 , \ r_{i-1}=r_{i+1} \right\} \right \vert.
\end{align} 
By introducing the notation 
 $x_2(\mathbf{r}) = \left \vert \left \{ i \in V \  \vert \ r_{i}=0 , \ r_{i-1} \neq r_{i+1} \right\} \right \vert$  
 for the number of other white vertices, 
as well as  
 $x_3(\mathbf{r}) = \left \vert \left \{ i \in V \  \vert \ r_{i}=1  \right\} \right \vert$ 
for the number of black vertices, we obtain the relation $x_1(\mathbf{r})+x_2(\mathbf{r})+x_3(\mathbf{r}) = n$.
Therefore, the $k$-body sector length $A_k$ is given by the cardinality of the set
\begin{align} \label{eq:SLRing_Xk}
\mathcal{X}_k = \left\{ \mathbf{r} \in \FF_2^n \ \big \vert \  x_1(\mathbf{r}) = n-k \right\} 
= \left\{ \mathbf{r} \in \FF_2^n \ \big \vert \  x_2 (\mathbf{r}) + x_3(\mathbf{r}) = k \right\} . 
\end{align} 
At each vertex $i\in V$ with $r_{i-1}=1$, $r_i=r_{i+1}=0$ there starts a path of $l_i\ge 2$ white vertices, i.e., $r_{i+2}=\ldots = r_{i+l_i}=0$ but $r_{i+l_i+1}=1$. The inner vertices contribute to $x_1(\mathbf{r})$ as they have two white neighbors. The two ends of the white path, however, contribute to $x_2(\mathbf{r}) = 2 m(\mathbf{r})$, where
 $m(\mathbf{r})$ is the number of white paths of length $l\ge 2$.
By sorting the color assignments  $\mathbf{r}\in \mathcal{X}_k$ by $m(\mathbf{r})$, we obtain the disjoint union
$\mathcal{X}_k = \bigcup_{m=0}^{\lfloor k/2 \rfloor} \mathcal{X}_k^{(m)}$ into the sets
\begin{align}
\mathcal{X}^{(m)}_k=  \left\{ \mathbf{r} \in \FF_2^n \ \big \vert \ x_2 (\mathbf{r}) =2m ,  \  x_3(\mathbf{r})= k-2m \right\}.
\end{align}
Note that $m$ only runs from 0 to $\lfloor k/2\rfloor$ because, otherwise, $x_2(\mathbf{r})$ or $x_3(\mathbf{r})$ would be negative.
By defining  $C_{n,b,m}$ as the \emph{number of color assignments of the $n$-vertex cycle graph with exactly $b\ge 0$ black vertices and exactly $m\ge 0$ white paths of length greater than or equal to 2}, we obtain the formal expression
\begin{align} \label{eq:SLRing_Ak_unsolved}
A_k = \sum_{m=0}^{\left\lfloor \frac{k}{2} \right \rfloor} C_{n,k-2m,m } ,
\end{align}
which will reduce to Eq.~\eqref{eq:Ak_RCL} once we have found explicit formulas for $C_{n,b,m}$.

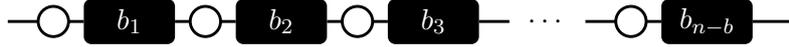
\begin{figure}[t]
  \begin{center}
\begin{tikzpicture}  
\draw[-, line width=.1em] (-0.2,0) -- (-0.6,0);      
\draw[-, line width=.1em] (0.2,0) -- (0.4,0);      
\draw[-, line width=.1em] (1.6,0) -- (1.8,0);
\draw[-, line width=.1em] (2.2,0) -- (2.4,0);      
\draw[-, line width=.1em] (3.6,0) -- (3.8,0);
\draw[-, line width=.1em] (4.2,0) -- (4.4,0);      
\draw[-, line width=.1em] (5.6,0) -- (6,0);

\draw[-, line width=.1em] (7,0) -- (7.4,0);
\draw[-, line width=.1em] (7.8,0) -- (8,0);
\draw[-, line width=.1em] (9.2,0) -- (9.8,0);
    
\draw[line width=.1em] (0,0) circle (0.2);               
\draw[line width=.1em] (2,0) circle (0.2);               
\draw[line width=.1em] (4,0) circle (0.2);               
\draw[line width=.1em] (7.6,0) circle (0.2);               
 
\fill[black, rounded corners=1mm] (.4, 0.3) rectangle (1.6, -0.3);    
\fill[black, rounded corners=1mm] (2.4, 0.3) rectangle (3.6, -0.3);    
\fill[black, rounded corners=1mm] (4.4, 0.3) rectangle (5.6, -0.3);    
\fill[black, rounded corners=1mm] (8, 0.3) rectangle (9.2, -0.3);    
\draw (1, 0) node[text=white]{$b_1$}; 
\draw (3, 0) node[text=white]{$b_2$}; 
\draw (5, 0) node[text=white]{$b_3$}; 
\draw (6.5, 0) node{$\cdots$};  
\draw (8.6, 0) node[text=white]{$b_{n-b}$}; 
 \end{tikzpicture}  
  \end{center} 
    
  \caption{If one distributes $b= b_1 + b_2 + \ldots + b_{n-b}$ black vertices into $n-b$ boxes in such a way that no box remains empty, one obtains a color assignment pattern $\mathbf{b}=(b_1,\ldots,b_{n-b}) \in \mathcal{B}_{b,n-b}$ of the cycle graph which contributes to $C_{n,b,0}$. By shifting the whole pattern to the left, where periodic boundary conditions are applied, one obtains $1+b_1$ different color assignments before there is again a white vertex on the leftmost position. 
  }
  \label{fig:SLRing_m0}
\end{figure}

Let us treat the easy case, $m=0$, first.
If $b=0$, all vertices must be white and we obtain the trivial sector length $A_0 = C_{n,0,0}=1$, which is fixed by normalization.
However, if there is at least one black vertex, $b\ge 1$,  the graph-theoretical problem from Thrm.~\ref{thrm:puzzle} from the main text can be restated into  \emph{``$C_{n,b,0}$ is the number of color assignments of the $n$-vertex cycle graph with exactly $b$ black vertices such that each white vertex has zero white neighbors''} because there are no white paths of length $l\ge 2$. 
To achieve this condition,  $b$ black vertices have to be distributed among the $n-b$ gaps between the white vertices, cf. 
Fig.~\ref{fig:SLRing_m0}.
The resulting set of possible patterns is given by 
\begin{align}\label{eq:SLRing_setB}
\mathcal{B}_{b,w} = \left\{ (b_1, \ldots, b_{w}) \in \ZZ^{w} \ \Bigg \vert \ \ \sum_{i=1}^{w} b_i = b, \  b_i\ge 1 \right \},
\end{align}
where $w=n-b$. 
We will need the notation introduced in Eq.~\eqref{eq:SLRing_setB} for general $w\ge1$ at a later stage. 
Via repeated shifts to the left, one obtains  $1+b_1$ different color assignments for each pattern $\mathbf{b}=(b_1,b_2, \ldots, b_{w})\in \mathcal{B}_{b,w}$. An additional shift would result in a color assignment which is already covered by the pattern  $(b_2,\ldots, b_{w},b_1)\in \mathcal{B}_{b,w} $. Therefore, we find
\begin{align} \label{eq:SLRing_Cnb0}
C_{n,b,0} = \sum_{\mathbf{b} \in \mathcal{B}_{b,n-b}} (1+b_1)  . 
\end{align}
By elementary combinatorics, there are $\vert \mathcal{B}_{b,w} \vert = \binom{b-1}{w-1}$ possibilities to distribute $b$ unlabelled balls into $w$ labelled boxes such that no box remains empty.
After a little algebra, we find the formula 
(needed later in full generality)
\begin{align} \label{eq:SLRing_sumB}
\sum_{\mathbf{b} \in \mathcal{B}_{b,w}} (x + b_1 y) = \binom{b-1}{w-1} \left(x+ \frac{by}{w} \right),
\end{align}
which holds for any  choice of $x,y\in \RR $ and integers $b\ge 0$, $m\ge1$.
Setting $w=n-b$ and $x=y=1$, we obtain 
$C_{n,b,0}= \frac{n}{b}\binom{b}{n-b}$
which appears as the first term in Eq.~\eqref{eq:Ak_RCL}. 

\begin{figure}[t]
  \begin{center}
\begin{tikzpicture}  
\draw[-, line width=.1em] (-0.2,0) -- (-0.8,0);      
\draw[-, line width=.1em] (0.2,0) -- (0.4,0);       
\draw[-, line width=.1em] (1.2,0) -- (1.4,0);       
\draw[-, line width=.1em] (1.8,0) -- (2.2,0);       
\draw[-, line width=.1em] (4.4,0) -- (4.8,0);
\draw[-, line width=.1em] (5.2,0) -- (5.4,0);       
\draw[-, line width=.1em] (6.2,0) -- (6.4,0);       
\draw[-, line width=.1em] (6.8,0) -- (7.1,0);        
\draw[-, line width=.1em] (7.9,0) -- (8.2,0);
\draw[-, line width=.1em] (8.6,0) -- (8.8,0);       
\draw[-, line width=.1em] (9.6,0) -- (9.8,0);       
\draw[-, line width=.1em] (10.2,0) -- (10.6,0);       
\draw[-, line width=.1em] (12.8,0) -- (13.4,0);       
    
\draw[line width=.1em] (0,0) circle (0.2);               
\draw (0.8, 0) node{$\cdots$}; 
\draw (0.8, -1) node{$l_1$};
\draw[line width=.1em] (1.6,0) circle (0.2);                
\draw[line width=.1em] (5,0) circle (0.2);
\draw (5.8, 0) node{$\cdots$}; 
\draw (5.8, -1) node{$l_2$};
\draw[line width=.1em] (6.6,0) circle (0.2);                
\draw (7.5, 0) node{$\cdots$};  
\draw[line width=.1em] (8.4,0) circle (0.2);
\draw (9.2, 0) node{$\cdots$};
\draw (9.2, -1) node{$l_m$}; 
\draw[line width=.1em] (10,0) circle (0.2);                
      
\draw[black, line width=.1em,  fill=Qlila, rounded corners=1mm] (2.2, 0.3) rectangle (4.4, -0.3);     
\draw (3.3, 0) node[text=LavenderBright]{$b_1+w_1$}; 
\draw[black, line width=.1em,  fill=Qlila, rounded corners=1mm] (10.6, 0.3) rectangle (12.8, -0.3);     
\draw (11.7, 0) node[text=LavenderBright]{$b_m+w_m$};

\draw[line width=.1em, decoration={brace, mirror, raise=0.5cm}, decorate] (-.3,0) -- (1.9,0);
\draw[line width=.1em, decoration={brace, mirror, raise=0.5cm}, decorate] (4.7,0) -- (6.9,0);
\draw[line width=.1em, decoration={brace, mirror, raise=0.5cm}, decorate] (8.1,0) -- (10.3,0);
 \end{tikzpicture}  
\vspace{8mm}
 
 \begin{tikzpicture}
\draw[black, line width=.1em, fill=Qlila, rounded corners=1mm] (-3.8, 0.3) rectangle (-1.6, -0.3);     
\draw (-2.7, 0) node[text=LavenderBright]{$b_i+w_i$};  
\draw[-, line width=.1em] (-4.2,0) -- (-3.8,0);
\draw[-, line width=.1em] (-1.6,0) -- (-1.2,0);      

\draw (-0.5,0) node{=};
 
\draw[-, line width=.1em] (0.2,0) -- (0.4,0);      
\draw[-, line width=.1em] (1.6,0) -- (1.8,0);
\draw[-, line width=.1em] (2.2,0) -- (2.4,0);      
\draw[-, line width=.1em] (3.6,0) -- (3.8,0);
\draw[-, line width=.1em] (4.2,0) -- (4.4,0);      
\draw[-, line width=.1em] (5.6,0) -- (6,0);

\draw[-, line width=.1em] (7,0) -- (7.4,0);
\draw[-, line width=.1em] (7.8,0) -- (8,0);
\draw[-, line width=.1em] (9.2,0) -- (9.8,0);
     
\draw[line width=.1em] (2,0) circle (0.2);               
\draw[line width=.1em] (4,0) circle (0.2);               
\draw[line width=.1em] (7.6,0) circle (0.2);               
 
\fill[black, rounded corners=1mm] (.4, 0.3) rectangle (1.6, -0.3);    
\fill[black, rounded corners=1mm] (2.4, 0.3) rectangle (3.6, -0.3);    
\fill[black, rounded corners=1mm] (4.4, 0.3) rectangle (5.6, -0.3);    
\fill[black, rounded corners=1mm] (8, 0.3) rectangle (9.2, -0.3);    
\draw (1, 0) node[text=white]{$b_{i,1}$}; 
\draw (3, 0) node[text=white]{$b_{i,2}$}; 
\draw (5, 0) node[text=white]{$b_{i,3}$}; 
\draw (6.5, 0) node{$\cdots$};  
\draw (8.6, 0) node[text=white]{$b_{i,w_i+1}$}; 

 \end{tikzpicture} 
  \end{center} 
    
\caption{Top: A pattern  contributing to $C_{n,b,m}$ consists of $m$ white paths of length $l_i \ge 2$,  which  are separated by mixed paths of length $b_i+w_i$ as depicted in blue.
By shifting the whole pattern to the left, where periodic boundary conditions are applied, one obtains $l_1+b_1+w_1$ different color assignments before there is again a white path (of length $l_2$) starting at the leftmost position. Bottom: In the $i$th mixed path, each of the $w_i$ white vertices needs two black neighbors. There are $\vert \mathcal{B}_{b_i,w_i+1} \vert = \binom{b_i -1}{w_i} $ choices to distribute the $b_{i}  = b_{i,1}+\ldots+ b_{i,w_i+1}$ black vertices into $w_i+1$ boxes such that no box remains empty.
  }
  \label{fig:SLRing_m1}
\end{figure}
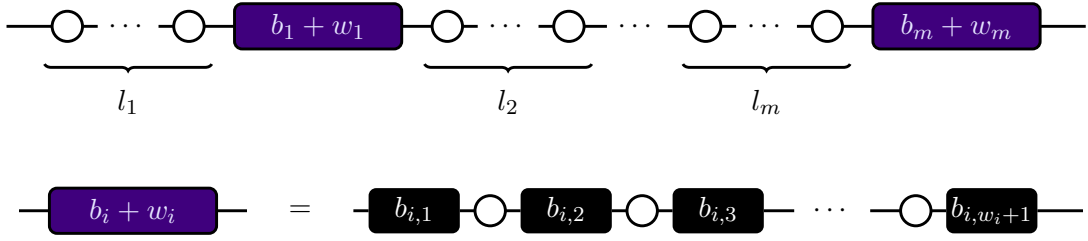

Now, we solve $C_{n,b,m}$ for $m\ge1 $ white paths of length $l_1,\ldots, l_m \ge 2$.
For each color assignment, the total number $l= l_1+\ldots + l_m$ of white vertices in the $m$ paths is somewhere in between $2m$ and $n-b$.
The number $w$ of isolated white vertices, i.e., white vertices with two black neighbors, is fixed by the relation $w=n-l-b$.
As it is depicted in Fig.~\ref{fig:SLRing_m1},
these isolated white vertices are part of the mixed paths which separate the white paths of length $l_i\ge 2$ from each other. 
The mixed path, which separates the white paths of length $l_i$ from the white path of length $l_{i+1}$, consists of $b_i$ black and $w_i$ white vertices.
Since the vertices at the end of the mixed path have to be black and the isolated white vertices are separated by at least one black vertex, there are $\vert \mathcal{B}_{b_i,w_i+1}\vert = \binom{b_i -1}{w_i}$ different mixed paths consisting of $b_i$ black and $w_i$ white vertices.
In analogy to Eq.~\eqref{eq:SLRing_setB}, we introduce the sets
\begin{align}
\mathcal{L}_{l,m} &= \left\{ (l_1, \ldots, l_{m}) \in \ZZ^{m} \hspace{3.15mm} \ \Bigg \vert \ \ \sum_{i=1}^{m} l_i = l, \  l_i\ge 2 \right \} \\
\text{and } \ \ \mathcal{W}_{w,m} &= \left\{ (w_1, \ldots, w_{m}) \in \ZZ^{m} \ \Bigg \vert \ \ \sum_{i=1}^{m} w_i = w, \  w_i\ge 0 \right \}.
\end{align}
The set $\mathcal{L}_{l,m}$ contains all possible lengths for the white paths of a fixed combined length $l\in\{2m,\ldots, n-b\}$.
The set $\mathcal{W}_{w,m}$ is used to parameterize the possibilities of distributing the remaining $w=n-b-l$ isolated white vertices among the mixed paths. 
Since there are $\binom{b_i-1}{w_i}$ mixed paths for each choice of $b_i$ and $w_i$, we obtain $\prod_{i=1}^m \binom{b_i-1}{w_i}$ different mixed-chain color assignments for each choice of $\mathbf{b} = (b_1,\ldots, b_m)\in \mathcal{B}_{b,m}$, $\mathbf{w}=(w_1,\ldots, w_m)\in \mathcal{W}_{w,m}$.
In analogy to our argumentation around the derivation of $C_{n,b,0}$ in Eq.~\eqref{eq:SLRing_Cnb0}, each pattern $(\mathbf{l},\mathbf{b},\mathbf{w}) \in \bigcup_{l=2m}^{n-b} \mathcal{L}_{l,m}\times \mathcal{B}_{b,m} \times \mathcal{W}_{n-l-b,m}$ gives rise to exactly $l_1+b_1+w_1$ color assignments if the mixed-chain color assignment is fixed.
Combining all of our arguments, we obtain the equation
\begin{align} \label{eq:SLRing_Cnbn}
C_{n,b,m} &=  \sum_{l=2m}^{n-b} \sum_{\mathbf{l} \in \mathcal{L}_{l,m}} \sum_{\mathbf{b} \in \mathcal{B}_{b,m}} \sum_{\mathbf{w} \in \mathcal{W}_{n-l-b,m}}
(l_1+b_1+w_1) \prod_{i=1}^m \binom{b_i-1}{w_i}.
\end{align}
To simplify this expression, we make use of the well-known Vandermonde identity
\begin{align} \label{eq:Vandermonde}
\sum_{\mathbf{w} \in \mathcal{W}_{n-l-b,m}}
\prod_{i=1}^m \binom{b_i-1}{w_i} = \binom{b-m}{n-b-l}
\end{align}
as well as one of its generalizations~\cite[Eq.~(8)]{mestrovic_several_generalization_2018}
\begin{align}\label{eq:Vandermonde+}
\sum_{\mathbf{w} \in \mathcal{W}_{n-l-b,m}}
w_1 \prod_{i=1}^m \binom{b_i-1}{w_i}
= \binom{b-m}{n-b-l} \frac{(n-b-l)(b_1-1)}{b-m} .
\end{align} 
By combining Eqs.~\eqref{eq:SLRing_Cnbn}--\eqref{eq:Vandermonde+}, we obtain
\begin{align}
C_{n,b,m} 
=&
  \sum_{l=2m}^{n-b} \binom{b-m}{n-b-l}  \sum_{\mathbf{l} \in \mathcal{L}_{l,m}} \sum_{\mathbf{b} \in \mathcal{B}_{b,m}}  \left(
(l_1 +b_1) + \frac{(n-b-l)(b_1-1)}{b-m}  \right) \\ 
= &
\sum_{l=2m}^{n-b} \binom{b-m}{n-b-l}  \sum_{\mathbf{l} \in \mathcal{L}_{l,m}} \sum_{\mathbf{b} \in \mathcal{B}_{b,m}}  \left( \left(l_1- \frac{n-b-l}{b-m}\right) + b_1\left(1+  \frac{n-b-l}{b-m}\right)  \right) \\
 \overset{\eqref{eq:SLRing_sumB}}{=} &
 \binom{b-1}{m-1}
\sum_{l=2m}^{n-b} \binom{b-m}{n-b-l}  \sum_{\mathbf{l} \in \mathcal{L}_{l,m}} \left( l_1 -  \frac{n-b-l}{b-m} + \frac{b}{m}\left(1+  \frac{n-b-l}{b-m}\right)  \right) \\
=& 
 \binom{b-1}{m-1 }
\sum_{l=2m}^{n-b} \binom{b-m}{n-b-l}  \sum_{\mathbf{l} \in \mathcal{L}_{l,m}} \left( l_1 + \frac{n-l}{m}   \right).
\end{align}
Similar to Eq.~\eqref{eq:SLRing_sumB}, we can simplify the last term,
\begin{align}
 \sum_{\mathbf{l} \in \mathcal{L}_{l,m}} \left(  \frac{n-l}{m} +l_1  \right)  
 = \binom{l-m-1}{m-1} \left( \frac{n-l}{m}   + \frac{l}{m} \right) 
 = \hspace{3mm}  \binom{l-m-1}{m-1} \frac{n}{m} .
\end{align}
Substituting $b=k-2m$ together with an index shift $l \mapsto l-2m$ yields
\begin{align}
C_{n,k-2m,m} &= \frac{n}{m} \binom{k-2m-1}{m-1 } \sum_{l=0}^{n-k} \binom{k-3m}{n-k-l} \binom{l+m-1}{l}. 
\end{align}
Inserting this into~Eq.~\eqref{eq:SLRing_Ak_unsolved}, 
 we finally arrive at Eq.~\eqref{eq:Ak_RCL} from the main text.
Note that it suffices if $m$  runs from 1 to  $\left \lfloor \frac{k-1}{2} \right \rfloor $ 
because $\binom{k-2m-1}{m-1} =0 $ if $m\ge \frac{k}{2}$.

\section{Lower bounds on entanglement noise thresholds of graph states}
\label{app:noise_thresholds}
 In this appendix, we apply Cor.~\ref{cor:purity_criterion_threshold} from the main text to derive lower bounds on the local-white-noise entanglement threshold for several families of stabilizer states.
We start with the qubit case in App.~\ref{app:qubit_robustness} and discuss higher-dimensional qudits in  App.~\ref{app:qudit_robustness}.

\subsection[Robustness of entanglement in qubit graph states against local noise]{Robustness of entanglement in qubit graph states against local white noise}
\label{app:qubit_robustness}

\begin{figure}[t]

\begin{flushright}
\begin{minipage}{0.03\textwidth}
\flushright
\vspace{-2em}

$p_\text{lb}$
\end{minipage}
\begin{minipage}{.95\textwidth}
\centering
\includegraphics[height = 12em]{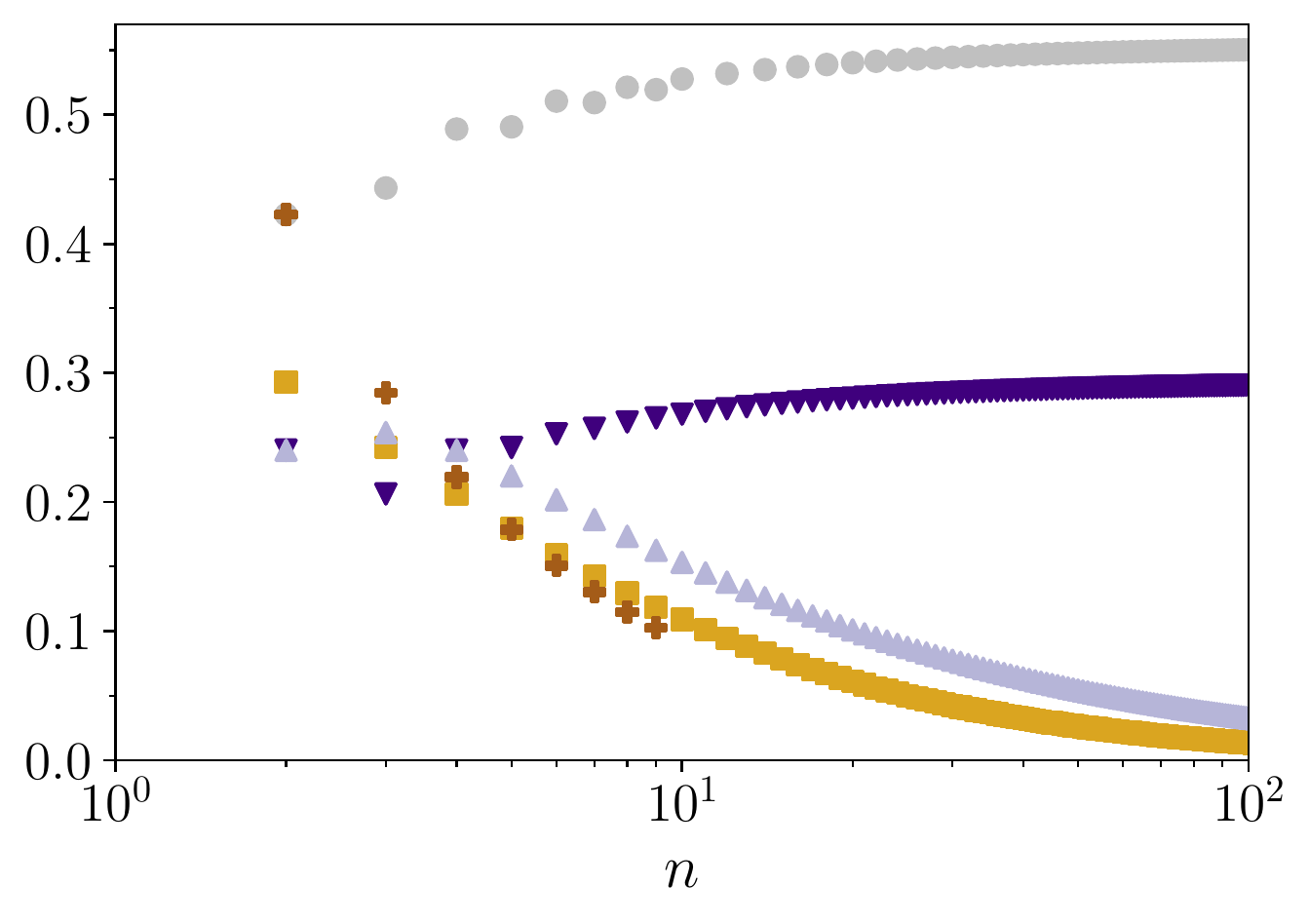}
\hspace{2mm}
\includegraphics[height = 12em]{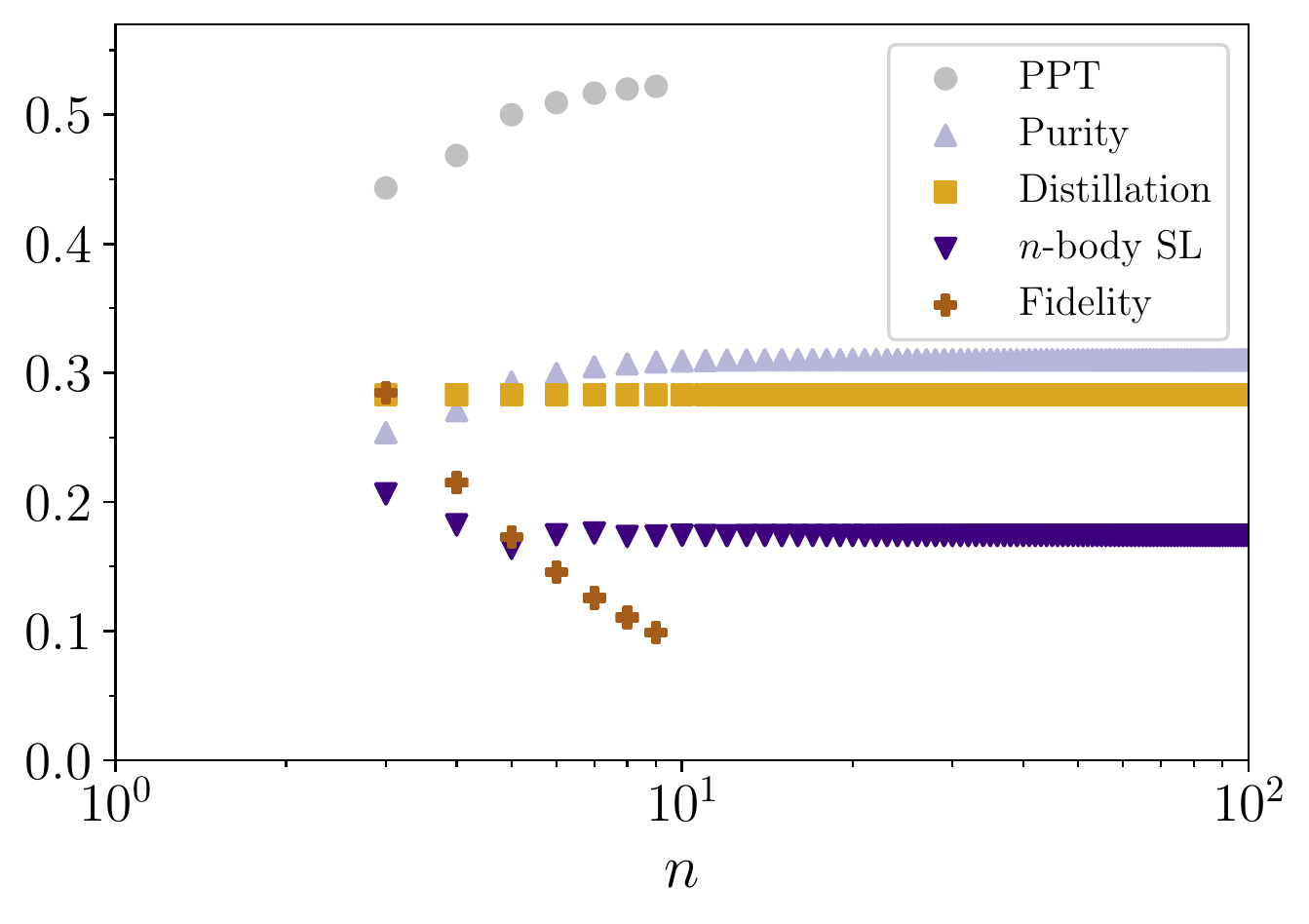} 
\end{minipage}
\end{flushright}
\caption{Lower bounds $p_\text{lb}$ on the local-white-noise entanglement threshold
$p_\text{crit}$
for $n$-qubit GHZ (left) and $\RCL$ states (right).
The noisy state $\mathcal{E}^{(p)}_\text{loc} \left[ \ket{\psi}\bra{\psi}\right ] $,
see Eq.~\eqref{eq:loc_white_noise},
is entangled iff $p< p_\text{crit}$, where $\ket{\psi}=\ket{\GHZn}$ and $\ket{\psi} = \ket{\RCLn}$, respectively.
}
\label{fig:Noise_thresholds}
\end{figure}

In Sec.~\ref{sec:3} of the main text, we show that the normalized SLD of a typical graph state is very close to a binomial distribution.
To cover both a special case and the generic case,  in Fig.~\ref{fig:Noise_thresholds} we plot for $n$-qubit GHZ (left) and $\RCL$ states (right) 
several lower bounds $p_\text{lb}$ on the local-white-noise entanglement threshold as a function of $n$.
Note that the largest value of $p_\text{lb}$ in Fig.~\ref{fig:Noise_thresholds} corresponds to the strongest entanglement criterion for a given state.
We see that the PPT criterion (gray circles) outperforms the other criteria in all cases for which it is available.
Whenever $n$ is even, the PPT criterion yields the lower bound,
\begin{align}  \label{eq:threshold_GHZ_local_PPT_qubit}
    p^{\GHZn}_\text{PPT,loc} =  1  - \frac{1}{\sqrt{2^{2-2/n}+1}},
\end{align}
on the entanglement noise threshold for $n$-qubit GHZ states~\cite{simon_robustness_of_2002}.
For GHZ states with odd $n\le 9$, we compute the PPT bound using a direct approach with exponential runtime.
To our knowledge, a result similar to Eq.~\eqref{eq:threshold_GHZ_local_PPT_qubit} is not available for $\RCL$ states; our direct approach yields $p^{\RCLn}_\text{PPT,loc}$ for all $n\in\{3,4,\ldots, 9\}$.
For $n\ge 10$, the best available lower bound on the noise threshold for $\ket{\RCLn}$ is based on the 
purity criterion (lavender upward triangles) and found via Cor.~\ref{cor:purity_criterion_threshold}.
Interestingly, the same criterion performs comparatively badly for GHZ states.
For them, the second best criterion (after PPT) is given by the $n$-body SL criterion (blue downward triangles) from Eq.~\eqref{eq:nSL_bound} of the main text.
For comparison, we also plot the bound (yellow squares),
\begin{align} \label{eq:loc_thresh_dist}
 p^\text{graph}_\text{distill,loc}  = 1- {2^{-{2}/\Big(2+ \max\limits_{\{i,j\}\in E} \{\deg(i)+\deg(j) \} \Big)}},
\end{align}
which is based on an entanglement distillation protocol for graph states $\ket{\Gamma}$ whose set of edges is denoted by $E$~\cite{hein_entanglement_properties_2005}. 
Finally, we determine the value of $p$ for which the fidelity of the noisy state with the target state, $\ket{\psi}$, is equal to $0.5$.
If the noise parameter is below this value (brown plusses), the noisy state is GME since the operator $W=\frac{\mathbbm 1}{2}- \ket{\psi}\bra{\psi}$ is a GME witness~\cite{terhal_bell_inequalities_2000, guehne_detection_of_2002, bourennane_experimental_detection_2004, guehne_entanglement_detection_2009}. 
As this direct approach also has exponential runtime, we can apply this fidelity criterion only for $n\le 9$.

In contrast to the unphysical case of global white noise from Eq.~\eqref{eq:threshold_stab_global_PPT},
all entanglement criteria considered in Fig.~\ref{fig:Noise_thresholds}
yield physically meaningful local-noise thresholds that are smaller than 1,
e.g., 
$\lim\limits_{n \rightarrow \infty} p^{\GHZn}_\text{PPT,loc}    = 1- 1/\sqrt{5}\approx 0.553$
and
$\lim\limits_{n \rightarrow \infty} p^{\GHZ}_{n\text{SL,loc}}  = 1- 1/\sqrt{2}\approx 0.293$.
The latter is larger than 
$\lim\limits_{n \rightarrow \infty} p^\RCLn_\text{nSL,loc} \approx 0.174$
because $A_n^{\GHZn} = 2^{n-1}+\delta_{n,\text{even}}$ exceeds
\begin{align}\label{eq:an_rcl}
    A_n^{\RCLn} = 1 + \sum_{k=1}^{\lfloor n/3 \rfloor } \binom{n-2k-1}{k-1} \frac{n}{k}.
\end{align}
Note that Eq.~\eqref{eq:an_rcl} is a simplified special case of Eq.~\eqref{eq:Ak_RCL} of the main text.
%
Next, we point out that for qubits, Thrm.~\ref{thrm:purity_criterion} simplifies to:
\begin{align}
    \sum_{k=0}^{\lfloor n/2\rfloor} (n- 2k) A_k[\rho] < \sum_{k=\lceil n/2\rceil}^n (2k-n) A_k[\rho]
    \hspace{3mm}
    \Longrightarrow
    \hspace{3mm}
    \rho \text{ is entangled}.
\end{align}
In other words, $\rho$ is entangled if the (slightly rescaled) $k$-body SLs with $k>n/2$ outperform those with $k<n/2$, which is intuitive because $A_k$ quantifies $k$-body correlations and entanglement is a strong form of correlation. 
For the GHZ state, the most important contribution to the $k$-body SLs with $k>n/2$ is $A_n$.
We attribute the observation that  $p_\text{pur,loc}^\GHZn $ from Cor.~\ref{cor:purity_criterion_threshold} converges to zero in Fig.~\ref{fig:Noise_thresholds} 
to the fact that  $A_n$ declines extremely fast, recall Eq.~\eqref{eq:Ak_p_loc}.
For $\RCL$ states, on the other hand, the normalized SLD is close to a binomial distribution centered at $3n/4$. 
Since the $k$-body SL decline around $k=3n/4$ is not as severe as for $k\approx n$, this causes the corresponding lower bound from Cor.~\ref{cor:purity_criterion_threshold} to numerically converge to
$\lim\limits_{n \rightarrow \infty} p^\RCLn_\text{pur,loc} \approx 0.310$,
which to our knowledge is the best available lower bound on the noise threshold for $\RCL$ states.\footnote{%
The minor differences between the two distributions in Fig.~\ref{fig:RCL_vs_binom}
that are discussed in Sec.~\ref{sec:2.3} 
would lead to an overestimation of the SL-based noise thresholds:
For a hypothetical state with $\mathbf{a}^{\text{hypo}} = \mathbf{b}(p=3/4)$, we would find 
$\lim_{n \rightarrow \infty} p^\text{hypo}_\text{pur,loc} \approx 0.423$ and
$\lim_{n \rightarrow \infty} p^\text{hypo}_{n\text{SL,loc}} \approx 0.184$.
For this reason, the approximation $\mathbf{a}^{\RCLn} \approx \mathbf{b}(3/4)$ and its analogues for other generic graph states should only be used with care.
}
It is closely followed by $p_\text{distill,loc}^\RCLn \approx 0.283$, which is constant because the degrees of the vertices in a cycle graph $C_n$ are independent of $n$~\cite{hein_entanglement_properties_2005}.
For the star graph $K_{1, n-1}$, on the other hand,
the degree of the central vertex is unbounded, which causes $p_\text{distill,loc}^\GHZn $ to converge to zero~\cite{hein_entanglement_properties_2005}.

Finally, consider the bound that is based on the fidelity criterion. 
Since this criterion can be used to certify GME, it is often employed in experiments~\cite{gong_genuine_12_2019, wei_verifying_multipartite_2020, mooney_generation_and_2021, song_generation_of_2019}.
As we can see in Fig.~\ref{fig:Noise_thresholds}, the level of noise that is required to successfully apply the fidelity criterion in such an experiment 
decreases in $n$.
This is bad news for benchmarkers of quantum processors in which errors are accurately modeled by local white noise because the qubits have to become less noisy (in the next generation of the quantum processor with an increased number of qubits) to verify GME via fidelity measurements.
This is very demanding for near-term quantum hardware.
The $n$-body SL for the verification of (possibly biseparable) entanglement, on the other hand, 
has critical noise thresholds well above ten percent, independent of the number of qubits.

\subsection[Robustness of entanglement in qudit graph states against local noise]{Robustness of entanglement in qudit graph states against local white noise}
\label{app:qudit_robustness}

For the sake of completeness, let us also apply the results developed in this paper to the general case of qudits in dimension $d\ge 2$.
The only bound on the local-white-noise threshold of an $n$-qudit state that we were able to find in the literature applies to the $n$-qudit GHZ state,
as defined in Eq.~\eqref{eq:GHZ_qudit}.
It is given by
\begin{align}\label{eq:threshold_GHZ_local_PPT_qudit}
p^{\text{GHZ}_d(n)}_\text{PPT,loc} 
=\frac{2d}{\sqrt[n]{4}+\sqrt[n]{2}\sqrt{4+\sqrt[n]{2d}}+2d}
\end{align} 
and holds for arbitrary $d$ and even $n$~\cite{liu_decay_of_2009}.
Note that Eq.~\eqref{eq:threshold_GHZ_local_PPT_qubit} is a special case of Eq.~\eqref{eq:threshold_GHZ_local_PPT_qudit}. 
\begin{figure}[t]

\begin{flushright}
\begin{minipage}{0.06\textwidth}
\flushright
\vspace{-2em}

$p_\text{lb}$
\end{minipage}
\begin{minipage}{.92\textwidth}
\centering
\includegraphics[height = 12em]{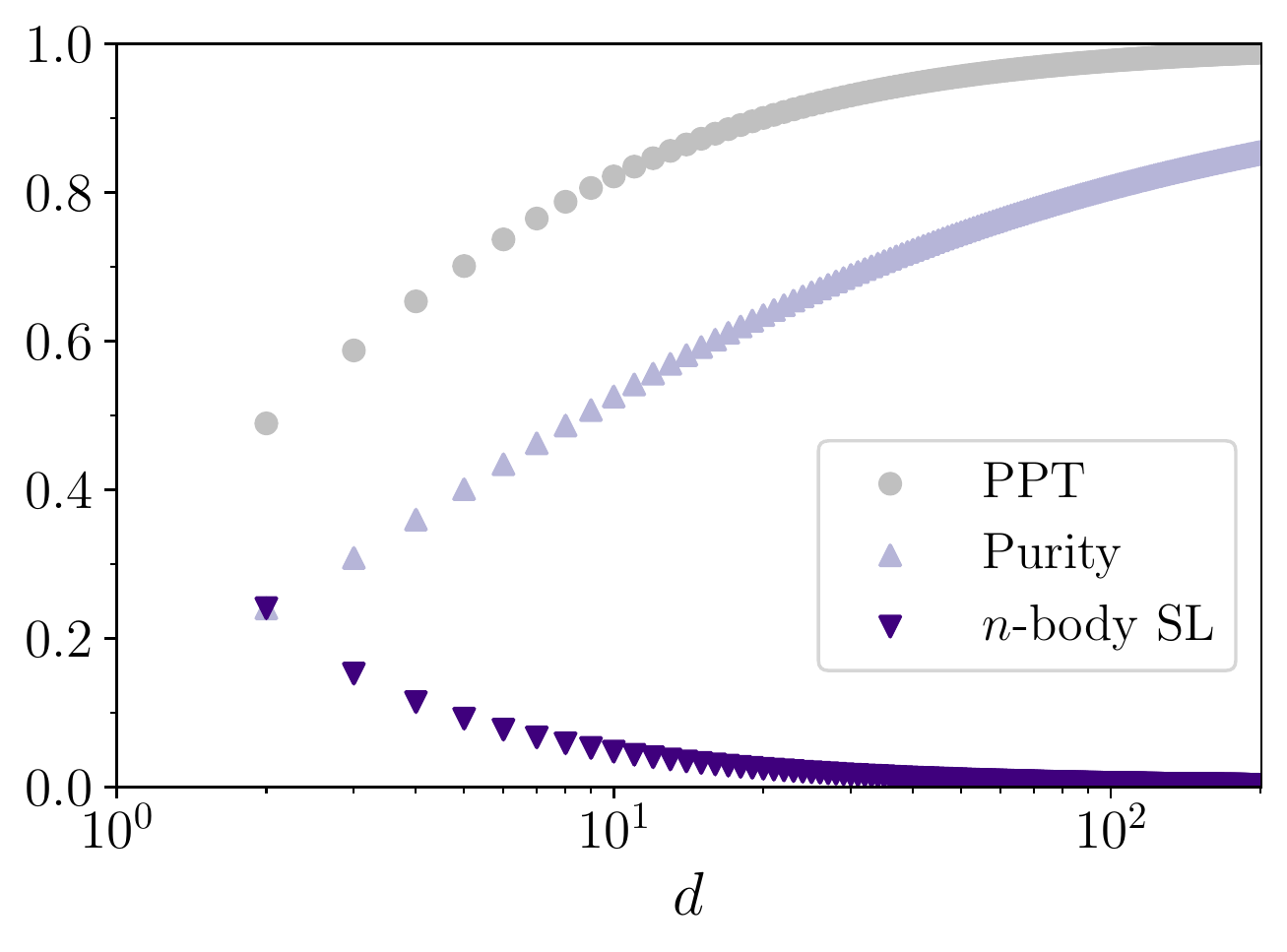}
\hspace{2mm}
\includegraphics[height = 12em]{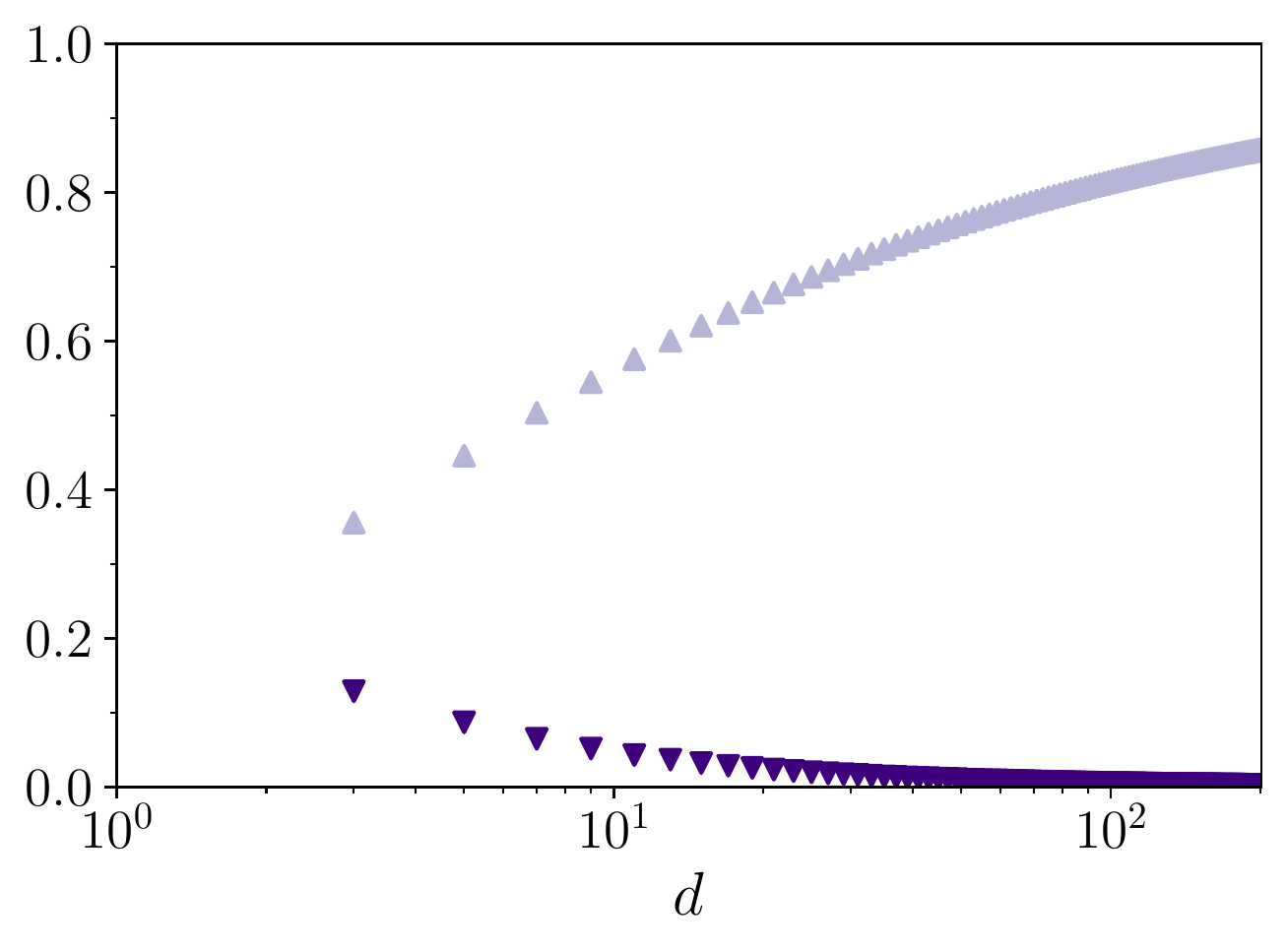} 
\end{minipage}
\end{flushright}
\caption{Lower bounds $p_\text{lb}$ on the local-white-noise entanglement threshold $p_\text{crit}$
for four-qudit GHZ (left) and $\AME$ states (right) as a function of the qudit dimension.
The noisy state $\mathcal{E}^{(p)}_\text{loc} \left[ \ket{\psi}\bra{\psi}\right]$
is entangled iff $p< p_\text{crit}$, where $\ket{\psi}=\ket{\GHZ_d(4)}$ or $\ket{\psi} = \ket{\text{RC}_d(4)}$.
  }
  \label{fig:Noise_thresholds_qudits}
\end{figure}
As we show in the left panel of Fig.~\ref{fig:Noise_thresholds_qudits} for the example of $n=4$ qubits, 
the PPT bound (gray circles) converges to 1 in the limit of $d\rightarrow \infty $.
The same trend is recovered by the bound based on the purity criterion (lavender upward triangles), which is obtained by applying Cor.~\ref{cor:purity_criterion_threshold} for the SLD of $\ket{\GHZ_d(n)}$ from Eq.~\eqref{eq:Ak_GHZ_qudit}.
We also display the bound based on the $n$-body SL criterion from Eq.~\eqref{eq:nSL_bound} of the main text (blue downward triangles), which converges to zero; this is consistent with the previous observation of the discrepancy between entanglement and $A_n$ in the qudit case~\cite{eltschka_maximum_nbody_2020}.
As in the qubit case, the PPT criterion always leads to the best lower bound on the entanglement noise threshold. 

For states where the PPT criterion is not solved, our SL-based approaches still work provided the SLD of the investigated state is known.
Consider, for example, the four-qudit AME state $\ket{\text{RC}_d(4)}$,
which is defined for odd $d$ in Eq.~\eqref{eq:ame4d}.
We present its SLD in Eq.~\eqref{eq:sld_ame4d} of App.~\ref{app:qudits} 
and plot the corresponding lower bounds on the local-white-noise threshold in the right panel of Fig.~\ref{fig:Noise_thresholds_qudits}.
As in the case of four-qudit GHZ states, we find that the bound based on the $n$-body SL criterion decreases, whereas 
the purity bound increases in $d$.
This demonstrates the usefulness of Cor.~\ref{cor:purity_criterion_threshold} in the case of higher-dimensional qudits.

\section{SLDs of $W$ states as an example of the non-stabilizer case}
\label{app:W}
   In the main text, we almost exclusively discuss SLDs of stabilizer states.
Since such states constitute only a finite subset of the $2^n$-dimensional state space of an $n$-qubit system, 
not all of our results apply in the general case.
In this appendix, 
we highlight some important differences using the example of $W$ states~\cite{duer_three_qubits_2000}.
The $n$-qubit $W$ state is defined as 
\begin{align} \label{eq:W_definition}
    \ket{\Wn} =  \frac{1}{\sqrt{n}} \sum_{i=1}^n\ket{\mathbf{e}_i},
\end{align}
where $\mathbf{e}_1=(1,0,\ldots,0)$, $\ldots$, $\mathbf{e}_n=(0,\ldots, 0,1)$ is the standard basis of $\FF_2^n$.
For $n\ge 3$, the $W$ state is not a (Pauli) stabilizer state.
While the definition of the $k$-body SL in Eq.~\eqref{def:Ak} is applicable to any $n$-qubit state, only in the case of a pure stabilizer state is $A_k$ guaranteed to be an integer.
The $k$-body SL of $\ket{\Wn}$, on the other hand,
is given by
\begin{align} \label{eq:Ak_W}
A_k^{\Wn} =  \binom{n}{k} \left( 1 + \frac{4k}{n^2} (2k-n-1)   \right),
\end{align}  
see Eq.~(18) of Ref.~\cite{aschauer_local_invariants_2004}
for the more general case of Dicke states. 
For $n\ge 3$, it can happen that $A_k^{\Wn}$ is not an integer, e.g., $\mathbf{A}^{W(5)} = (1, 1.8, 3.6, 10, 11.4, 4.2)$.
However, there are also non-stabilizer states for which the SLD takes integer values, e.g., $\mathbf{A}^{W(4)} = (1,1,3,7,4)$.
Interestingly, $\ket{W(4)}$ has exactly the same SLD as the only  stabilizer state (up to LU-equivalence)  for which $A_1=I=1$, namely $\ket{+}\otimes \ket{\text{GHZ}(3)}$.

For pure stabilizer states, we establish in Eq.~\eqref{eq:a1} that $A_1^\text{stab}$ is the number of qubits that are disentangled from all other qubits.
In particular, $A_1^\text{stab}\ge1$ implies that the stabilizer state is separable.
The $W$ state, however, has a 1-body SL of $A_1^{\Wn} = {(n-2)^2}/{n}$.
In particular, we have $A_1^{\Wn}\ge 1$ 
for all $n\ge4$, despite 
$\ket{\Wn}$ being GME.

\begin{figure}[t]
\centering
\includegraphics[width = .75\textwidth]{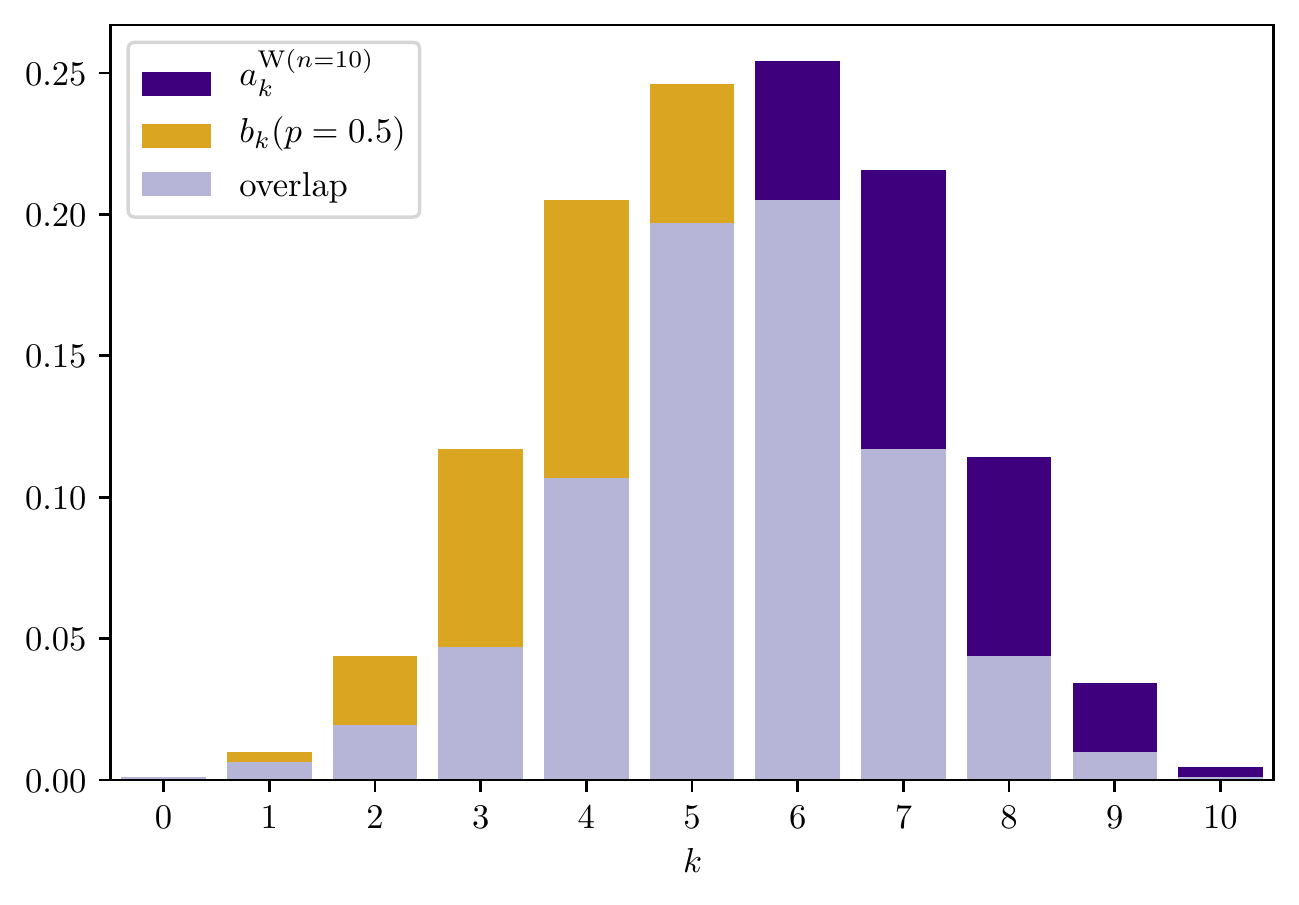}
\caption{Comparison of the normalized SLD $a_k^\Wn =2^{-n} A_k^\Wn $ of the $W$ state from Eq.~\eqref{eq:Ak_W} to the symmetric binomial distribution $b_k(p=0.5)=\binom{n}{k}2^{-n}$  for $n=10$.
The latter coincides with the normalized SLD of a fully separable state, i.e., $\mathbf{b}(p=0.5)=\mathbf{a}^{\fullysepn}$.
  }
  \label{fig:W_vs_binom}
\end{figure}
Note that the mean $\langle k \rangle_{\mathbf{a}^{\Wn}} = (n^2+2n-4)/2n $ can still be inferred from $A_1^{\Wn}$ via Eq.~\eqref{eq:purity1} from the main text because MacWilliams identities hold for arbitrary pure states.
While GME stabilizer states obey
$\langle k \rangle_{\mathbf{a}^{\text{stab}}} /n = 3/4$ for all $n$,
we find $\langle k \rangle_{\mathbf{a}^{\Wn}} /n \rightarrow 1/2$ for $n\rightarrow\infty$.
In Fig.~\ref{fig:W_vs_binom}, we plot the normalized SLD $\mathbf{a}^{\Wn}$ (blue) and the symmetrical binomial distribution $\mathbf{b}$ (yellow) with $b_k = \frac{1}{2^n} \binom{n}{k}$ for the example of $n=10$ qubits.
Although the mean $\langle k \rangle_{\mathbf{a}^{W(10)}} = 5.9$ of $\mathbf{a}^{\Wn}$ is still notably larger than 
$\langle k \rangle_{\mathbf{b}} = 5$,
the two distributions exhibit a considerable overlap (lavender).
In the main text, Sec.~\ref{sec:3.1}, we plot the TVD between $\mathbf{a}^{\Wn}$ and $\mathbf{b}$ as a function of $n$ (lavender curve in Fig.~\ref{fig:TVD_VS_n}).
There, we observe that the curve decreases with $n$.
This is unsurprising because the difference of the normalized Bloch vectors of $\ket{W(n)}$ and $\ket{0}^{\otimes n}$ converges to zero in the limit of $n\rightarrow \infty$; see supplemental material of Ref.~\cite{flammia_direct_fidelity_2011} for the Bloch vector components of $\ket{W(n)}$.

Let us rigorously show that $\TVD(\mathbf{a}^{\Wn}, \mathbf{b})$ converges to zero.
For simplicity, we write $\mathbf{a} = \mathbf{a}^{\Wn}$.
From Eq.~\eqref{eq:Ak_W}, we can see that $ a_k \ge b_k$  is equivalent to $2k \ge n+1$.
Thus, we can split the sum in Eq.~\eqref{eq:TVD} to avoid absolute values, which yields
\begin{align} \label{eq:tvd_w_1}
    \TVD(\mathbf{a}, \mathbf{b})  
&    = \frac{1}{2}\left( \sum_{k=0}^{\left\lfloor \tfrac{n}{2}\right\rfloor-1 } (b_k -a_k) + \sum_{k= \left\lfloor \tfrac{n}{2}\right\rfloor+1}^n  (a_k - b_k )  \right)  + \frac{b_{n/2}-a_{n/2}}{2} \delta_{n,\text{even}} .
\end{align}
After substituting $k\mapsto n-k$ in the second sum and exploiting the fact that all terms of the form $b_k - b_{n-k}$ vanish,
we can rewrite Eq.~\eqref{eq:tvd_w_1} as
\begin{align} \label{eq:tvd_w_2}
\TVD(\mathbf{a}, \mathbf{b})  
&= \frac{1}{2}  \sum_{k=0}^{\left\lfloor \tfrac{n}{2}\right\rfloor-1 } (a_{n-k} - a_k )  + \frac{1}{2^n n} \binom{n}{\tfrac{n}{2}} \delta_{n,\text{even}},
\end{align}
where we also used $a_{n/2}-b_{n/2}= \tfrac2{2^n n} \binom{n}{n/2} $ for the case of $n$ even.
Next, we find
\begin{align}
    a_{n-k} - a_k  = \frac{n-1}{2^{n-2} n^2} \binom{n}{k} (n-2k)
\end{align}
by exploiting Eq.~\eqref{eq:Ak_W}.
In combination with $\sum_{k=0}^{\left\lfloor {n}/{2}\right\rfloor-1}  (n-2k)
    = \left\lfloor \frac{n}{2}\right\rfloor  \binom{n}{\left\lfloor {n}/{2}\right\rfloor}$,
this yields
\begin{align} \label{eq:tvd_w_3}
    \TVD(\mathbf{a}, \mathbf{b})  
&    =  \frac{n-1}{2^{n-1}n^2}
    \left\lfloor \frac{n}{2}\right\rfloor  \binom{n}{\left\lfloor \tfrac{n}{2}\right\rfloor}
  + \frac{1}{2^n n} \binom{n}{\tfrac{n}{2}} \delta_{n,\text{even}} .
\end{align}
In the case of even $n$, 
Stirling's formula allows us to rewrite 
Eq.~(\ref{eq:tvd_w_3})
as
\begin{align}
    \TVD(\mathbf{a}, \mathbf{b}) 
  = 
  \hspace{3mm} 
  \frac1{2^{n}}\binom{n}{\frac n2}
\hspace{3mm} \propto \hspace{3mm} 
\frac{1}{\sqrt{n}}
\hspace{3mm}
\overset{n\rightarrow \infty }{ \xrightarrow{\hspace*{1cm}} }
\hspace{3mm}
 0.
\end{align}
Similarly, Eq.~\eqref{eq:tvd_w_3} simplifies to
$\TVD(\mathbf{a}, \mathbf{b}) = \frac{n^2-1}{2^{n} n^2}\binom{n}{(n-1)/2}
\propto 1/\sqrt{n}$
if $n$ is odd.
This establishes that $\TVD(\mathbf{a}^{\Wn},\mathbf{b}(p=0.5))$ converges to zero as $1/\sqrt{n}$.
Recall from Fig.~\ref{fig:TVD_VS_n}, that we have strong numerical evidence that $\TVD(\mathbf{a}^{\RCLn},\mathbf{b}(p=0.75))$ features the same behavior.
Conducting a more detailed study of such convergence effects could be a worthwhile endeavor as it may further strengthen our understanding of SLDs.
   
\section{Expected SLD for random graph states}  
\label{app:sld_random}
Here, we derive Eq.~\eqref{eq:Ak_random_unsolved_3} from the main text.
The expected $k$-body SL of an Erdős-Rényi graph state with $n$ vertices and edge-probability $q$
is given by
\begin{align} \label{eq:Ak_random_unsolved}
    \langle A_k \rangle_q = \sum_{\Gamma \in \mathcal{G} } \Pr[\Gamma_n^{(q)}= \Gamma] A_k \left [\ket{\Gamma}\bra{\Gamma}\right],
\end{align}
where $\mathcal{G}\subset \FF_2^{n \times n}$ denotes the set of all adjacency matrices.
Inserting Eq.~\eqref{eq:Ak_coloring} into Eq.~\eqref{eq:Ak_random_unsolved} allows us to write
\begin{align}\label{eq:Ak_random_unsolved1}
    \langle A_k \rangle_q = \sum_{b=0}^k \sum_{\mathbf{r}\in \mathcal{B}_b} \sum_{\Gamma \in \mathcal{G} } \Pr[\Gamma_n^{(q)}= \Gamma]  \delta_{\swt(\mathbf{r},\Gamma \mathbf{r}), k}
\end{align}
as a sum over all color assignments $\mathbf{r}\in \FF_2^n$ with an increasing number $b\le k$ of black vertices.
As $\Pr[\Gamma_n^{(q)}= \Gamma]$ is invariant under renumeration of the vertices of $\Gamma$, we can replace $\mathbf{r}$ by the color assignment $\mathbf{r}_b=(1,\ldots,1,0,\ldots,0)$ for which the first $b$ vertices are black, while the other $n-b$ vertices are white. 
Since there are $\vert \mathcal{B}_b\vert = \binom{n}{b}$ color assignments with exactly $b$ black vertices, we can restate Eq.~\eqref{eq:Ak_random_unsolved1} as
\begin{align} \label{eq:Ak_random_unsolved_2}
       \langle A_k \rangle_q = \sum_{b=0}^k \binom{n}{b} p(n,q,b,k),
\end{align}
where 
$
p (n,q,b,k) =  \sum_{\Gamma \in \mathcal{G} } \Pr[\Gamma_n^{(q)}= \Gamma] \delta_{\swt(\mathbf{r}_b, \Gamma\mathbf{r}_b),k}
$
denotes the probability that a graph with $b$ black vertices has exactly $n-k$ white vertices with an even number of black neighbors; recall Thrm.~\ref{thrm:puzzle}.
For any given white vertex, the probability of having an even number of black neighbors is given by 
\begin{align} \label{eq:p_even}
    p_\text{even}(q,b) = \sum_{\substack{a = 0\\a \text{ even}}}^b \binom{b}{a} q^a (1-q)^a = \frac{1+(1-2q)^b}{2}
\end{align}
because every edge (between the given white vertex and any of the black vertices) is present with probability $q$.
Since this probability is independently the same for each of the $n-b$ white vertices,
the probability that exactly $n-k$ of them have the desired property follows as
\begin{align} \label{eq:p_contribute}
    p(n,q,b,k) = \binom{n-b}{n-k} p_\text{even}(n,q,b)^{n-k}(1-p_\text{even}(n,q,b))^{k-b}.
\end{align}
Inserting Eqs.~\eqref{eq:p_even} and~\eqref{eq:p_contribute} into Eq.~\eqref{eq:Ak_random_unsolved_2} yields Eq.~\eqref{eq:Ak_random_unsolved_3} from the main text.

\section{Graph-theoretical treatment of SLDs of qudit graph states}
\label{app:qudits}
In this appendix, we discuss what information about the SLD of a qudit graph state $\ket{\Gamma}$, as defined in Eq.~\eqref{eq:graph_state_qudit} of the main text,
one can directly infer from the graph $\Gamma$.
Since the stabilizer group 
\begin{align}
    \mathcal{S} = \left\{ X_d^\mathbf{r} Z_d^{\Gamma \mathbf{r}} \omega_d^{\sum_{i<j} r_i \gamma_{i,j} r_j } \ \bigg \vert \ \mathbf{r} \in (\ZdZ)^n   \right\}
\end{align}
of $\ket{\Gamma}$ is parameterized by the color assignments $\mathbf{r}\in (\ZdZ)^n$ of $\Gamma$ (with $d$ colors), we can state 
\begin{align} \label{eq:Ak_graph_qudit}
A_k = \left\vert \left\{ \mathbf{r} \in (\ZdZ)^n \ \vert \ \swt_d(\mathbf{r}, \Gamma\mathbf{r})=k  \right\} \right \vert,
\end{align}
which generalizes Eq.~\eqref{eq:Ak_graph} from the main text.
However, only if $d$ and $n$ are small enough, it is feasible to iterate through all $d^n$ color assignments to compute the SLD $\mathbf{A}=(A_0, \ldots, A_n)$ via Eq.~\eqref{eq:Ak_graph_qudit}.
Since only color assignments with $b\le k$ non-white ($r_i \neq 0$) vertices $i$ can contribute to $A_k$, we find for $n$-qudit graph states the bound
\begin{align} \label{eq:coarse_bound_qudit}
    A_k \le \sum_{b=1}^k \binom{n}{k} (d-1)^b,
\end{align}
which generalizes Eq.~\eqref{eq:coarse_bound} from the main text, where $k\ge1$.
Since SLDs are convex, Eq.~\eqref{eq:coarse_bound_qudit} also holds for mixtures of graph states.
If $d$ is prime, every stabilizer state is LU-equivalent to a graph state~\cite{bahramgiri_graph_states_2006}.
This implies the validity of the bound in Eq.~\eqref{eq:coarse_bound_qudit} for all mixtures of arbitrary $n$-qudit stabilizer states.

For qudits in prime dimension $d$, one can relate $A_1$ and $A_2$ to graph-theoretical notions:
The $1$-body SL is equal to the number of color assignments where exactly one vertex $j\in\{1,\ldots,n\}$ obeys $r_j\neq0$ and all other vertices $i\neq j$ fulfill $\gamma_{i,j}r_j=0$.
Here, $\ZdZ = \FF_d$ is a field. 
Thus, $\gamma_{i,j}r_j=0$ is equivalent to $\gamma_{i,j}=0$, i.e., $i$ is an isolated vertex.
Since there are $d-1$ choices for $r_j\neq0$, we find $A_1 = (d-1)I$, as mentioned in Eq.~\eqref{eq:a1_qudit_prime} of the main text.
For the $2$-body SL, two types of color assignments $\mathbf{r}\in \FF_d^n$ can contribute:
either one or two vertices are not colored white.
If there is exactly one vertex $j\in\{1,\ldots,n\}$ with $r_j\neq0$,
there has to be exactly one other vertex $i\neq j$ with $\gamma_{i,j}r_j\neq0$.
Since $\FF_d$ is a field, the latter is equivalent to $\gamma_{i,j}\neq 0$.
Therefore, such a color assignment contributes iff $j$ has exactly one neighbor, i.e., $j$ is a leaf.
This yields exactly $(d-1)L$ color assignments of the first type that contribute to $A_2$, where $L$ is the number of leaves.
For the second type of color assignments $\mathbf{r}\in \FF_d^n$, which have exactly two vertices $j\neq j'$ with $r_j,r_{j'}\neq 0$,
the property in Eq.~\eqref{eq:puzzle_qudit} simplifies to:
A color assignment $\mathbf{r}$ contributes to $A_2$ iff
\begin{align} \label{eq:puzzle_qudit_prime_A2}
    \gamma_{i,j} r_j = - \gamma_{i,j'} r_{j'}
\end{align}
for all $i\in\{1,\ldots,n\}\backslash \{j,j'\} $.
If $\gamma_{i,j}\neq 0$, for a given vertex $i$, then Eq.~\eqref{eq:puzzle_qudit_prime_A2} can only be fulfilled if $\gamma_{i,j'}\neq 0$ because $\FF_d$ is a field.
This implies that only \emph{twin pairs}, i.e., pairs of vertices $(j,j')$ with the same neighborhood, have the potential to contribute to $A_2$; note that $\gamma_{i,j} \neq \gamma_{i,j'}$ is allowed here.
We denote the number of twin pairs with exactly $m\in\{0,\ldots, n-2\}$ common neighbors  as 
\begin{align}
    T_m = \left\vert \left \{  \{j,j'\}\subset\{1,\ldots,n\} \ \Big \vert \  j\neq j', 
    m = \left\vert\left\{ i\in\{1,\ldots,n\}\backslash\{ j,j'\} \, \big \vert \, \gamma_{i,j}, \gamma_{i,j'}\neq 0   \right\} \right\vert
    \right\} \right\vert.
\end{align}
Every twin pair with exactly $m=0$ common neighbors, 
of which there are $T_0 = \binom{I}{2}$,
contributes with $(d-1)^2$ color assignments because Eq.~\eqref{eq:puzzle_qudit_prime_A2} is trivially fulfilled for every choice of $(r_j,r_{j'})\in (\ZdZ)^2$.
Every twin pair with exactly $m=1$ common neighbors contributes with exactly $d-1$ color assignments because one can freely pick $r_j \neq 0$ but $r_{j'} = - \gamma_{i,j} r_j /\gamma_{i,j'}$ is fully determined by the other parameters.
All of these considerations lead to
the lower bound on $A_2$ in Eq.~\eqref{eq:a2_qudit_prime_both_bounds} of the main text,
which is tight if $T_2=\ldots = T_{n-2}=0$, e.g., if $\Gamma$ does not feature any cycles of length $4$.
In general, however, $T_m \ge 0$ for $m\ge 2$.
Then, $r_{j'} = - \gamma_{i,j} r_j /\gamma_{i,j'}$ is again fully determined by $r_j$, where $i$ is one of the shared neighbors of $j$ and $j'$.
However, the color assignment $\mathbf{r}$ only contributes to $A_2$ if Eq.~\eqref{eq:puzzle_qudit_prime_A2} is fulfilled \emph{for all} common neighbors $i$ of $j$ and $j'$; this may or may not happen.
If it happens, there are again $d-1$ color assignments per valid twin pair.
This establishes the upper bound on $A_2$ in Eq.~\eqref{eq:a2_qudit_prime_both_bounds} of the main text.

Consider, for example, the qudit generalization $\ket{\text{RC}_d (n=4) }$
of the four-qubit ring cluster state for which the adjacency matrix is given by
\begin{align} \label{eq:ame4d}
\text{RC}_d(4) =
    \begin{bmatrix}0&+1&0&-1 \\ +1&0&+1&0\\ 0&+1&0&+1 \\ -1&0&+1&0 \end{bmatrix} .
\end{align} 
This graph has $I=0$ isolated vertices, $L=0$ leaves, $T_2=2$ twin pairs with two common neighbors, and $T_m=0$ twin pairs with $m\neq2$ common neighbors.
For every choice of $d\ge2$ (prime or not), all edges have invertible weights.
Hence, the only possibility for a color assignment $\mathbf{r}=(r_1,r_2,r_3,r_4)\in \ZdZ$ to contribute to $A_2^{\text{RC}_d(4)}$ is if
one pair of twin pairs is white, while the other is not; without loss of generality, $r_1,r_3\neq 0$ and $r_2=r_4=0$.
Additionally, the induced $Z$-operators on the qudits 2 and 4 must cancel, i.e., $r_1+r_3=0$ and $r_1-r_3=0$, which can only be solved if $d$ is even.
Therefore, $A_2^{\text{RC}_d(4)}=0$ for all odd $d$. 
Since $I=0$ implies that $A_1^{\text{RC}_d(4)}=0$ also, we recover the well-known fact 
that all $2$-body marginals of $\ket{\text{RC}_d(4)}$ are maximally mixed, i.e., $\ket{\text{RC}_d(4)}$ is  absolutely maximally entangled (AME)~\cite{helwig_absolutely_maximally_2013}.
In Prop.~11 of Ref.~\cite{miller_small_quantum_2019}, we computed the SLD of this AME state for arbitrary odd $d$,
\begin{align} \label{eq:sld_ame4d}
    \mathbf{A}^{\text{RC}_d(4)}= \left(1,\ 0,\ 0,\ 4(d^2-1),\  d^4- 4(d^2-1) -1 \right).
\end{align}

From now on, consider the case of a general $n$-qudit graph state for which the dimension $d$ is a composite number.
Then, $\ZdZ$ contains zerodivisors, which 
causes the SLD problem to change from being mainly graph-theoretical to mainly algebraic in nature.
For simplicity,
we limit our discussion to the case $k=1$, where only color assignments with $n-1$ white vertices contribute to $A_1$.
For the vertex $j$ which is not white, there are $d-1$ choices for $r_j\neq 0$ and, if $j$ is isolated, all of them contribute.
Thus, we obtain a lower bound for arbitrary $d$,
\begin{align} \label{eq:a1_qudit_lower_bound}
    A_1 \ge (d-1)I,
\end{align}
which is a generalization of Eq.~\eqref{eq:a1_qudit_prime}.
For composite $d$, the bound in Eq.~\eqref{eq:a1_qudit_lower_bound} is not necessarily tight, e.g., if $d=6$, $n=2$, and $\gamma_{1,2}=2$, then there are $I=0$ isolated vertices but the color assignment $\mathbf{r}=(3,0)$ still contributes to $A_1=2$ because $r_1\gamma_{1,2}=0$ modulo~6, and similarly for $\mathbf{r}=(0,3)$.
For an arbitrary $n$-qudit graph state, we find 
for general $d$,
\begin{align}
    A_1 = \sum_{i=1}^n \left\vert \left\{ r \in \ZdZ \ \big \vert\ r\neq 0 \text{ and } (r\gamma_{1,i},\ldots,r\gamma_{n,i}) = (0,\ldots,0)  \right \}\right \vert,
\end{align}
or, using notions from commutative algebra~\cite{eisenbud_commutative_algebra_1995},
\begin{align}
    A_1 = \sum_{i=1}^n
     \vert \text{Ann}_{\ZdZ} (\Gamma \mathbf{e}_i) \,\backslash \, \{0\} \vert ,
\end{align}
where $\text{Ann}_R(x)$ denotes the annihilator of an element $x\in M$ in a module $M$ over a ring $R$,
and $\mathbf{e}_i$ is the $i$-th standard basis element of $(\ZdZ)^n$, i.e., $\Gamma \mathbf{e}_i\in (\ZdZ)^n$ is the $i$-th column of the adjacency matrix $\Gamma$.
If $x\neq 0$,
then $\text{Ann}_R(x)$ cannot contain any invertible elements; we find $\vert \text{Ann}_{\ZdZ} (\Gamma \mathbf{e}_i) \vert  \le d - \varphi(d)$,
where $\varphi$ is Euler's totient function, i.e.,  $\varphi(d)$ is the number of invertible elements in $\ZdZ$.
This yields an upper bound 
for general $d$,
\begin{align} \label{eq:a1_qudit_improved_bound}
    A_1 \le I(d-1) + (n-I)(d-1-\varphi(d)).
\end{align}
Note that the second term in Eq.~\eqref{eq:a1_qudit_improved_bound} vanishes if $d$ is a prime. 
In this case, the bounds in Eq.~\eqref{eq:a1_qudit_lower_bound} and~\eqref{eq:a1_qudit_improved_bound} combine to the result stated in Eq.~\eqref{eq:a1_qudit_prime} of the main text.

\section{Proof of Theorem~\ref{thrm:purity_criterion}}  
\label{app:proof_purity}

In this appendix,
we derive a new family of SLD-based entanglement criteria.
The purity criterion \cite{horodecki_quantum_entanglement_2009}
states that every fully separable $n$-qudit state $\rho$  obeys
\begin{align}   \label{eq:maj_unsolved}
    \Tr[\rho^2] \leq \Tr\left[\Tr_J[\rho]^2 \right],
\end{align}  
where $J\subset \{ 1,\ldots, n\}$ is a subset of parties and $\Tr_J[\rho]$ is the reduced density matrix of the complementary system $J^\mathrm{C}=\{1,\ldots,n\}\backslash J$. 
Using the expansion in Eq.~\eqref{eq:PauliDecomposition_qudit}, the marginal state can be written as
\begin{align}
\Tr_J[\rho] &= \sum_{\mathbf{l} \in (\ZdZ)^m}  \frac{1}{d^n} \sum_{\mathbf{r},\mathbf{s} \in (\ZdZ)^n} \rho_{\mathbf{r},\mathbf{s}}  \bra{\mathbf{l}}_J X_d^\mathbf{r}Z_d^\mathbf{s} \ket{\mathbf{l}}_J \\
&= \frac{1}{d^n} \sum_{\mathbf{r},\mathbf{s} \in (\ZdZ)^n} \rho_{\mathbf{r},\mathbf{s}}  X_d^{\mathbf{r}\vert_J}Z_d^{\mathbf{s}\vert_J} \underbrace{\sum_{\mathbf{l} \in (\ZdZ)^m} \omega_d ^{\mathbf{s}\vert_ J \cdot \mathbf{l}}
\left \langle \mathbf{l} \  \Big\vert \  \mathbf{r}\vert_J+\mathbf{l}\right\rangle}_{= \delta_{\mathbf{r}\vert_J ,0}  \delta_{\mathbf{s}\vert_J ,0} d^m} \\
&= \frac{1}{d^{n-m}} \sum_ {\substack{\mathbf{r,s}\in (\ZdZ)^n\\  
 \forall j \in   J: \  r_j=s_j=0}} \rho_{\mathbf{r},\mathbf{s}} X_d^{\mathbf{r}\vert_J}Z_d^{\mathbf{s}\vert_J} ,
\end{align} 
where $\mathbf{r}\vert _J =(r_{j_1},\ldots,r_{j_m})$ denotes the restriction of $\mathbf{r}$ to $J=\{j_1,\ldots, j_m\}$ and likewise for $\mathbf{s}\vert _J$.
Consequently, the right-hand side of Eq.~\eqref{eq:maj_unsolved} takes the form
\begin{align}
    \Tr\left[\Tr_{J}[\rho]^2\right] =  \frac{1}{d^{n-m}}
\sum_ {\substack{\mathbf{r,s}\in (\ZdZ)^n\\  
 \forall j \in   J: \  r_j=s_j=0}} \vert \rho_{\mathbf{r},\mathbf{s}} \vert^2.
\end{align}
In the next step of our derivation, we sum up Eq.~\eqref{eq:maj_unsolved} for all choices of $J\subset\{1,\ldots,n\}$ with $\vert J \vert = m$.
The right-hand side becomes
\begin{align}
    \sum_{\substack{J\subset \{1,\ldots, n\}\\ \vert J \vert = m}} \Tr\left[ \Tr_J[\rho]^2 \right] &= 
    \frac{1}{d^{n-m}} 
    \sum_{\substack{J\subset \{1,\ldots, n\}\\ \vert J \vert = m}} 
\sum_ {\substack{\mathbf{r,s}\in (\ZdZ)^n\\  
 \forall j \in   J: \  r_j=s_j=0}} \vert \rho_{\mathbf{r},\mathbf{s}} \vert^2 \\
 &= 
    \frac{1}{d^{n-m}} \sum_{k=0}^{n-m}
    \sum_{\substack{J\subset \{1,\ldots, n\}\\ \vert J \vert = m}} 
\sum_ {\substack{\mathbf{r,s}\in (\ZdZ)^n\\  
 \forall j \in   J: \  r_j=s_j=0 \\ \swt_d(\mathbf{r},\mathbf{s})=k}} \vert \rho_{\mathbf{r},\mathbf{s}} \vert^2.
 \label{eq:derivation_maj}
\end{align}
For each choice of $k$ and $J$, there are
\begin{align}
\left\vert\left \{\mathbf{r,s}\in (\ZdZ)^n \ \big \vert \   
 \forall j \in   J: \  r_j=s_j=0, \swt_d(\mathbf{r},\mathbf{s})=k \right\} \right\vert 
 = \binom{n-m}{k}(d^2-1)^k
\end{align} 
terms in the inner sum of Eq.~\eqref{eq:derivation_maj}
as there are $\binom{n-m}{k}$ choices for $k$ indices $i\in \{1,\ldots,n\}\backslash J$ with $(r_i,s_i)\neq (0,0)$ and, for each such choice, there are $(d^2-1)^k$ choices for the values of the nonzero entries of $(\mathbf{r},\mathbf{s})$.
Hence, there are $\binom{n}{m} \binom{n-m}{k}(d^2-1)^k = \binom{n-k}{m} \binom{n}{k}(d^2-1)^k$ terms in the inner two sums of Eq.~\eqref{eq:derivation_maj} for each $k\in\{0,\ldots,n-m\}$.
In contrast, there are only $\binom{n}{k} (d^2-1)^k$ terms in the sum of Eq.~\eqref{def:Ak_qudit}. 
Due to our symmetrization, we thus find
\begin{align} \label{eq:symmetrization_rhs}
    \sum_{\substack{J\subset \{1,\ldots, n\}\\ \vert J \vert = m}} \Tr\left[ \Tr_J[\rho]^2 \right] 
    = \frac{1}{d^{n-m}} \sum_{k=0}^{n} \binom{n-k}{m} A_k,
\end{align}
where, again, we have used the convention $\binom{n-k}{m} = 0 $ for all $m > n-k$.
A similar calculation leads to
\begin{align} \label{eq:symmetrization_lhs}
\sum_{\substack{J\subset \{1,\ldots, n\}\\ \vert J \vert = m}} \Tr[\rho^2] 
=\binom{n}{m} 
\frac{1}{d^n} \sum_{k=0}^n A_k.
\end{align}
Since, for a fully separable state, the expression in Eq.~\eqref{eq:symmetrization_rhs} is always larger than the one in Eq.~\eqref{eq:symmetrization_lhs},
it follows that
every state $\rho$ with
\begin{align} \label{eq:maj_solved}
\sum_{k=0}^n \left(
\binom{n}{m} - d^{m} \binom{n-k}{m}
\right)A_k[\rho] >0
\end{align}
is entangled. 
As we explain next, we have numerical evidence that the criterion in Eq.~\eqref{eq:maj_solved} is strongest if $m=1$.
In this case, Eq.~\eqref{eq:maj_solved} simplifies to Thrm.~\ref{thrm:purity_criterion}.

Our numerical evidence is as follows.
We draw  $n$-qubit states 
at random from the Haar distribution.
Then, we compute its SLD via Eq.~\eqref{def:Ak}.
For every $m\in \{1,\ldots, n-1\}$, we compute the noise threshold below which Ineq.~\eqref{eq:maj_solved} is satisfied using a straightforward generalization of Cor.~\ref{cor:purity_criterion_threshold}.
We run this test for 1000 random per qubit numbers for all $n\in\{3,4,5,6,7,8\}$.
In every single case, we find that the noise threshold is a strictly decreasing function of $m$.


\newpage

\begin{thebibliography}{10}

\bibitem{shor_quantum_analog_1997}
Peter Shor and Raymond Laflamme.
\newblock ``{Quantum Analog of the MacWilliams Identities for Classical Coding
  Theory}''.
\newblock \href{https://dx.doi.org/10.1103/PhysRevLett.78.1600}{Phys. Rev.
  Lett. {\bf 78}, 1600}~(1997).

\bibitem{gottesman_phd_thesis_1997}
Daniel Gottesman.
\newblock ``{Stabilizer Codes and Quantum Error Correction}''~(1997).

\bibitem{scott_multipartite_entanglement_2004}
Andrew~J. Scott.
\newblock ``{Multipartite entanglement, quantum-error-correcting codes, and
  entangling power of quantum evolutions}''.
\newblock \href{https://dx.doi.org/10.1103/PhysRevA.69.052330}{Phys. Rev. A
  {\bf 69}, 052330}~(2004).

\bibitem{PhysRevA.106.062424}
Zahra Raissi, Adam Burchardt, and Edwin Barnes.
\newblock ``General stabilizer approach for constructing highly entangled graph
  states''.
\newblock \href{https://dx.doi.org/10.1103/PhysRevA.106.062424}{Phys. Rev. A
  {\bf 106}, 062424}~(2022).

\bibitem{horodecki_quantum_entanglement_2009}
Ryszard Horodecki, Pawe\l{} Horodecki, Micha\l{} Horodecki, and Karol
  Horodecki.
\newblock ``Quantum entanglement''.
\newblock \href{https://dx.doi.org/10.1103/RevModPhys.81.865}{Rev. Mod. Phys.
  {\bf 81}, 865}~(2009).

\bibitem{aschauer_local_invariants_2004}
Hans Aschauer, John Calsamiglia, Marc Hein, and Hans~J. Briegel.
\newblock ``Local invariants for multi-partite entangled states allowing for a
  simple entanglement criterion''.
\newblock \href{https://dx.doi.org/10.48550/arXiv.quant-ph/0306048}{Quantum
  Inf. Comput. {\bf 4}, 383}~(2004).

\bibitem{de_vincente_multipartite_entanglement_2011}
Julio~I. de~Vicente and Marcus Huber.
\newblock ``Multipartite entanglement detection from correlation tensors''.
\newblock \href{https://dx.doi.org/10.1103/PhysRevA.84.062306}{Phys. Rev. A
  {\bf 84}, 062306}~(2011).

\bibitem{kloeckl_characterizing_multipartite_2015}
Claude Kl\"ockl and Marcus Huber.
\newblock ``Characterizing multipartite entanglement without shared reference
  frames''.
\newblock \href{https://dx.doi.org/10.1103/PhysRevA.91.042339}{Phys. Rev. A
  {\bf 91}, 042339}~(2015).

\bibitem{tran_quantum_entanglement_2015}
Minh~Cong Tran, Borivoje Daki\ifmmode~\acute{c}\else \'{c}\fi{}, Fran\c{c}ois
  Arnault, Wies\l{}aw Laskowski, and Tomasz Paterek.
\newblock ``Quantum entanglement from random measurements''.
\newblock \href{https://dx.doi.org/10.1103/PhysRevA.92.050301}{Phys. Rev. A
  {\bf 92}, 050301}~(2015).

\bibitem{tran_correlations_between_2016}
Minh~Cong Tran, Borivoje Daki\ifmmode~\acute{c}\else \'{c}\fi{}, Wies\l{}aw
  Laskowski, and Tomasz Paterek.
\newblock ``Correlations between outcomes of random measurements''.
\newblock \href{https://dx.doi.org/10.1103/PhysRevA.94.042302}{Phys. Rev. A
  {\bf 94}, 042302}~(2016).

\bibitem{eltschka_maximum_nbody_2020}
Christopher Eltschka and Jens Siewert.
\newblock ``Maximum {$N$}-body correlations do not in general imply genuine
  multipartite entanglement''.
\newblock \href{https://dx.doi.org/10.22331/q-2020-02-10-229}{{Quantum} {\bf
  4}, 229}~(2020).

\bibitem{wyderka_characterizing_quantum_2020}
Nikolai Wyderka and Otfried Gühne.
\newblock ``Characterizing quantum states via sector lengths''.
\newblock \href{https://dx.doi.org/10.1088/1751-8121/ab7f0a}{J. Phys. A: Math.
  Theor. {\bf 53}, 345302}~(2020).

\bibitem{hein_multiparty_entanglement_2004}
Marc Hein, Jens Eisert, and Hans~J. Briegel.
\newblock ``Multiparty entanglement in graph states''.
\newblock \href{https://dx.doi.org/10.1103/PhysRevA.69.062311}{Phys. Rev. A
  {\bf 69}, 062311}~(2004).

\bibitem{hein_entanglement_in_2006}
Marc Hein, Wolfgang Dür, Jens Eisert, Robert Raussendorf, Maarten~Van den
  Nest, and Hans~J. Briegel.
\newblock ``{Entanglement in Graph States and its Applications}''~(2006).
\newblock
  url:~\href{https://doi.org/10.48550/arXiv.quant-ph/0602096}{doi.org/10.48550/arXiv.quant-ph/0602096}.

\bibitem{bouchet_recognizing_locally_1993}
André Bouchet.
\newblock ``{Recognizing locally equivalent graphs}''.
\newblock \href{https://dx.doi.org/10.1016/0012-365X(93)90357-Y}{Discrete Math.
  {\bf 114}, 75}~(1993).

\bibitem{vandennest_graphical_description_2004}
Maarten Van~den Nest, Jeroen Dehaene, and Bart De~Moor.
\newblock ``{Graphical description of the action of local Clifford
  transformations on graph states}''.
\newblock \href{https://dx.doi.org/10.1103/PhysRevA.69.022316}{Phys. Rev. A
  {\bf 69}, 022316}~(2004).

\bibitem{ji_the_lulc_2010}
Zhengfeng Ji, Jianxin Chen, Zhaohui Wei, and Mingsheng Ying.
\newblock ``{The LU-LC conjecture is false}''.
\newblock \href{https://dx.doi.org/10.48550/arXiv.0709.1266}{Quantum Inf Comput
  {\bf 10}, 97}~(2010).

\bibitem{tsimakuridze_graph_states_2017}
Nikoloz Tsimakuridze and Otfried Gühne.
\newblock ``{Graph states and local unitary transformations beyond local
  Clifford operations}''.
\newblock \href{https://dx.doi.org/10.1088/1751-8121/aa67cd}{J. Phys. A Math.
  Theor. {\bf 50}, 195302}~(2017).

\bibitem{miller_hardware_tailored_2022}
Daniel Miller, Laurin~E. Fischer, Igor~O. Sokolov, Panagiotis~Kl. Barkoutsos,
  and Ivano Tavernelli.
\newblock ``{Hardware-Tailored Diagonalization Circuits}''~(2022).
\newblock
  url:~\href{https://doi.org/10.48550/arXiv.2203.03646}{doi.org/10.48550/arXiv.2203.03646}.

\bibitem{huber_some_ulams_2018}
Felix Huber and Simone Severini.
\newblock ``{Some Ulam's reconstruction problems for quantum states}''.
\newblock \href{https://dx.doi.org/10.1088/1751-8121/aadd1e}{J. Phys. A: Math.
  Theor. {\bf 51}, 435301}~(2018).

\bibitem{huber_bounds_on_2018}
Felix Huber, Christopher Eltschka, Jens Siewert, and Otfried Gühne.
\newblock ``Bounds on absolutely maximally entangled states from shadow
  inequalities, and the quantum {MacWilliams} identity''.
\newblock \href{https://dx.doi.org/10.1088/1751-8121/aaade5}{J. Phys. A Math.
  Theor. {\bf 51}, 175301}~(2018).

\bibitem{greenberger_going_beyond_1989}
Daniel~M. Greenberger, Michael~A. Horne, and Anton Zeilinger.
\newblock ``{Going Beyond Bell's Theorem}''.
\newblock {Bell's Theorem, Quantum Theory, and Conceptions of the Universe, M.
  Kafatos (Ed.), Kluwer, Dordrecht, 69, ISBN:978-90-481-4058-9}~(1989).
\newblock
  url:~\href{https://doi.org/10.48550/arXiv.0712.0921}{doi.org/10.48550/arXiv.0712.0921}.

\bibitem{miller_graphstatevis_interactive_2021}
Matthias Miller and Daniel Miller.
\newblock ``{GraphStateVis: Interactive Visual Analysis of Qubit Graph States
  and their Stabilizer Groups}''.
\newblock \href{https://dx.doi.org/10.1109/QCE52317.2021.00057}{IEEE Trans.
  Quantum Eng. {\bf 1}, 378}~(2021).

\bibitem{jungnitsch_entanglement_witnesses_2011}
Bastian Jungnitsch, Tobias Moroder, and Otfried G\"uhne.
\newblock ``{Entanglement witnesses for graph states: General theory and
  examples}''.
\newblock \href{https://dx.doi.org/10.1103/PhysRevA.84.032310}{Phys. Rev. A
  {\bf 84}, 032310}~(2011).

\bibitem{raussendorf_measurement_based_2003}
Robert Raussendorf, Daniel~E. Browne, and Hans~J. Briegel.
\newblock ``Measurement-based quantum computation on cluster states''.
\newblock \href{https://dx.doi.org/10.1103/PhysRevA.68.022312}{Phys. Rev. A
  {\bf 68}, 022312}~(2003).

\bibitem{briegel_measurement_based_2009}
Hans~J. Briegel, Daniel~E. Browne, Wolfgang Dür, Robert Raussendorf, and
  Maarten~Van den Nest.
\newblock ``{Measurement-based quantum computation}''.
\newblock \href{https://dx.doi.org/10.1038/nphys1157}{Nat. Phys. {\bf 5},
  19}~(2009).

\bibitem{miller_small_quantum_2019}
Daniel Miller.
\newblock ``{Small quantum networks in the qudit stabilizer formalism}''.
\newblock {Master's Thesis}~(2019).
\newblock
  url:~\href{https://doi.org/10.48550/arXiv.1910.09551}{doi.org/10.48550/arXiv.1910.09551}.

\bibitem{royle_graph_that_2020}
Gordon Royle.
\newblock ``{Graph that minimizes the number of b/w colorings where white
  vertices have an odd number of black}''.
\newblock
  url:~\href{https://mathoverflow.net/q/376673}{mathoverflow.net/q/376673}.
\newblock Accessed on 22.06.2022.

\bibitem{erdos_on_the_1960}
Paul Erdős and Alfred Rényi.
\newblock ``On the evolution of random graphs''.
\newblock \href{https://dx.doi.org/10.1515/9781400841356.38}{Publ. Math. Inst.
  Hungary. Acad. Sci. {\bf 5}, 17}~(1960).

\bibitem{knill_non_binary_1996}
Emanuel Knill.
\newblock ``{Non-binary unitary error bases and quantum codes}''.
\newblock {LANL report LAUR-96-2717}~(2019).
\newblock
  url:~\href{https://doi.org/10.48550/arXiv.quant-ph/9608048}{doi.org/10.48550/arXiv.quant-ph/9608048}.

\bibitem{gheorghiu_standard_form_2014}
Vlad Gheorghiu.
\newblock ``Standard form of qudit stabilizer groups''.
\newblock
  \href{https://dx.doi.org/https://doi.org/10.1016/j.physleta.2013.12.009}{Physics
  Letters A {\bf 378}, 505}~(2014).

\bibitem{grassl_graphs_quadratic_2002}
Markus Grassl, Andreas Klappenecker, and Martin Rotteler.
\newblock ``Graphs, quadratic forms, and quantum codes''~(2002).
\newblock
  url:~\href{https://doi.org/10.48550/arXiv.quant-ph/0703112}{doi.org/10.48550/arXiv.quant-ph/0703112}.

\bibitem{bahramgiri_graph_states_2006}
Mohsen Bahramgiri and Salman Beigi.
\newblock ``{Graph States Under the Action of Local Clifford Group in
  Non-Binary Case}''~(2006).
\newblock
  url:~\href{https://doi.org/10.48550/arXiv.quant-ph/0610267}{doi.org/10.48550/arXiv.quant-ph/0610267}.

\bibitem{looi_tripartite_entanglement_2011}
Shiang~Yong Looi and Robert~B. Griffiths.
\newblock ``Tripartite entanglement in qudit stabilizer states and application
  in quantum error correction''.
\newblock \href{https://dx.doi.org/10.1103/PhysRevA.84.052306}{Phys. Rev. A
  {\bf 84}, 052306}~(2011).

\bibitem{kaszlikowski_quantum_correlation_2008}
Dagomir Kaszlikowski, Aditi Sen(De), Ujjwal Sen, Vlatko Vedral, and Andreas
  Winter.
\newblock ``{Quantum Correlation without Classical Correlations}''.
\newblock \href{https://dx.doi.org/10.1103/PhysRevLett.101.070502}{Phys. Rev.
  Lett. {\bf 101}, 070502}~(2008).

\bibitem{miller_propagation_of_2018}
Daniel Miller, Timo Holz, Hermann Kampermann, and Dagmar Bru\ss{}.
\newblock ``{Propagation of generalized Pauli errors in qudit Clifford
  circuits}''.
\newblock \href{https://dx.doi.org/10.1103/PhysRevA.98.052316}{Phys. Rev. A
  {\bf 98}, 052316}~(2018).

\bibitem{quek_exponentially_tighter_2022}
Yihui Quek, Daniel~Stilck França, Sumeet Khatri, Johannes~Jakob Meyer, and
  Jens Eisert.
\newblock ``{Exponentially tighter bounds on limitations of quantum error
  mitigation}''~(2022).
\newblock
  url:~\href{https://doi.org/10.48550/arXiv.2210.11505}{doi.org/10.48550/arXiv.2210.11505}.

\bibitem{graselli_quantum_cryprography_2020}
Federico Grasselli.
\newblock ``{Quantum Cryptography: From Key Distribution to Conference Key
  Agreement}''.
\newblock \href{https://dx.doi.org/10.1007/978-3-030-64360-7}{Springer Cham,
  Switzerland}. ~(2020).

\bibitem{peres_separability_criterion_1996}
Asher Peres.
\newblock ``{Separability Criterion for Density Matrices}''.
\newblock \href{https://dx.doi.org/10.1103/PhysRevLett.77.1413}{Phys. Rev.
  Lett. {\bf 77}, 1413}~(1996).

\bibitem{horodecki_separability_of_1996}
Michał Horodecki, Paweł Horodecki, and Ryszard Horodecki.
\newblock ``Separability of mixed states: necessary and sufficient
  conditions''.
\newblock \href{https://dx.doi.org/10.1016/S0375-9601(96)00706-2}{Phys. Lett. A
  {\bf 223}, 1}~(1996).

\bibitem{bombin_optimal_resources_2007}
Héctor Bombin and Miguel~A. Martin-Delgado.
\newblock ``{Optimal resources for topological two-dimensional stabilizer
  codes: Comparative study}''.
\newblock \href{https://dx.doi.org/10.1103/PhysRevA.76.012305}{Phys. Rev. A
  {\bf 76}, 012305}~(2007).

\bibitem{bravyi_correcting_coherent_2018}
Sergey Bravyi, Matthias Englbrecht, Robert König, and Nolan Peard.
\newblock ``Correcting coherent errors with surface codes''.
\newblock \href{https://dx.doi.org/10.1038/s41534-018-0106-y}{Npj Quantum Inf.
  {\bf 4}, 55}~(2018).

\bibitem{fowler_surface_codes_2012}
Austin~G. Fowler, Matteo Mariantoni, John~M. Martinis, and Andrew~N. Cleland.
\newblock ``{Surface codes: Towards practical large-scale quantum
  computation}''.
\newblock \href{https://dx.doi.org/10.1103/PhysRevA.86.032324}{Phys. Rev. A
  {\bf 86}, 032324}~(2012).

\bibitem{vandennest_finite_set_2005}
Maarten Van~den Nest, Jeroen Dehaene, and Bart De~Moor.
\newblock ``{Finite set of invariants to characterize local Clifford
  equivalence of stabilizer states}''.
\newblock \href{https://dx.doi.org/10.1103/PhysRevA.72.014307}{Phys. Rev. A
  {\bf 72}, 014307}~(2005).

\bibitem{cabello_compact_set_2009}
Ad\'an Cabello, Antonio~J. L\'opez-Tarrida, Pilar Moreno, and Jos\'e~R.
  Portillo.
\newblock ``Compact set of invariants characterizing graph states of up to
  eight qubits''.
\newblock \href{https://dx.doi.org/10.1103/PhysRevA.80.012102}{Phys. Rev. A
  {\bf 80}, 012102}~(2009).

\bibitem{nielsen_separable_states_2001}
Michael~A. Nielsen and Julia Kempe.
\newblock ``{Separable States Are More Disordered Globally than Locally}''.
\newblock \href{https://dx.doi.org/10.1103/PhysRevLett.86.5184}{Phys. Rev.
  Lett. {\bf 86}, 5184}~(2001).

\bibitem{hein_entanglement_properties_2005}
Marc Hein, Wolfgang D\"ur, and Hans~J. Briegel.
\newblock ``Entanglement properties of multipartite entangled states under the
  influence of decoherence''.
\newblock \href{https://dx.doi.org/10.1103/PhysRevA.71.032350}{Phys. Rev. A
  {\bf 71}, 032350}~(2005).

\bibitem{navascues2008convergent}
Miguel Navascu{\'e}s, Stefano Pironio, and Antonio Ac{\'\i}n.
\newblock ``A convergent hierarchy of semidefinite programs characterizing the
  set of quantum correlations''.
\newblock \href{https://dx.doi.org/10.1088/1367-2630/10/7/073013}{New J. Phys.
  {\bf 10}, 073013}~(2008).

\bibitem{adesso2007entanglement}
Gerardo Adesso and Fabrizio Illuminati.
\newblock ``Entanglement in continuous-variable systems: recent advances and
  current perspectives''.
\newblock \href{https://dx.doi.org/10.1088/1751-8113/40/28/S01}{J. Phys. A:
  Math. Theor. {\bf 40}, 7821}~(2007).

\bibitem{sun2012entanglement}
Qingqing Sun and M.~Suhail Zubairy.
\newblock ``{Entanglement Criteria for Continuous-Variable Systems. In: Cohen,
  L., Poor, H., Scully, M. (eds) Classical, Semi-classical and Quantum
  Noise.}''.
\newblock \href{https://dx.doi.org/10.1007/978-1-4419-6624-7_17}{Pages
  249--258}.
\newblock Springer US. New York, NY~(2012).

\bibitem{rossi_quantum_hypergraph_2013}
Matteo Rossi, Marcus Huber, Dagmar Bru{\ss}, and Chiara Macchiavello.
\newblock ``Quantum hypergraph states''.
\newblock \href{https://dx.doi.org/10.1088/1367-2630/15/11/113022}{New J. Phys.
  {\bf 15}, 113022}~(2013).

\bibitem{mestrovic_several_generalization_2018}
Romeo Meštrović.
\newblock ``{Several generalizations and variations of Chu-Vandermonde
  identity}''~(2018).
\newblock
  url:~\href{https://doi.org/10.48550/arXiv.1807.10604}{doi.org/10.48550/arXiv.1807.10604}.

\bibitem{simon_robustness_of_2002}
Christoph Simon and Julia Kempe.
\newblock ``Robustness of multiparty entanglement''.
\newblock \href{https://dx.doi.org/10.1103/PhysRevA.65.052327}{Phys. Rev. A
  {\bf 65}, 052327}~(2002).

\bibitem{terhal_bell_inequalities_2000}
Barbara~M. Terhal.
\newblock ``Bell inequalities and the separability criterion''.
\newblock \href{https://dx.doi.org/10.1016/S0375-9601(00)00401-1}{Phys. Lett. A
  {\bf 271}, 319}~(2000).

\bibitem{guehne_detection_of_2002}
Otfried G\"uhne, Philipp Hyllus, Dagmar Bru\ss{}, Artur Ekert, Maciej
  Lewenstein, Chiara Macchiavello, and Anna Sanpera.
\newblock ``Detection of entanglement with few local measurements''.
\newblock \href{https://dx.doi.org/10.1103/PhysRevA.66.062305}{Phys. Rev. A
  {\bf 66}, 062305}~(2002).

\bibitem{bourennane_experimental_detection_2004}
Mohamed Bourennane, Manfred Eibl, Christian Kurtsiefer, Sascha Gaertner, Harald
  Weinfurter, Otfried G\"uhne, Philipp Hyllus, Dagmar Bru\ss{}, Maciej
  Lewenstein, and Anna Sanpera.
\newblock ``{Experimental Detection of Multipartite Entanglement using Witness
  Operators}''.
\newblock \href{https://dx.doi.org/10.1103/PhysRevLett.92.087902}{Phys. Rev.
  Lett. {\bf 92}, 087902}~(2004).

\bibitem{guehne_entanglement_detection_2009}
Otfried Gühne and Géza Tóth.
\newblock ``Entanglement detection''.
\newblock \href{https://dx.doi.org/10.1016/j.physrep.2009.02.004}{Phys. Rep.
  {\bf 474}, 1}~(2009).

\bibitem{gong_genuine_12_2019}
Ming Gong, Ming-Cheng Chen, Yarui Zheng, Shiyu Wang, Chen Zha, Hui Deng,
  Zhiguang Yan, Hao Rong, Yulin Wu, Shaowei Li, Fusheng Chen, Youwei Zhao,
  Futian Liang, Jin Lin, Yu~Xu, Cheng Guo, Lihua Sun, Anthony~D. Castellano,
  Haohua Wang, Chengzhi Peng, Chao-Yang Lu, Xiaobo Zhu, and Jian-Wei Pan.
\newblock ``{Genuine 12-Qubit Entanglement on a Superconducting Quantum
  Processor}''.
\newblock \href{https://dx.doi.org/10.1103/PhysRevLett.122.110501}{Phys. Rev.
  Lett. {\bf 122}, 110501}~(2019).

\bibitem{wei_verifying_multipartite_2020}
Ken~X. Wei, Isaac Lauer, Srikanth Srinivasan, Neereja Sundaresan, Douglas~T.
  McClure, David Toyli, David~C. McKay, Jay~M. Gambetta, and Sarah Sheldon.
\newblock ``{Verifying multipartite entangled Greenberger-Horne-Zeilinger
  states via multiple quantum coherences}''.
\newblock \href{https://dx.doi.org/10.1103/PhysRevA.101.032343}{Phys. Rev. A
  {\bf 101}, 032343}~(2020).

\bibitem{mooney_generation_and_2021}
Gary~J. Mooney, Gregory A.~L. White, Charles~D. Hill, and Lloyd C.~L.
  Hollenberg.
\newblock ``{Generation and verification of 27-qubit
  Greenberger-Horne-Zeilinger states in a superconducting quantum computer}''.
\newblock \href{https://dx.doi.org/10.1088/2399-6528/ac1df7}{J. Phys. Commun.
  {\bf 5}, 095004}~(2021).

\bibitem{song_generation_of_2019}
Chao Song, Kai Xu, Hekang Li, Yu-Ran Zhang, Xu~Zhang, Wuxin Liu, Qiujiang Guo,
  Zhen Wang, Wenhui Ren, Jie Hao, Hui Feng, Heng Fan, Dongning Zheng, Da-Wei
  Wang, Haohua Wang, and Shi-Yao Zhu.
\newblock ``{Generation of multicomponent atomic Schrödinger cat states of up
  to 20 qubits}''.
\newblock \href{https://dx.doi.org/10.1126/science.aay0600}{Science {\bf 365},
  574}~(2019).

\bibitem{liu_decay_of_2009}
Zhao Liu and Heng Fan.
\newblock ``Decay of multiqudit entanglement''.
\newblock \href{https://dx.doi.org/10.1103/PhysRevA.79.064305}{Phys. Rev. A
  {\bf 79}, 064305}~(2009).

\bibitem{duer_three_qubits_2000}
Wolfgang D\"ur, Guifré Vidal, and J.~Ignacio Cirac.
\newblock ``Three qubits can be entangled in two inequivalent ways''.
\newblock \href{https://dx.doi.org/10.1103/PhysRevA.62.062314}{Phys. Rev. A
  {\bf 62}, 062314}~(2000).

\bibitem{flammia_direct_fidelity_2011}
Steven~T. Flammia and Yi-Kai Liu.
\newblock ``{Direct Fidelity Estimation from Few Pauli Measurements}''.
\newblock \href{https://dx.doi.org/10.1103/PhysRevLett.106.230501}{Phys. Rev.
  Lett. {\bf 106}, 230501}~(2011).

\bibitem{helwig_absolutely_maximally_2013}
Wolfram Helwig.
\newblock ``{Absolutely Maximally Entangled Qudit Graph States}''~(2013).
\newblock
  url:~\href{https://doi.org/10.48550/arXiv.1306.2879}{doi.org/10.48550/arXiv.1306.2879}.

\bibitem{eisenbud_commutative_algebra_1995}
David Eisenbud.
\newblock ``{Commutative Algebra with a View Toward Algebraic Geometry}''.
\newblock \href{https://dx.doi.org/10.1007/978-1-4612-5350-1}{Springer,
  Graduate Texts in Mathematics (volume 150)}. ~(1995).

\end{thebibliography}
\end{document}